\documentclass[longauth]{aa} %referee

\usepackage[varg]{txfonts}
\usepackage{graphicx}
\usepackage{hyperref}

\bibpunct{(}{)}{;}{a}{}{,} % to follow the A&A style

\begin{document}

\title{JWST/CEERS sheds light on dusty star-forming galaxies: Forming bulges, lopsidedness, and outside-in quenching at cosmic noon}

\author{Aur{\'e}lien Le Bail\inst{1} \and Emanuele Daddi\inst{1} \and David Elbaz\inst{1} \and Mark Dickinson\inst{2} \and Mauro Giavalisco\inst{3} \and Benjamin Magnelli\inst{1} \and Carlos G{\'o}mez-Guijarro\inst{1} \and Boris S. Kalita\inst{4,5,6} \and Anton M. Koekemoer\inst{7} \and Benne W. Holwerda\inst{8} \and Fr{\'e}d{\'e}ric Bournaud\inst{1} \and Alexander de la Vega\inst{9} \and Antonello Calabr{\`o}\inst{10} \and Avishai Dekel\inst{11} \and Yingjie Cheng\inst{3} \and Laura Bisigello\inst{12,13} \and Maximilien Franco\inst{14} \and Luca Costantin\inst{15} \and Ray A. Lucas\inst{7} \and Pablo G. P\'erez-Gonz\'alez\inst{15} \and Shiying Lu\inst{16,1,17} \and Stephen M.~Wilkins\inst{18,19} \and Pablo Arrabal Haro\inst{2} \and Micaela B. Bagley\inst{14} \and Steven L. Finkelstein\inst{14} \and Jeyhan S. Kartaltepe\inst{20} \and Casey Papovich\inst{21,22} \and Nor Pirzkal\inst{23} \and L. Y. Aaron Yung\inst{24}\thanks{NASA Postdoctoral Fellow}}

\institute{Universit{\'e} Paris-Saclay, Universit{\'e} Paris Cit{\'e}, CEA, CNRS, AIM, 91191, Gif-sur-Yvette, France \and NSF's National Optical-Infrared Astronomy Research Laboratory, 950 N. Cherry Ave., Tucson, AZ 85719, USA \and University of Massachusetts Amherst, 710 North Pleasant Street, Amherst, MA 01003-9305, USA \and Kavli Institute for the Physics and Mathematics of the Universe, The University of Tokyo, Kashiwa, 277-8583, Japan \and Kavli Institute for Astronomy and Astrophysics, Peking University, Beijing 100871, People's Republic of China \and Center for Data-Driven Discovery, Kavli IPMU (WPI), UTIAS, The University of Tokyo, Kashiwa, Chiba 277-8583, Japan \and Space Telescope Science Institute, 3700 San Martin Drive, Baltimore, MD 21218, USA \and Physics \& Astronomy Department, University of Louisville, 40292 KY, Louisville, USA \and Department of Physics and Astronomy, University of California, 900 University Ave, Riverside, CA 92521, USA \and INAF - Osservatorio Astronomico di Roma, via di Frascati 33, 00078 Monte Porzio Catone, Italy \and Racah Institute of Physics, The Hebrew University of Jerusalem, Jerusalem 91904, Israel \and Dipartimento di Fisica e Astronomia "G.Galilei", Universit\'a di Padova, Via Marzolo 8, I-35131 Padova, Italy \and INAF--Osservatorio Astronomico di Padova, Vicolo dell'Osservatorio 5, I-35122, Padova, Italy \and Department of Astronomy, The University of Texas at Austin, Austin, TX, USA \and Centro de Astrobiolog\'{\i}a (CAB), CSIC-INTA, Ctra. de Ajalvir km 4, Torrej\'on de Ardoz, E-28850, Madrid, Spain \and School of Astronomy and Space Science, Nanjing University, Nanjing, 210093, China \and Key Laboratory of Modern Astronomy and Astrophysics (Nanjing University), Ministry of Education, Nanjing, 210093, China \and Astronomy Centre, University of Sussex, Falmer, Brighton BN1 9QH, UK \and Institute of Space Sciences and Astronomy, University of Malta, Msida MSD 2080, Malta \and Laboratory for Multiwavelength Astrophysics, School of Physics and Astronomy, Rochester Institute of Technology, 84 Lomb Memorial Drive, Rochester, NY 14623, USA \and Department of Physics and Astronomy, Texas A\&M University, College Station, TX, 77843-4242 USA \and George P.\ and Cynthia Woods Mitchell Institute for Fundamental Physics and Astronomy, Texas A\&M University, College Station, TX, 77843-4242 USA \and ESA/AURA Space Telescope Science Institute \and Astrophysics Science Division, NASA Goddard Space Flight Center, 8800 Greenbelt Rd, Greenbelt, MD 20771, USA}

%\date{Received / Accepted}

\abstract
{We investigate the morphology and resolved physical properties of a sample of 22 IR-selected dusty star-forming galaxies at cosmic noon using the \textit{James Webb Space Telescope} NIRCam images obtained in the EGS field for the CEERS survey. The exceptional resolution of the NIRCam images allowed us to spatially resolve these galaxies up to $4.4\mu$m and identify their bulge or core even when very extinguished by dust.}
{The goal of this study is to obtain a better understanding of the formation and evolution of FIR-bright galaxies by spatially resolving their properties using JWST in order to look through the dust and bridge the gap between the compact FIR sources and the larger optical star-forming galaxies.}
{Based on red-green-blue images from the F115W, F200W, and F444W filters, we divided each galaxy into several uniformly colored regions, fit their respective SEDs, and measured physical properties. After classifying each region as star forming or quiescent, we assigned galaxies to three classes depending on whether active star formation is located in the core, in the disk, or in both.}
{(i) We find that the galaxies at a higher redshift tend to have a fragmented disk with a low core mass fraction. They are at an early stage of bulge formation. When moving toward a lower redshift, the core mass fraction increases, and the bulge growth is associated with a stabilization of the disk, which translates into less patches and clumps. The NIRCam data clearly point toward bulge formation in preexisting disks.
(ii) Lopsidedness is a very common feature of DSFGs. It has been wrongly overlooked for a long time and could have a major impact on the evolution of DSFGs. 
(iii) Twenty-three percent of the galaxies have a star-forming core embedded in a quiescent disk. They seem to be undergoing outside-in quenching, often facilitated by their strong lopsidedness inducing instabilities.
(iv) We show that half of our galaxies with star formation concentrated in their core are good sub-millimeter galaxy near-IR counterpart candidates, demonstrating that compact SMGs are usually surrounded by a larger, less obscured disk.
(v) Finally,  we found surprising evidence for clump-like substructures being quiescent or residing in quiescent regions.}
{This work demonstrates the major impact JWST/NIRCam has on understanding the complexity of the evolution of distant massive galaxies regarding bulge formation and quenching mechanisms.}

%\end{abstract}

\keywords{Galaxies: Bulges -- Formation -- Evolution -- Star-formation -- Structure}

\titlerunning{JWST view of DSFGs at cosmic noon}
\authorrunning{Le Bail et al.}

\maketitle
\nolinenumbers

\section{Introduction} \label{sec:intro}

%The majority of star-forming (SF) galaxies are observed to follow a correlation in the stellar mass ($M_{*}$) versus star formation rate (SFR) plane (e.g., \citealt{}). 
Until recently, the existence of the so-called galaxy main sequence (MS), a correlation that the majority of star-forming galaxies follow in the stellar mass ($M_{*}$) versus star formation rate (SFR) plane up to a redshift of three (e.g., \citealt{daddi_multiwavelength_2007,elbaz_reversal_2007,noeske_star_2007,schreiber_herschel_2015}) and its tight scatter have been interpreted as evidence that star formation in most galaxies is a fairly ordered process (\citealt{schreiber_star-forming_2020}). 
The ‘‘consensus’’ is that galaxies on the MS are forming stars in a quasi steady state inside gas-rich stellar disks (e.g., \citealt{sancisi_cold_2008,dekel_formation_2009}), whereas galaxies above the MS undergo a starburst driven by stochastic processes, such as major mergers, whose typical signature is compact star formation (e.g., \citealt{tacconi_submillimeter_2008}).

However, recent studies at $z \sim 1 - 3$ have shown that some massive ($M_{*} \geq 10^{11} M_{\odot}$) MS galaxies have a stellar distribution typical of late type galaxies but where the star formation only occurs in a compact nucleus (\citealt{elbaz_starbursts_2018,puglisi_main_2019,puglisi_sub-millimetre_2021, tadaki_rotating_2017, tadaki_structural_2020,franco_goods-alma_2020,gomez-guijarro_goods-alma_2022,jimenez-andrade_radio_2019,jimenez-andrade_vla_2021}). The origin of these compact star-forming sub-millimeter galaxies (SMGs) observed with the \textit{Atacama Large Millimeter Array} (ALMA) is yet to be fully understood. The three main scenarios to form the compact sub-millimeter nucleus are (1) gas fueled to the core via violent disk instabilities (VDI) and clump migration, (2) a starburst induced by a major merger, or (3) accretion  and/or minor mergers (e.g., \citealt{gomez-guijarro_goods-alma_2022-1}). These compact star-forming nuclei could be an indication of an early quenching phase (\citealt{puglisi_main_2019,franco_goods-alma_2020,puschnig_unveiling_2023}).

Apart from the compact nucleus, high-\textit{z} star-forming galaxies have been observed to have giant star-forming clumps (radius $\sim 1$kpc). The origin of these clumps has been investigated by many studies (\citealt{puschnig_unveiling_2023,fensch_role_2021,hodge_alma_2019,rujopakarn_alma_2019,mandelker_population_2014,wuyts_smoother_2012,elmegreen_supercloud_1994,elmegreen_molecular_1989}).  \cite{mandelker_population_2014} suggested that they could be defined as ex situ clumps originating from minor mergers; in that case they would be old and characterized by a low gas fraction and a low specific star-formation rate (sSFR). Or they could be defined as in situ clumps originating from VDI (e.g., \citealt{noguchi_early_1999, elmegreen_gravitational_2011}). In this case, they would be younger and star forming. A recent simulation showed that inducing enough VDI to form such long-lived giant clumps would require a gas fraction of at least 50\%  (\citealt{fensch_role_2021}). These clumps would then migrate toward the center of the galaxy, creating a strong gas inflow and triggering an evolution of the structure of the galaxy leading to a morphological evolution (\citealt{fensch_role_2021}). This scenario is also favored by some observations (\citealt{forster-schreiber_constraints_2011, guo_rest-frame_2012}). More recently, \cite{puschnig_unveiling_2023} studied a local galaxy as a proxy for high-\textit{z} galaxies, confirming that the giant star-forming clumps mostly originate from a fragmentation of the disk induced by VDI and not accretion or minor mergers. With its high spatial resolution, the \textit{James Webb Space Telescope}'s (JWST) near-IR Camera (NIRCam) should be able to better resolve such giant star-forming clumps and could help in constraining this scenario.
It is thus becoming clear that the galaxies within the MS scatter are not all largely unperturbed gas-rich disks. The compact star-forming cores as well as the giant clumps, independently of their formation histories, imply complex phenomenology at play that is much different from the local star-forming galaxies in the MS that are typically well-behaved spirals.

Recently, emphasis has been brought onto other kinds of asymmetries characterising high redshift star-forming galaxies. 
\cite{kalita_bulge_2022} discovered strong lopsidedness affecting the three massive star-forming galaxies in a $z=2.91$ group core. They suggested a link between the lopsidedness of a galaxy in a dense environment and gas accretion and minor mergers. Following \cite{bournaud_lopsided_2005}, who investigated the origins of lopsidedness in simulated galaxies, the lopsidedness would then be a marker of the point of impact of the accretion stream . Their conclusion was that it is very unlikely that the lopsidedness is the result of internal mechanisms. It is, however, more likely to be linked to the assembly history and the environment of the galaxy, including asymmetric gas accretion, minor mergers, and interactions with neighboring galaxies. This is also the conclusion of studies on the lopsidedness of galaxies in the local Universe (\citealt{jog_lopsided_2009,zaritsky_origin_2013}).
%Lopsided sub-structures have been observed in \textit{ALMA} sources (\cite{rujopakarn_alma_2019, hodge_alma_2019}), forming a disk-like dust structure embedded in a larger system. 
\cite{rujopakarn_jwst_2023} studied a galaxy in a dense environment with off-center star-forming substructures. They interpreted them as either forming spiral arms following a minor merger, as the result of interactions with a neighboring galaxy, or as a lopsided structure resulting from the point of impact of a cold gas accretion stream. 
\cite{colina_uncovering_2023} reported JWST/MIRI observations of GN20, an extremely luminous sub-millimeter galaxy residing in a $z=4.05$ protocluster (\citealt{daddi_co_2009}). They have revealed a massive extended disk surrounding the sub-millimeter compact nucleus and displaying strong lopsidedness.
As of today, the lopsidedness has only been studied in dense environments and serendipitously. Observing a lopsided disk in a less crowded environment and inferring the prevalence of such disks in complete samples could shed further light on their presumed origin from interactions and accretion and clarify whether a massive hosting dark matter halo is (or is not) required. 

By probing the rest-frame in optical to near-infrared (near-IR) at cosmic noon, JWST/NIRCam has a unique ability to fill in the gap between the sub-millimeter compact nuclei observed with ALMA and the larger galactic disks observed in the optical and to help critically examine the competing scenarios. As an example,  \cite{rujopakarn_jwst_2023} recently studied substructures within a dusty star forming galaxy (DSFG) at $z \sim 3$ imaged with both ALMA and JWST. From the NIRCam images, they showed that the ALMA substructures are also visible at $4\mu$m, demonstrating the direct link that one can draw between near-IR and sub-millimeter emissions. This suggests that the long wavelength channel of the NIRCam might be a good tracer of compact obscured star formation in MS DSFGs. 

This study is part of the Cosmic Evolution and Epoch of Re-ionization Survey (CEERS\footnote{\url{https://ceers.github.io}}; ERS 1345, PI: S. Finkelstein), which is one of the Early Release Science (ERS) programs of JWST (\citealt{gardner_james_2023}) that observed a part of the Extended Groth Strip (EGS) \textit{Hubble Space Telescope} (HST) field with the NIRCam (\citealt{rieke_performance_2023}). The EGS is too far north to be observed with ALMA, and there is no high resolution imaging with the \textit{Northern Extended Millimeter Array} (NOEMA) yet. However, the high sensitivity and exquisite spatial resolution of the NIRCam toward 5$\mu$m can be used as a surrogate to identify the most obscured and massive regions within galaxies, hence those that are most likely vigorously star forming.

Understanding how DSFGs form and evolve is crucial  to obtain a larger picture of galaxy formation and evolution, and it could be a key element to explain the quenching of galaxies at and after cosmic noon. To this aim, JWST/CEERS enables a major step forward. Indeed, \cite{kartaltepe_ceers_2023} have already shown that JWST reveals the diversity of morphologies of galaxies at high redshift. With its high spatial resolution and sensitivity, JWST is able to detect faint disks that were previously undetectable with HST. Moreover, a recent study by \cite{kamieneski_are_2023} uses JWST/NIRCam to probe the dust attenuation and sSFR of a lensed DSFG at $z = 2.3$. They have demonstrated the power of JWST/NIRCam to precisely measure these properties at sub-galactic scales, which allowed them to conclude that despite a more dust attenuated bulge, the color gradient of \textit{El Anzuelo} galaxy is mainly driven by an early stage of inside-out quenching. This makes JWST/NIRCam the best instrument to investigate the morphological evolution of DSFGs around cosmic noon in terms of compact star formation, giant clumps, and galaxy structure.

The paper is organized as follows. In Sect. \ref{sec:data}, we present the data used in this study and the sample selection process. In Sect. \ref{sec:analysis}, we detail the methods used to analyze each galaxy individually. In Sect. \ref{sec:results}, we outline the main results of the analysis. Finally, in Sect. \ref{sec:discussion}, we discuss the possible implications of the results in terms of formation and evolution of DSFGs at cosmic noon.
In this work, we adopt $H_{0}$ = 70km s$^{-1}$ Mpc$^{-1}$, $\Omega_M$ = 0.3, $\Lambda_0$ = 0.7, and a Chabrier IMF (\citealt{chabrier_galactic_2003}). When necessary, we converted stellar masses and SFR from Salpeter IMF (\citealt{salpeter_luminosity_1955}) to Chabrier IMF by subtracting 0.24dex.

\section{Data} \label{sec:data}

\subsection{CEERS imaging} \label{subsec:nircam_img}

For the purpose of this study, we used the NIRCam imaging of CEERS, reduced using a customized pipeline by the CEERS collaboration (\citealt{bagley_ceers_2022}). It includes images in 7 filters: F115W, F150W, F200W, F277W, F356W, F410M and F444W for an average $5\sigma$ depth of 28.6 AB mag (See Table 3 of \cite{bagley_ceers_2022} for more details, each filter and pointing as a slightly different depth). The Point-Spread-Function (PSF) Full-Width at Half-Maximum (FWHM) of those filters range from 0.040'' to 0.145'' for F115W and F444W respectively\footnote{\url{https://jwst-docs.stsci.edu/jwst-near-infrared-camera/nircam-performance/nircam-point-spread-functions}}. For this study, we used the CEERS imaging from the June 2022 pointings, which represent 40\% of the total area covered by the NIRCam for CEERS between June and December 2022. We used the background subtracted images as we wanted to measure precise photometry.
As we later needed to extract galaxy properties based on spectral energy distributions (SEDs), we decided to complement shorter wavelengths by taking advantage of the existing HST imaging in the field. 
We used the publicly available HST data products version 1.9, available through CEERS. These mosaics were derived from HST archival data, but with improved calibration compared to the default pipeline products, and have astrometry tied to Gaia-EDR3 (\citealt{lindegren_gaia_2021}). As described in the accompanying data release, the mosaics were created from the combination of HST programs 10134, 12063, 12099, 12167, 12177, 12547, 13063, and 13792, and the reduction and calibration followed a similar procedure to those described in \cite{koekemoer_candels_2011}.
We used two ACS filters; F606W and F814W with a PSF FWHM of 0.115'' and 0.110'' respectively (\citealt{koekemoer_candels_2011}). We did not use the HST/WFC3 images as these bands are redundant for bright galaxies, as they are covered by JWST/NIRCam images which are deeper and with better spatial resolution.

\subsection{The ``super-deblended" FIR catalog} \label{subsec:SD_select}

The goal of this paper is to study the morphology and star-forming activity of DSFGs. We selected galaxies based on their IR detection in the state-of-the-art super-deblended far-IR (FIR) catalog of the EGS field (Henry et al., in preparation).
%Due to the high level of dust attenuation, DSFGs have a low luminosity in the UV/optical wavelengths, but are bright in the infra-red part of the spectrum. Moreover, 
FIR emission is a secure tracer of star formation (once the AGN components are removed), while optical and near-IR classification of star-forming galaxies is subject to larger uncertainties especially in the presence of dust. Hence, our FIR selection ensures the galaxies under scrutiny are truly highly SF.

The super-deblending is based on a well-established technique (\citealt{liu_super-deblended_2018,jin_super-deblended_2018}). It is a multi-wavelength fitting technique meant to optimize the number of priors fitted at each band to extract the deepest reachable information. They used images from \textit{Spitzer} ($24\mu$m (FIDEL, \citealt{dickinson_far-infrared_2007})),  \textit{Herschel} ($100\mu$m and $160\mu$m (PEP, \citealt{lutz_pacs_2011}), $250\mu$m, $350\mu$m, $500\mu$m (HerMES, \citealt{oliver_herschel_2012})),  SCUBA2 ($850\mu$m (S2CLS, \citealt{geach_scuba-2_2017}), $450\mu$m and $850\mu$m from \cite{zavala_scuba-2_2017}) and  AzTEC (1.1mm from \cite{aretxaga_lmt_2015}). The key of the technique is to obtain an adaptive balance as a function of wavelength between the density of priors fitted, the quality of the fit, and the achievable deblending given the PSF sizes. They started with the deepest images and fitted band after band toward shallower images. Extensive Monte-Carlo simulations ensured that the uncertainties associated to the flux measurements were “quasi-Gaussian” (see \citealt{liu_super-deblended_2018,jin_super-deblended_2018}; A. Henry et al. in preparation). 

\subsection{Sample definition} \label{subsec:reject}

\begin{figure}[htb]
    \centering
    \resizebox{\hsize}{!}{\includegraphics{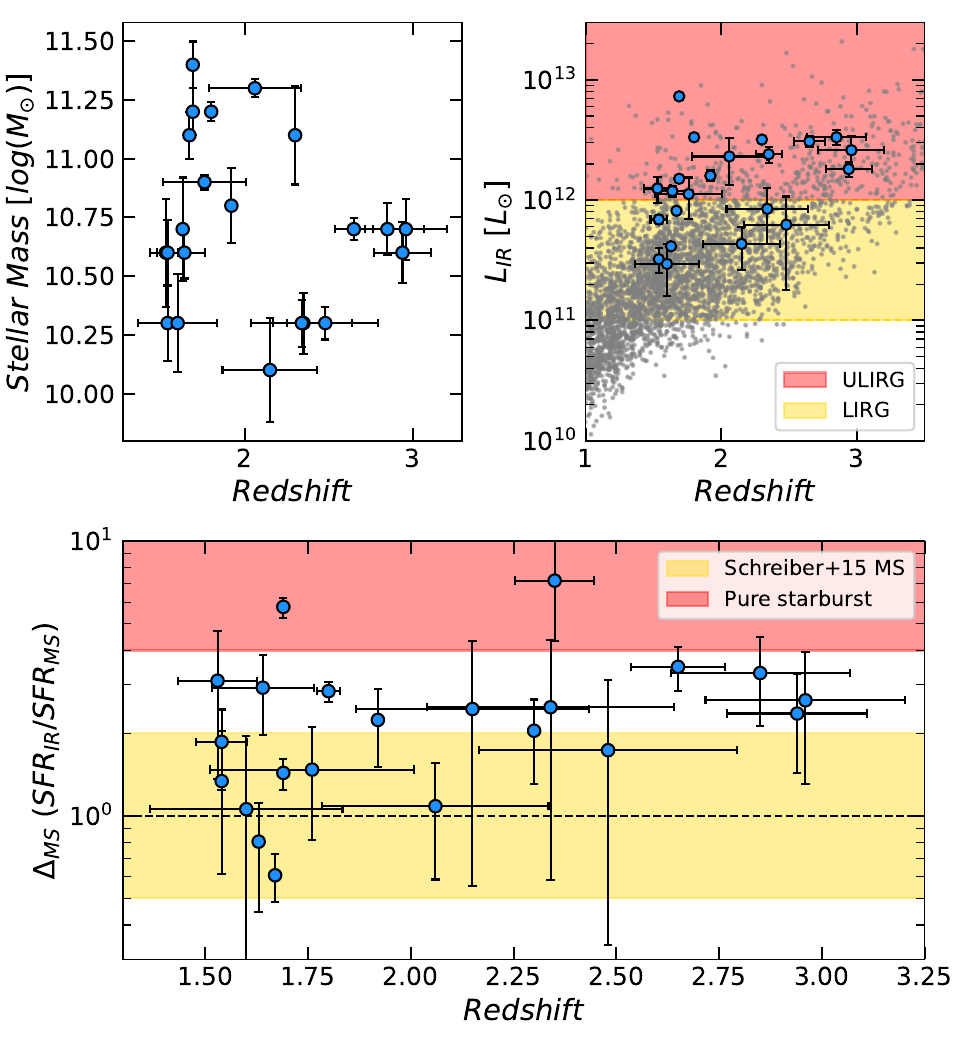}}
    \caption{Stellar mass (upper-left panel), total IR luminosity (upper-right panel) (Henry et al. in preparation), and distance from the MS (lower panel) of the galaxies in the selected sample versus their redshift. The colors on the upper-right panel delimits the luminous IR galaxies (LIRG; in yellow) and ultra-LIRG (ULIRG; in red) local regimes for information. In the lower panel, the yellow shaded region illustrates the MS from \cite{schreiber_herschel_2015}, while the red shaded region illustrates the pure starburst regime (\citealt{liu_super-deblended_2018}). On the upper-right panel, the blue filled circle markers are the selected DSFG candidates, while the grey dots are the sources from the ``super-deblended" catalog.}
    \label{fig:general_prop}
\end{figure}

We selected all sources securely detected in the FIR catalog (see Sect. \ref{subsec:SD_select}) that fell in the CEERS/NIRCam regions observed in June 2022. Since short wavelength channels have a slightly different field of view than long wavelength channels, we checked that the sources are observed in all of them and that they were not too close to the edge of the image so that there were not partially cut. 
In detail, we required the galaxies to have S/N$_{FIR} > 5$, where S/N$_{FIR}$ is the FIR S/N added in quadrature from $100\mu$m to $1.1$mm (Henry et al. in preparation), and to have at least one detection (S/N$ > 3$) in one of the three \textit{Herschel}/SPIRE bands after deblending (this is required to reliably measure star-forming components in case of AGNs).
The implication of the IR selection is that we don’t have a stellar mass complete sample of star-forming galaxies (e.g., complete above some mass threshold), and we have instead something closer to a (redshift-dependent) SFR limit. Indeed, as one can see on the upper-right panel of Figure \ref{fig:general_prop}, the higher the redshift, the higher the infrared luminosity ($L_{IR}$) has to be for the source to be in the `super-deblended' catalog. As $L_{IR}$ is directly linked to $SFR_{IR}$ ($L_{IR} = 10^{10}\times SFR_{IR}$), this limit translates into a (redshift-dependent) SFR limit (around $10-20 M_{\odot}yr^{-1}$ for $z = 1.5$ and around $100 M_{\odot}yr^{-1}$ for $z = 3.0$). We are aware that we are missing star-forming galaxies below our IR detection threshold, as we wish to focus on highly (and securely) star-forming galaxies. However, all our sources have an optical counterpart.

We also limited the sample to  galaxies within $1.5<z<3.0$, as we are willing to focus on galaxies at ``cosmic noon,'' as recalled in the introduction.
To get accurate redshift estimates, we used the recent redshift compilation produced by \cite{kodra_optimized_2022}, which includes photometric redshifts based on CANDELS (\citealt{grogin_candels_2011,koekemoer_candels_2011}) as well as grism-based redshifts from 3D-HST (\citealt{momcheva_3d-hst_2016}) and spectroscopic redshifts from the MOSDEF survey (\citealt{kriek_mosfire_2015}). 

This sample comprised a total of 26 IR-detected sources. From these, 4 had to be rejected after a clean up. After close inspection, three galaxies were in a blended region or close to a much brighter IR source, making the \textit{Herschel} measurements less reliable, despite the deblending. The last rejected source hosted an AGN (clear radio excess, $\sim 10\times$ brighter than what is expected for the radio continuum based on IR emissions, and X-ray detected: ID15327, RA = 215.82825, Dec = 52.80844, $z_{phot} = 1.61$, $log_{10}(L_{AGN}/L_{\odot}) \gtrsim 11.3$), hence the majority of its IR luminosity does not come from star-forming regions which are the main objects of this study.

This left us with a clean sample of 22 FIR-bright DSFGs around cosmic noon. We illustrate in Fig. \ref{fig:general_prop} the distribution of the sample in terms of stellar mass estimated in the pre-JWST era (\citealt{stefanon_candels_2017}) and total IR luminosity (Henry et al. in preparation, calculated based on the equations in \cite{press_numerical_1992}) versus redshift (\citealt{kodra_optimized_2022}). We also show the distance from the MS (\citealt{schreiber_herschel_2015}) with a 0.6 dex total scatter (\citealt{rodighiero_lesser_2011}) defined as $\Delta _{MS} = SFR_{IR}/SFR_{MS}$. \cite{schreiber_herschel_2015} uses a Salpeter IMF (\citealt{salpeter_luminosity_1955}), we converted stellar masses and SFRs from Salpeter IMF to Chabrier IMF by subtracting 0.24 dex. The red shaded region corresponds to the pure starburst region as defined in \cite{liu_super-deblended_2018} ($log_{10}(SFR_{IR}/SFR_{MS}) > 0.6$ dex), we have two galaxies in our sample classified as pure starburst. The rest is mostly either within the scatter of the MS, but above its average trend, that is, above the MS but below the starburst regime.

\begin{figure*}[htb]
    \centering
    \includegraphics[width=5.0cm]{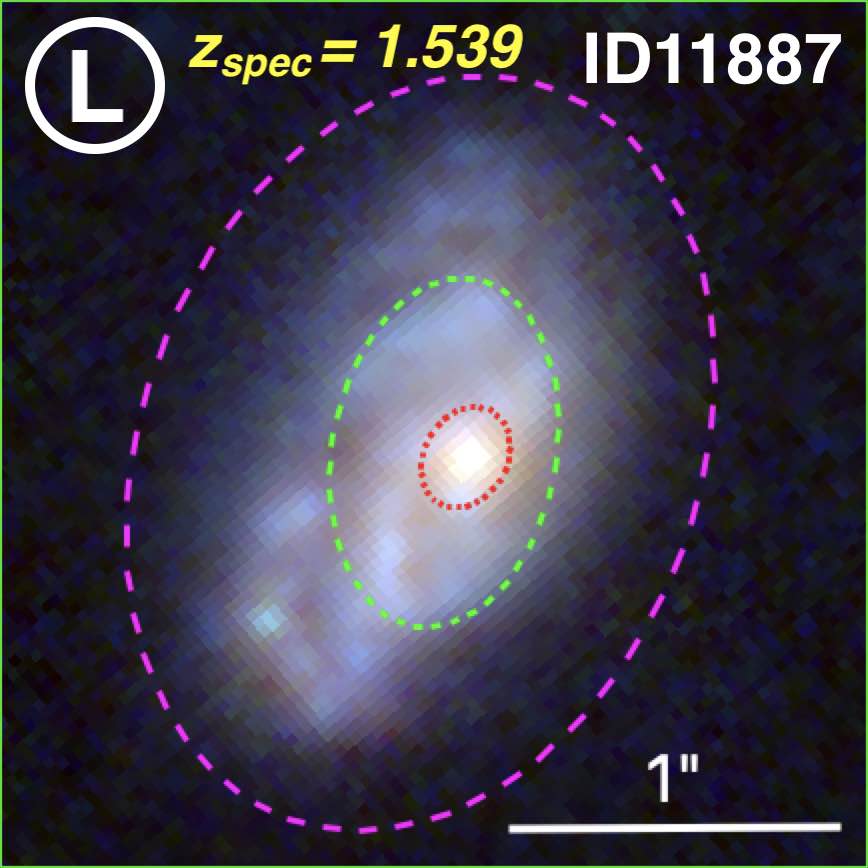}
    \includegraphics[width=5.0cm]{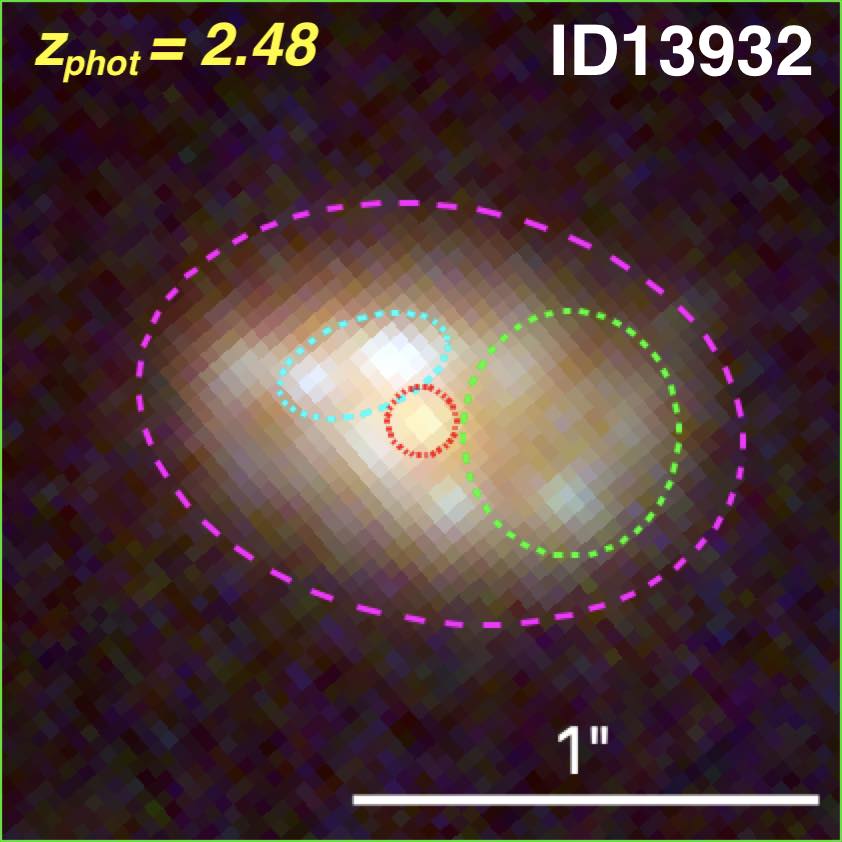}
    \includegraphics[width=5.0cm]{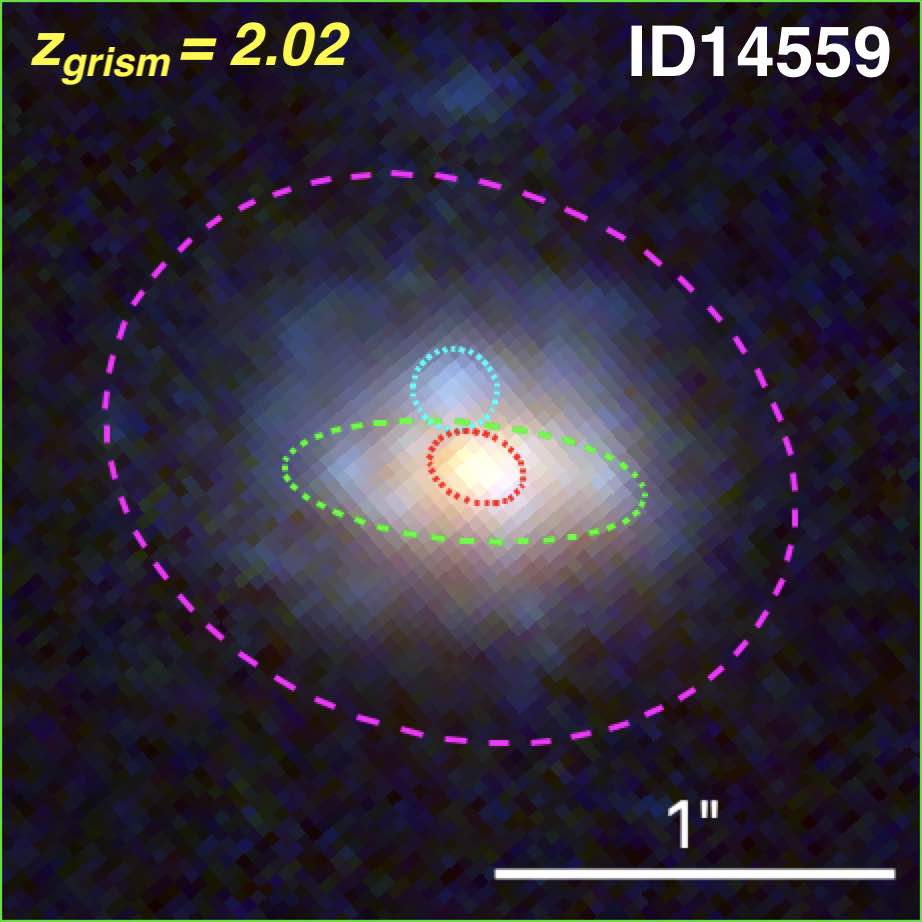}%}
    %\resizebox{\hsize}{!}{}
    \includegraphics[width=5.0cm]{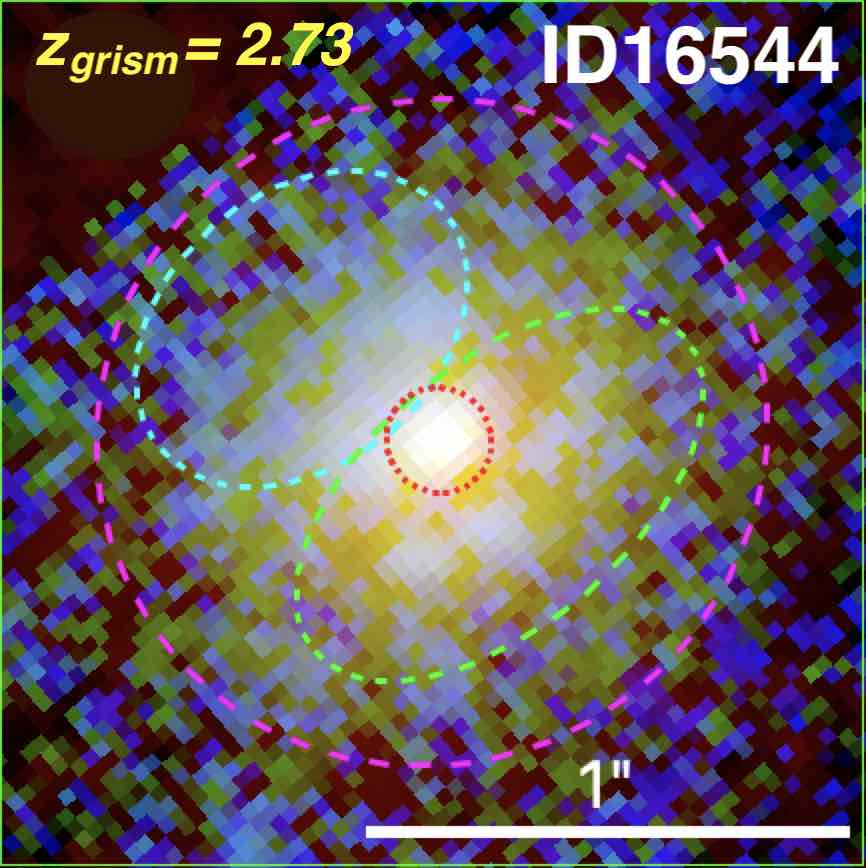}
    \includegraphics[width=5.0cm]{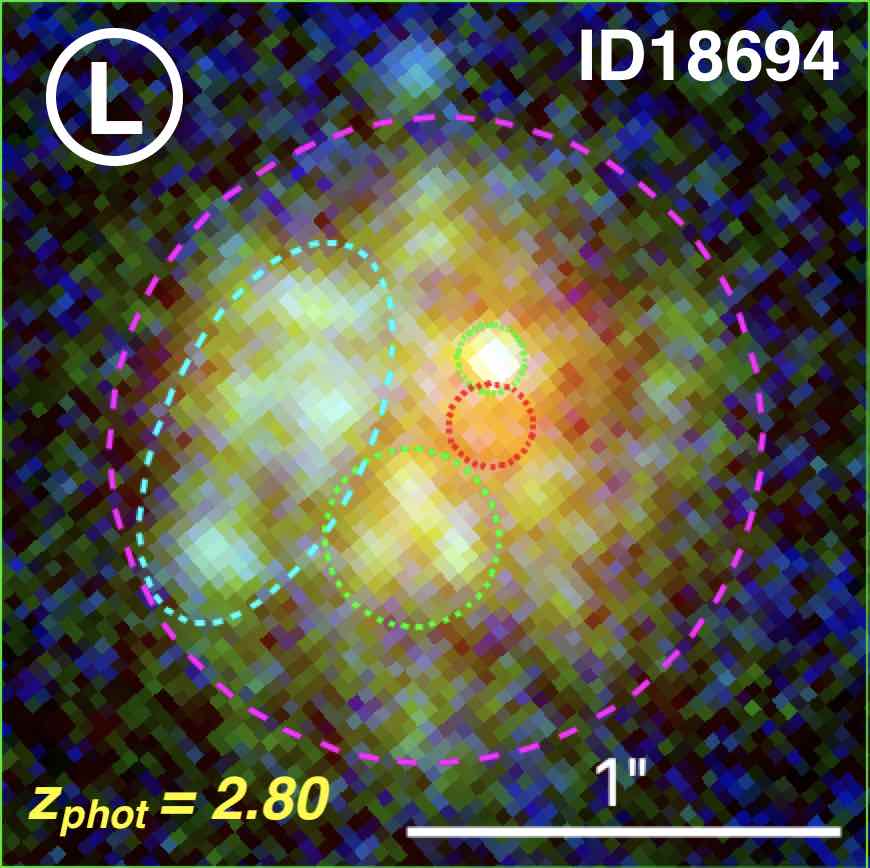}
    \includegraphics[width=5.0cm]{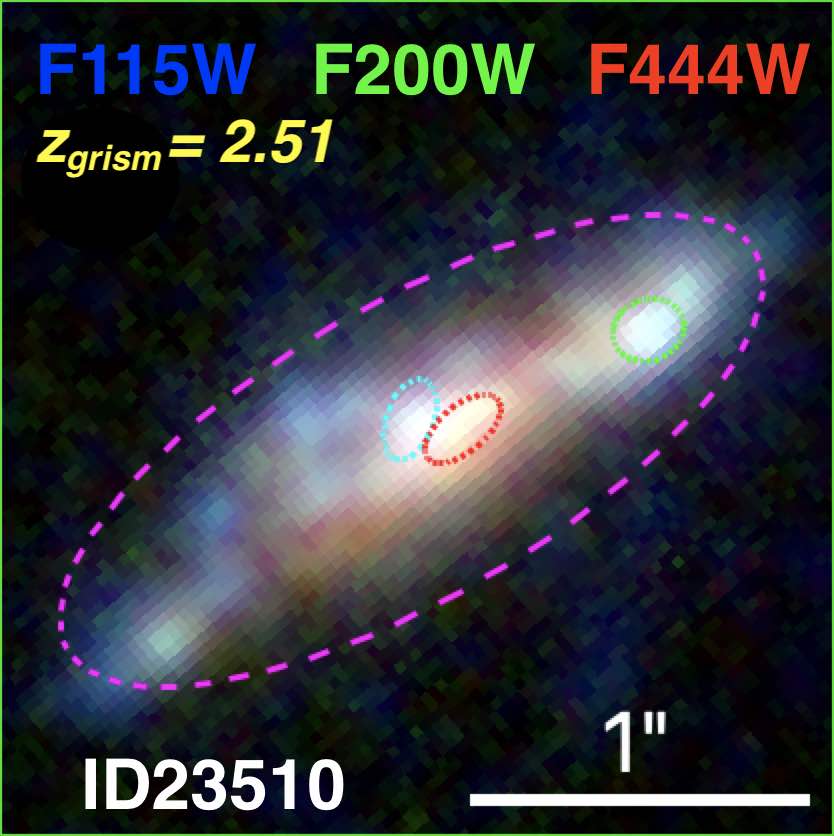}%}
    \includegraphics[width=5.0cm]{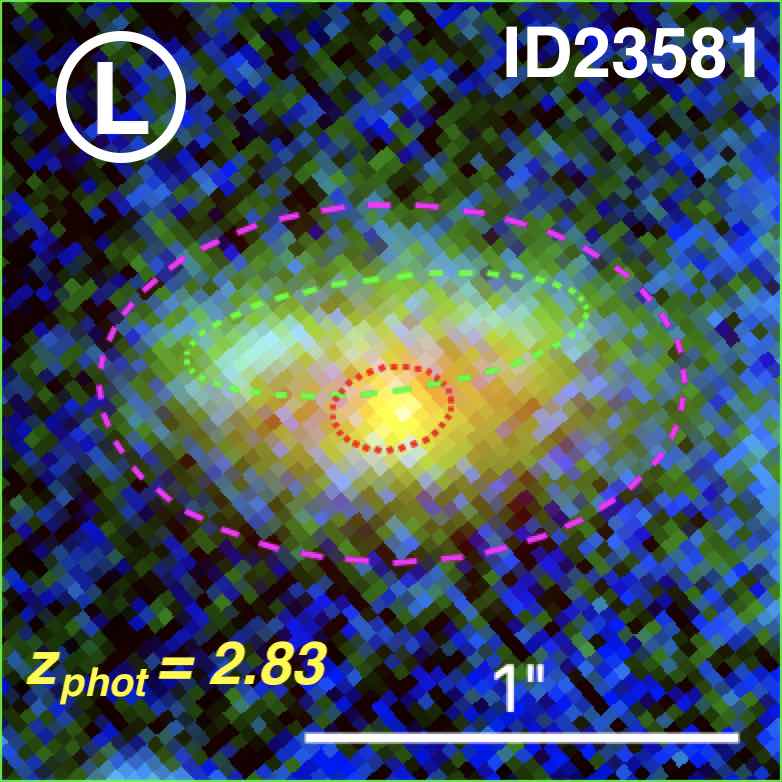}
    \includegraphics[width=5.0cm]{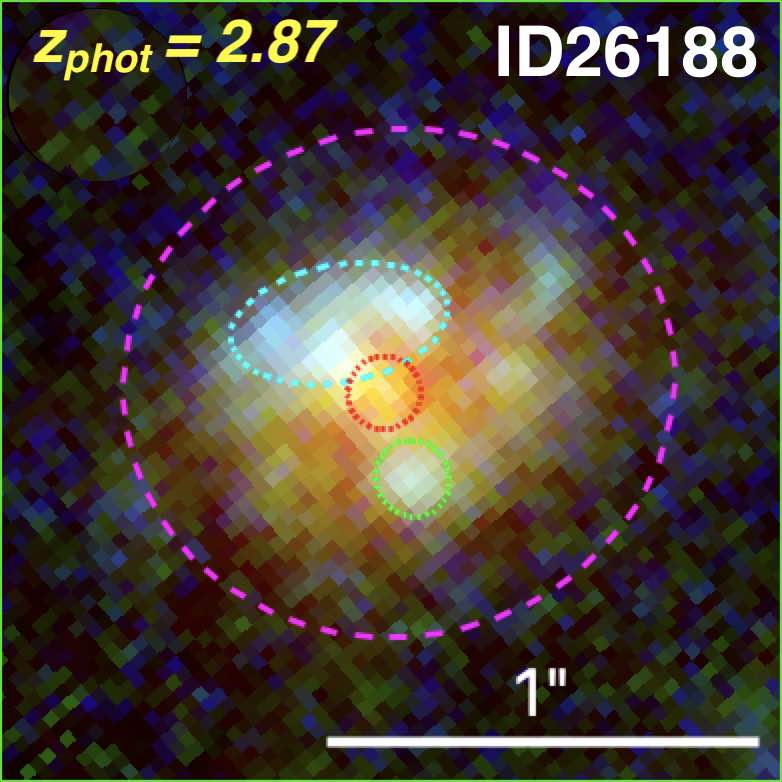}
    \includegraphics[width=5.0cm]{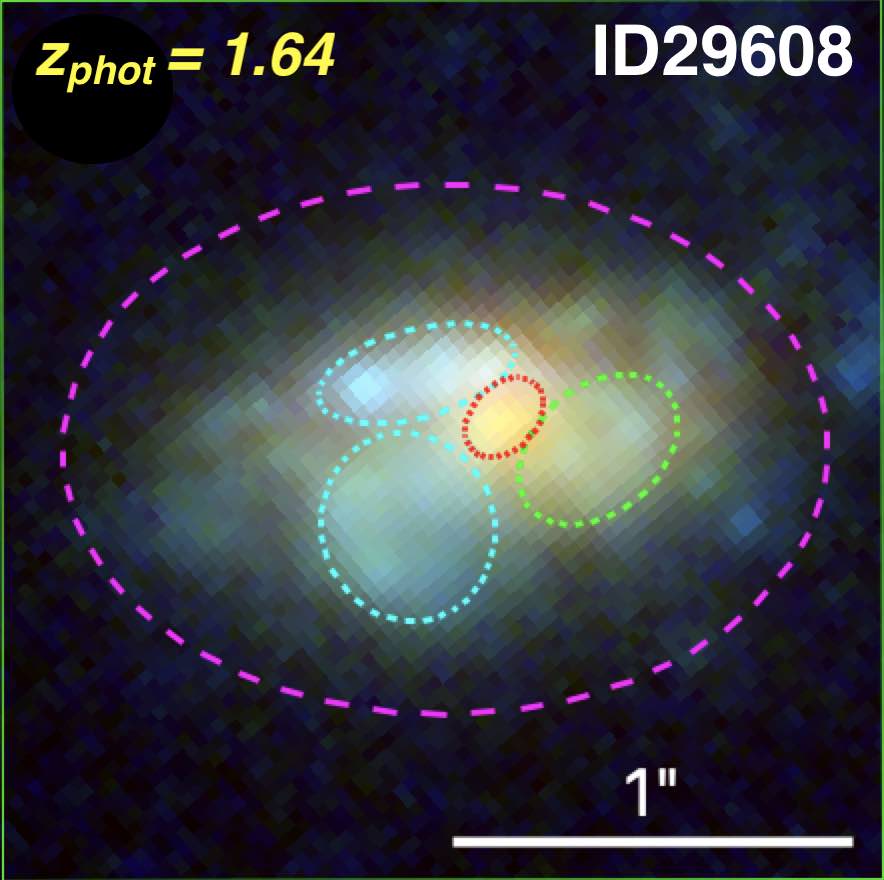}
    \includegraphics[width=5.0cm]{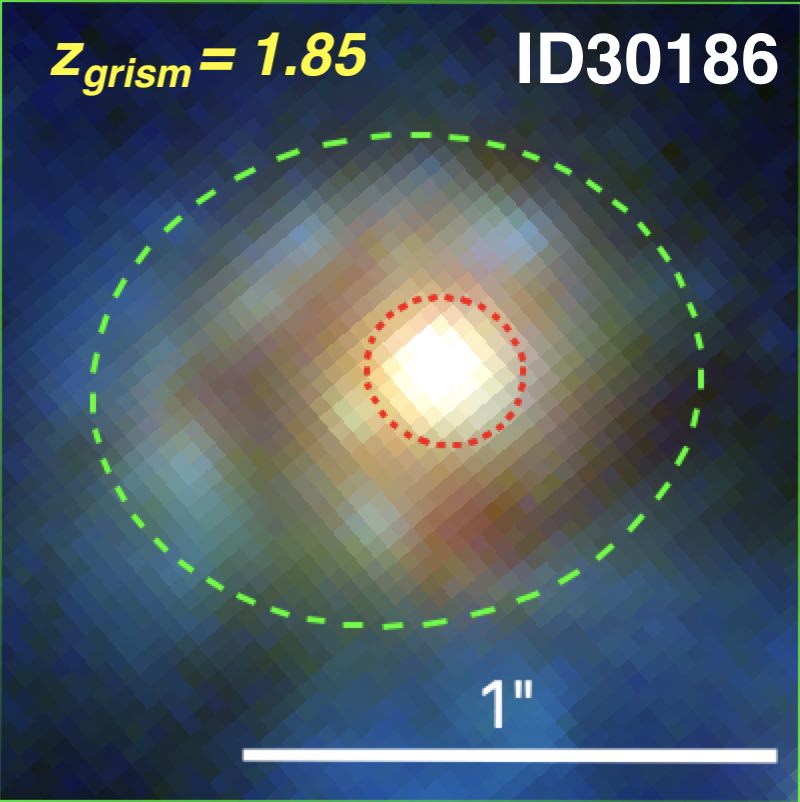}
    \caption{Type I: Star-forming disks with a red star-forming core (see Sect. \ref{subsec:class}). RGB (F115W, F200W, F444W) non-PSF-matched cutouts. In each cutout, the white bar defines the scale of the image. The dotted regions correspond to the different studied components, the core and bulges are shown in red, while the disks are shown in green, blue or magenta and divided in several components showing color patches when present, see Sect. \ref{subsec:flux}. The galaxies with a "\textcircled{\scriptsize L}" are the most lopsided galaxies of our sample (see Sect. \ref{subsec:asym}).}
    \label{fig:cutouts_I}
\end{figure*}

\begin{figure*}
    \centering
    %\resizebox{\hsize}{!}{
    \includegraphics[width=6.1cm]{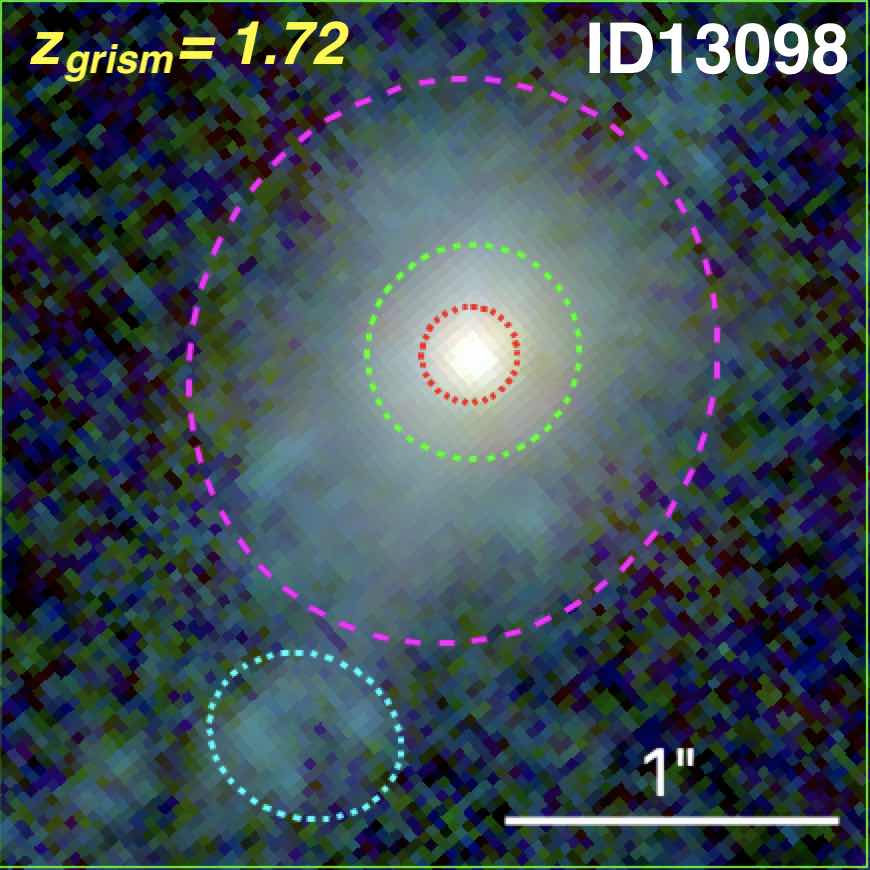}
    \includegraphics[width=6.1cm]{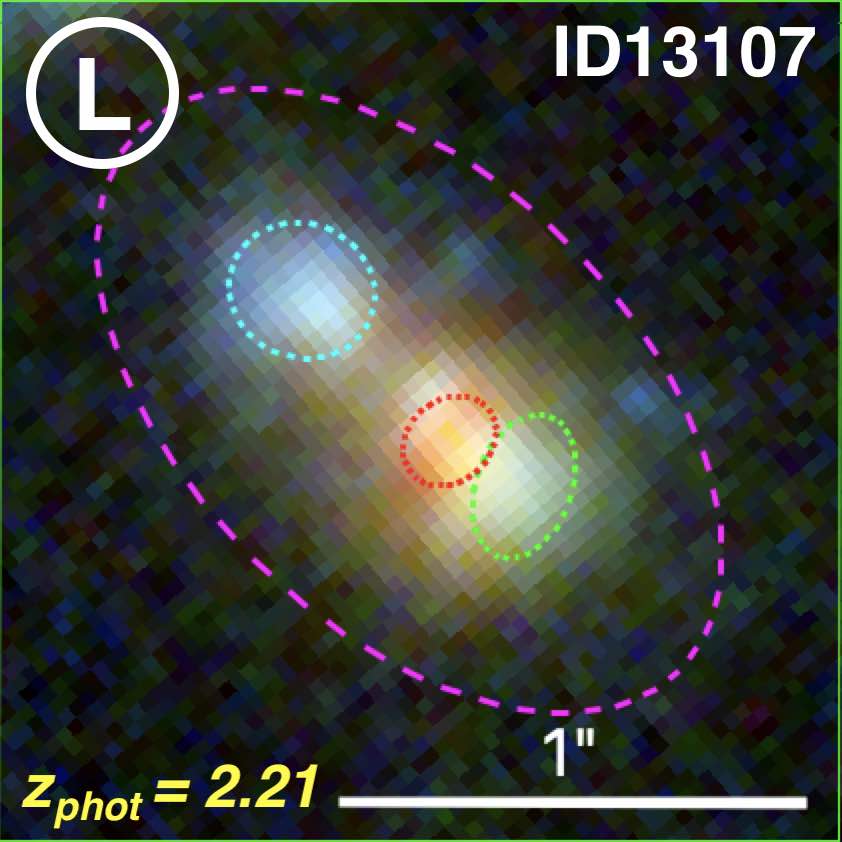}
    \includegraphics[width=6.1cm]{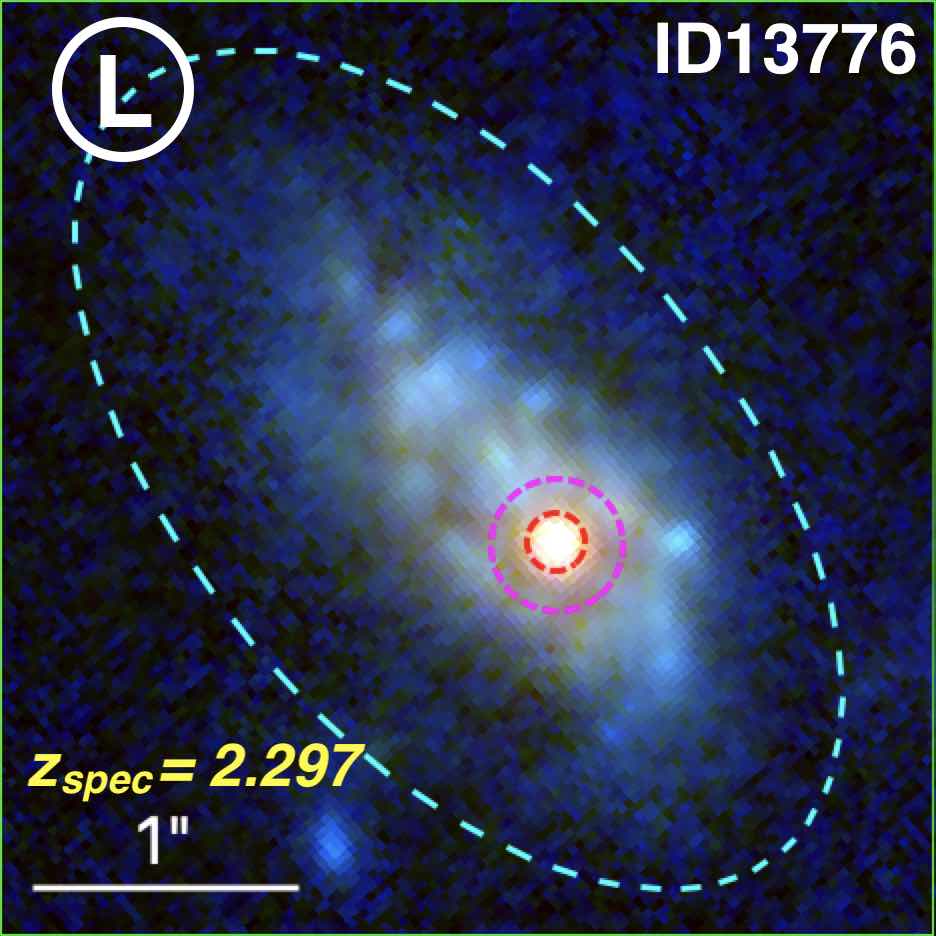}%}
    %\resizebox{0.66\hsize}{!}{
    \includegraphics[width=6.1cm]{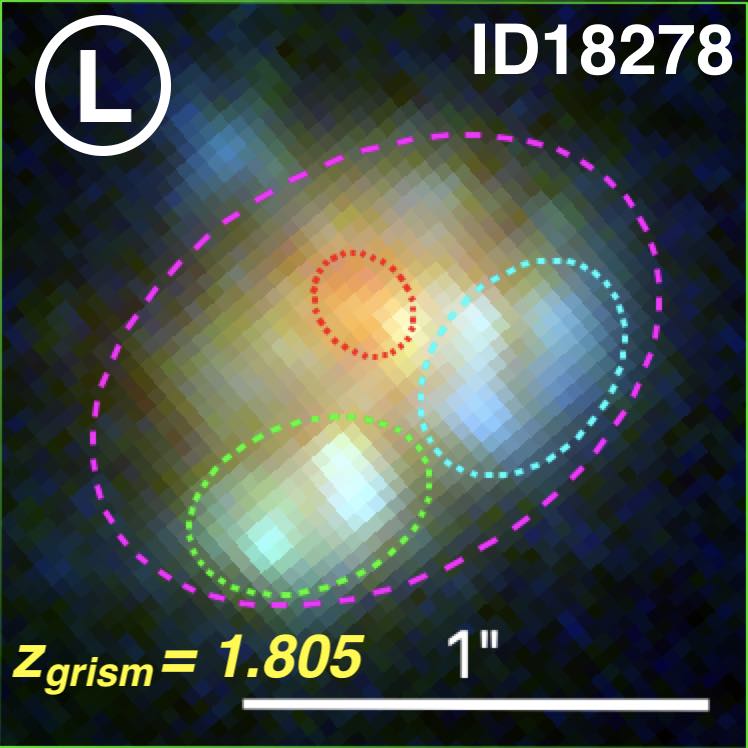}
    \includegraphics[width=6.1cm]{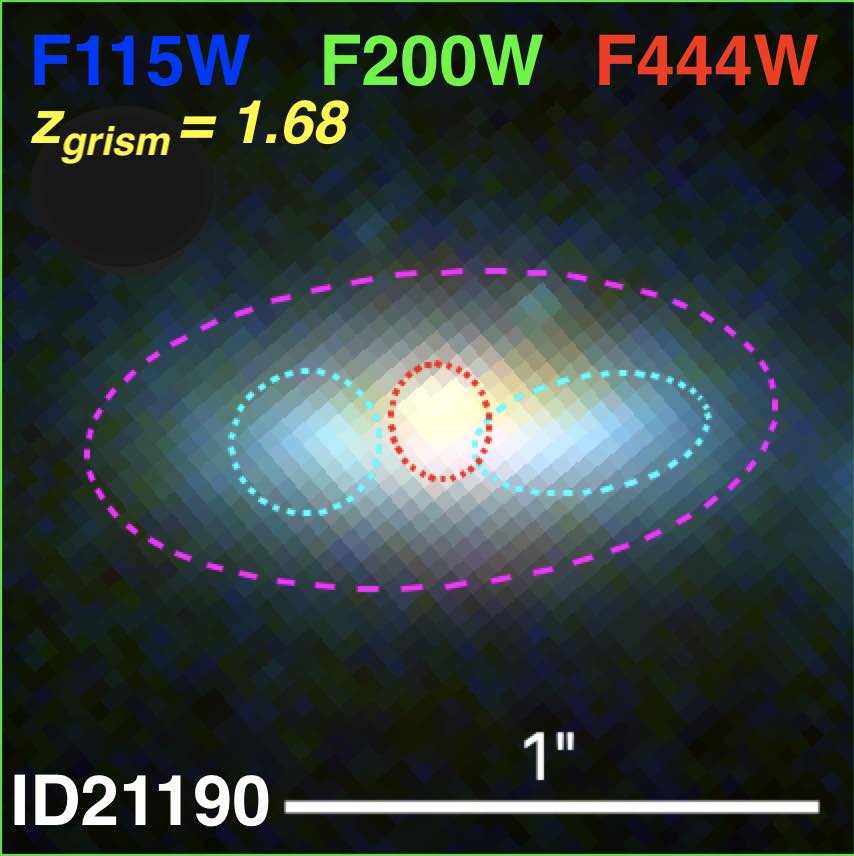}%}
    \caption{Type II: Quenched disks with a star-forming core (see Sect. \ref{subsec:class}). Similar to Fig. \ref{fig:cutouts_I}.}
    \label{fig:cutouts_II}
\end{figure*}

\begin{figure*}
    \centering
    %\resizebox{\hsize}{!}{
    \includegraphics[width=6.1cm]{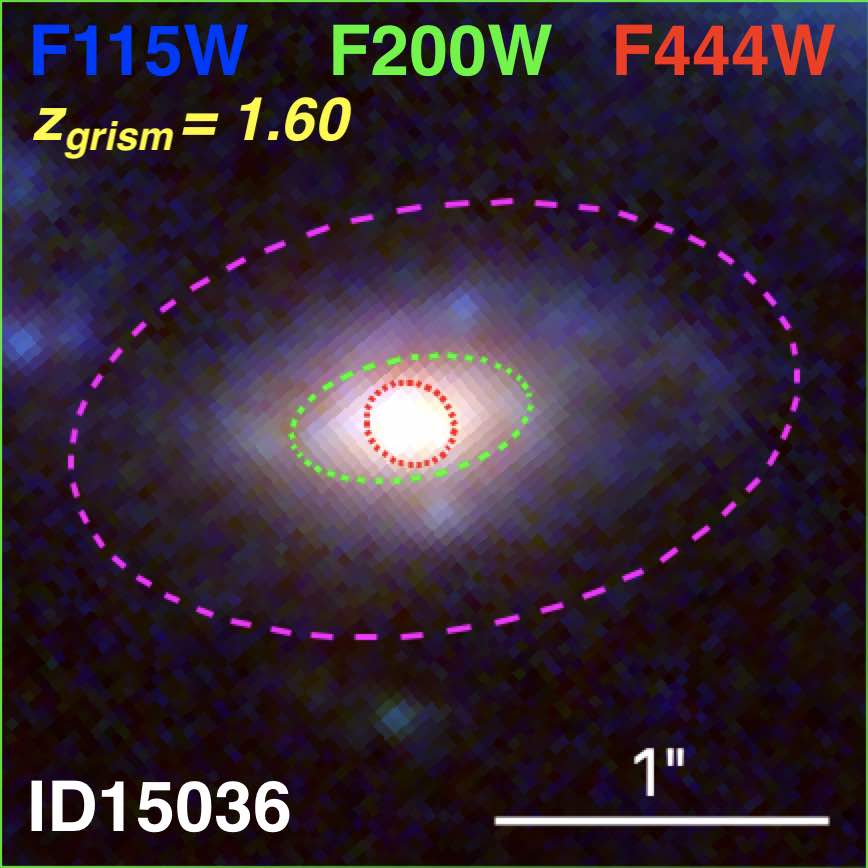} %0.32
    \includegraphics[width=6.1cm]{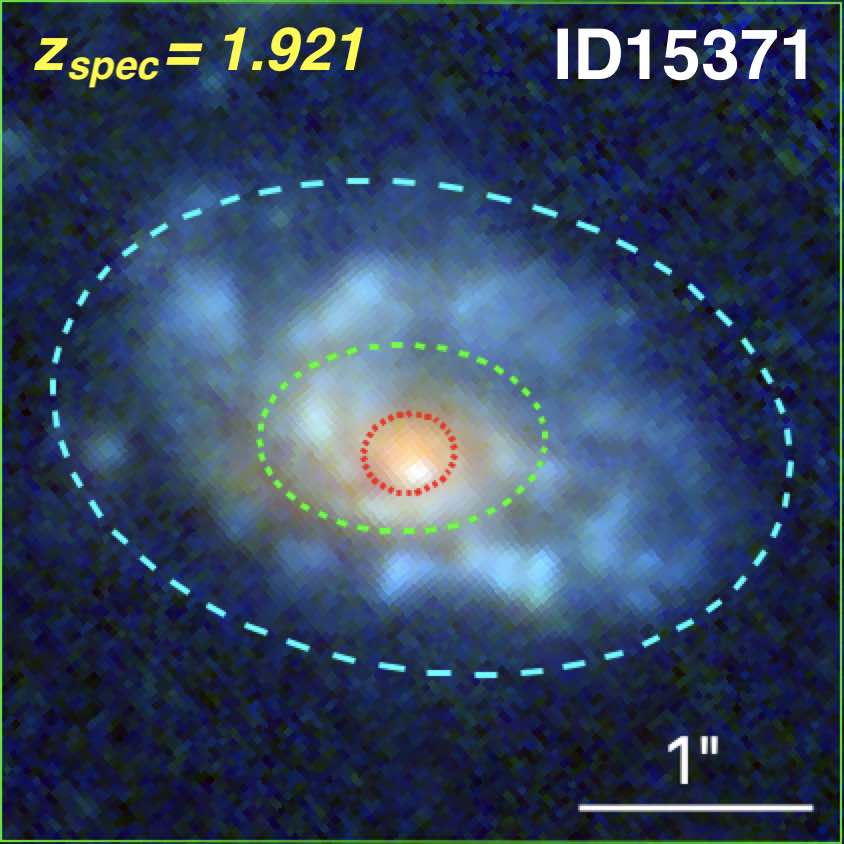}
    \includegraphics[width=6.1cm]{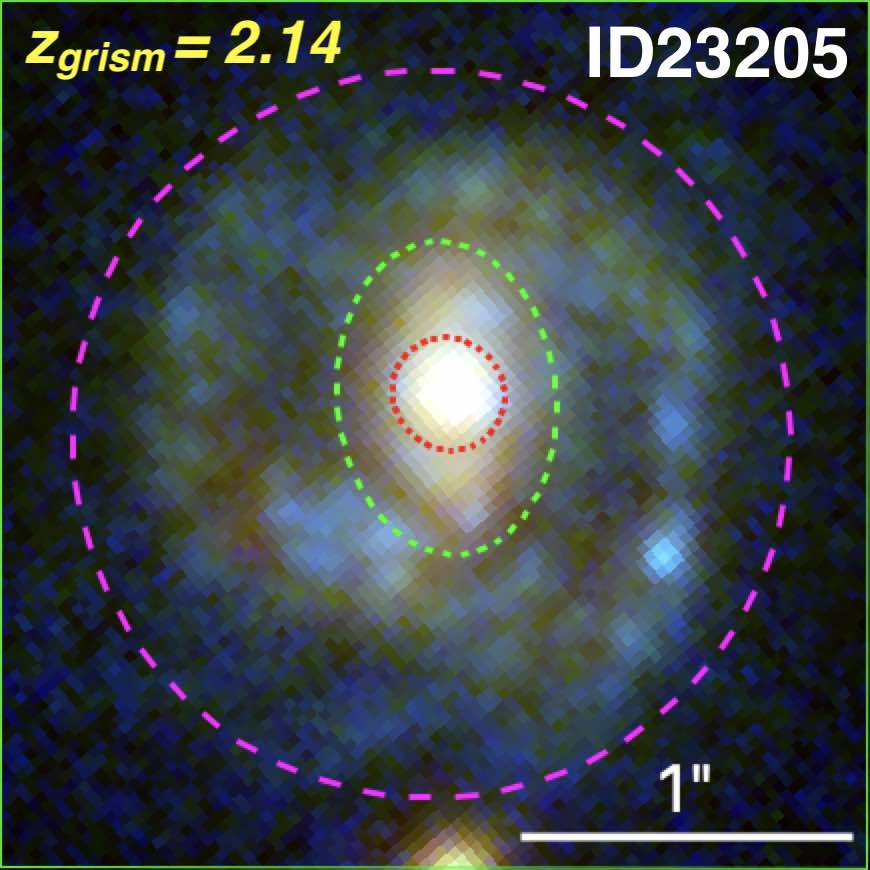}%}
    %\resizebox{\hsize}{!}{
    \includegraphics[width=6.1cm]{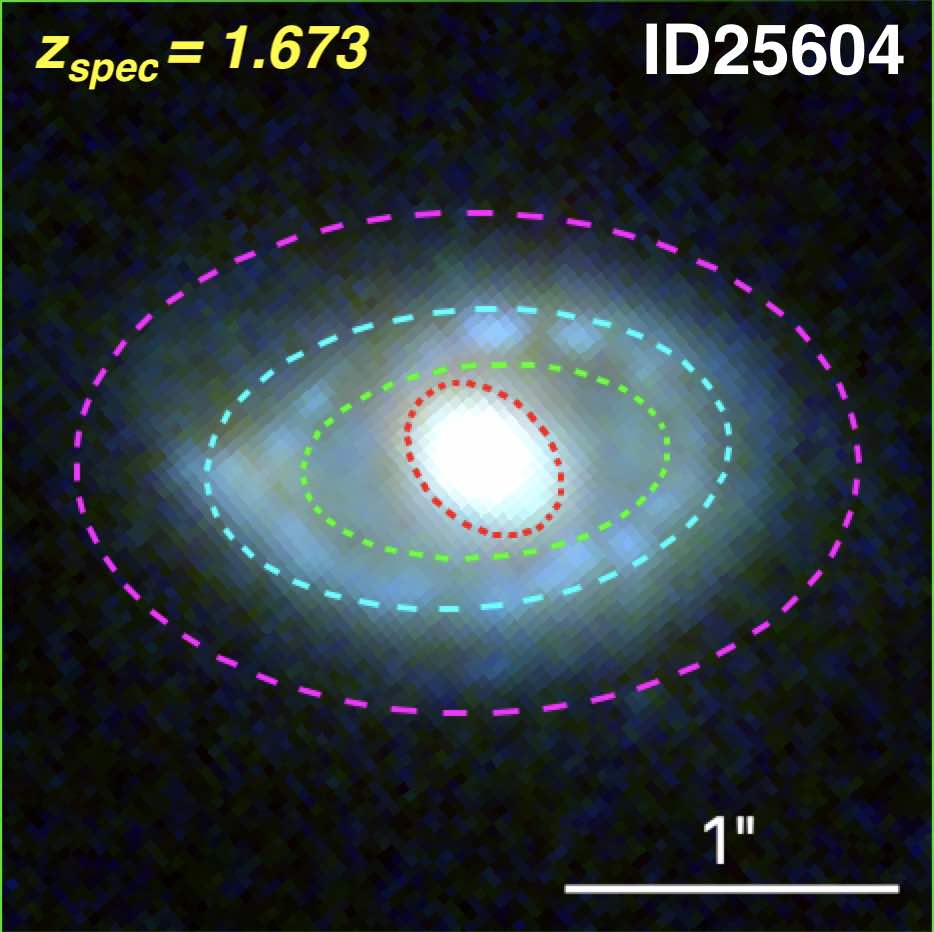}
    \includegraphics[width=6.1cm]{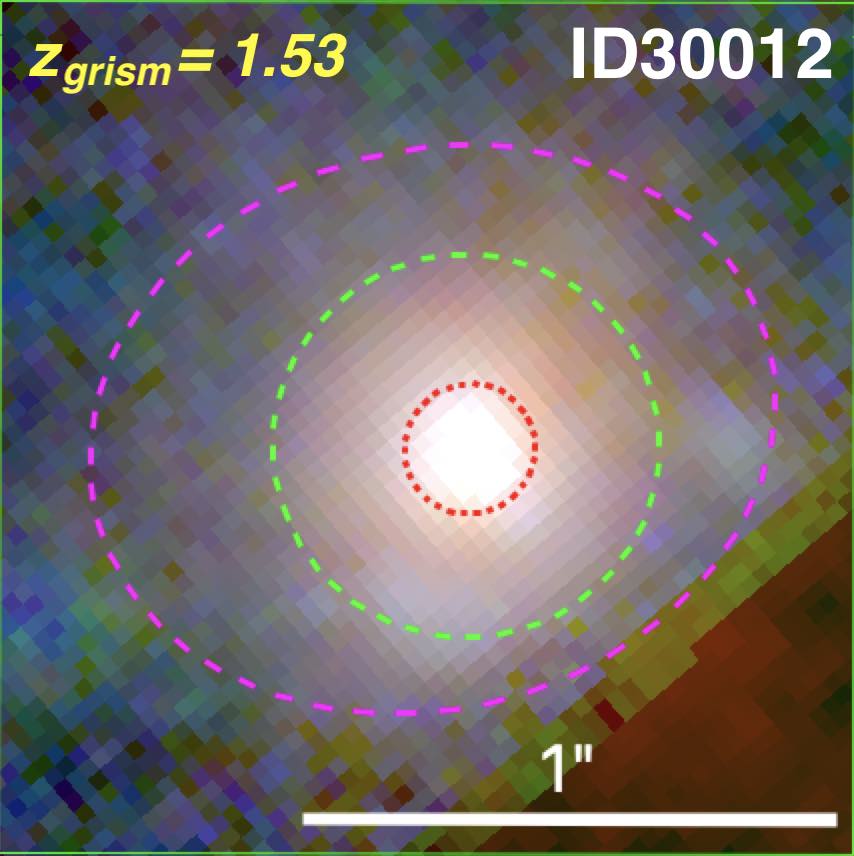}
    \includegraphics[width=6.1cm]{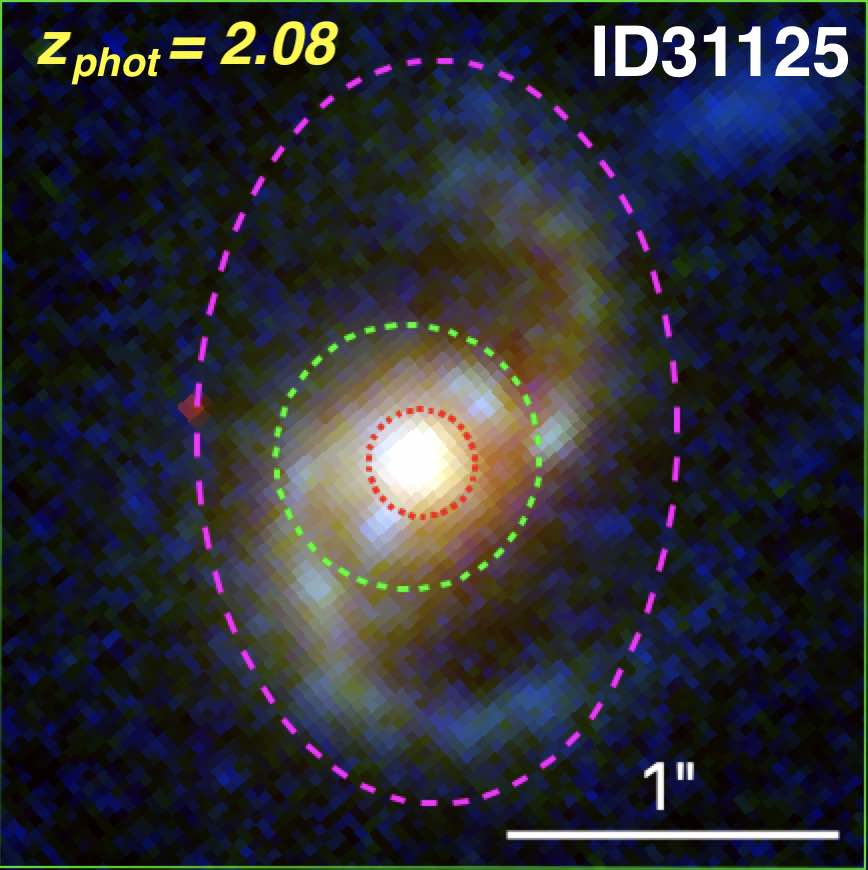}%}
    %\resizebox{0.33\hsize}{!}{
    \includegraphics[width=6.1cm]{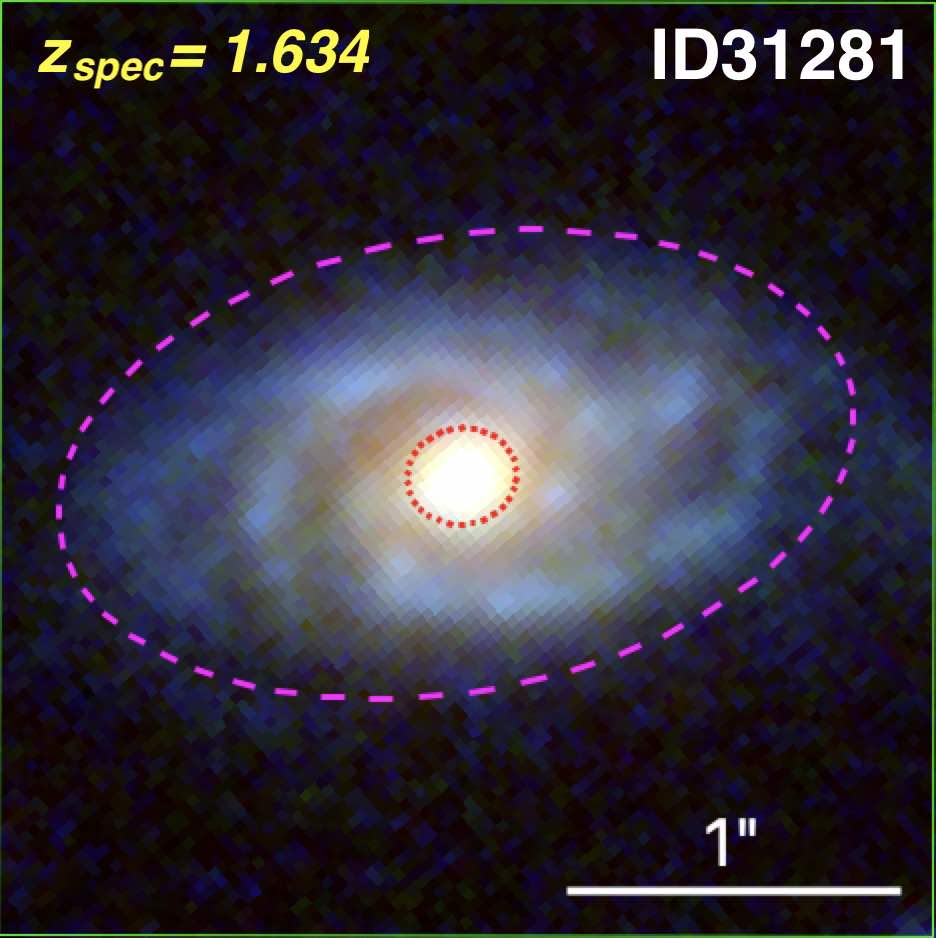}%}
    \caption{Type III: Star-forming disks with a quenched bulge (see Sect. \ref{subsec:class}). Similar to Fig. \ref{fig:cutouts_I} and \ref{fig:cutouts_II}.}
    \label{fig:cutouts_III}
\end{figure*}

In Figs. \ref{fig:cutouts_I}, \ref{fig:cutouts_II} and \ref{fig:cutouts_III}, we show RGB cutouts of our sample of galaxies using the F115W, F200W and F444W filters of the NIRCam. The galaxies are separated in three classes, as discussed in the next Section.

\section{Methods} \label{sec:analysis}

\begin{figure*}[htb]
    \centering
    \includegraphics[width=3.35cm]{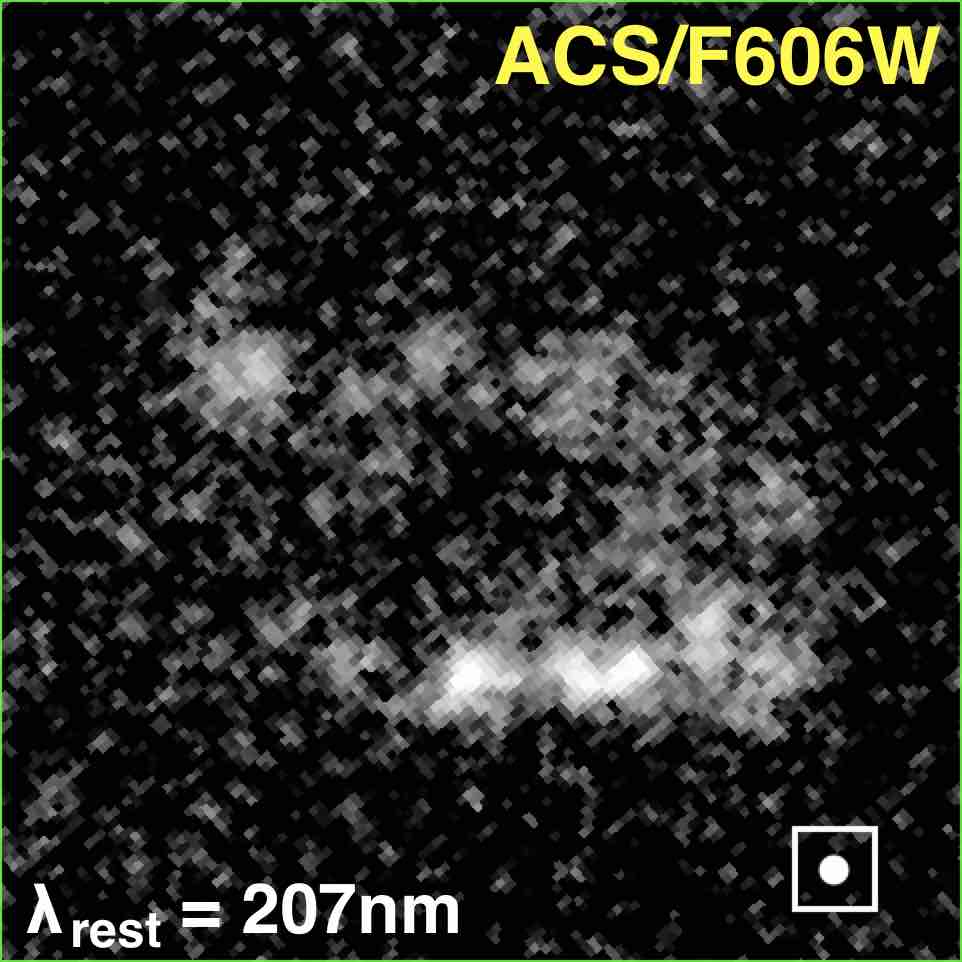}
    \includegraphics[width=3.35cm]{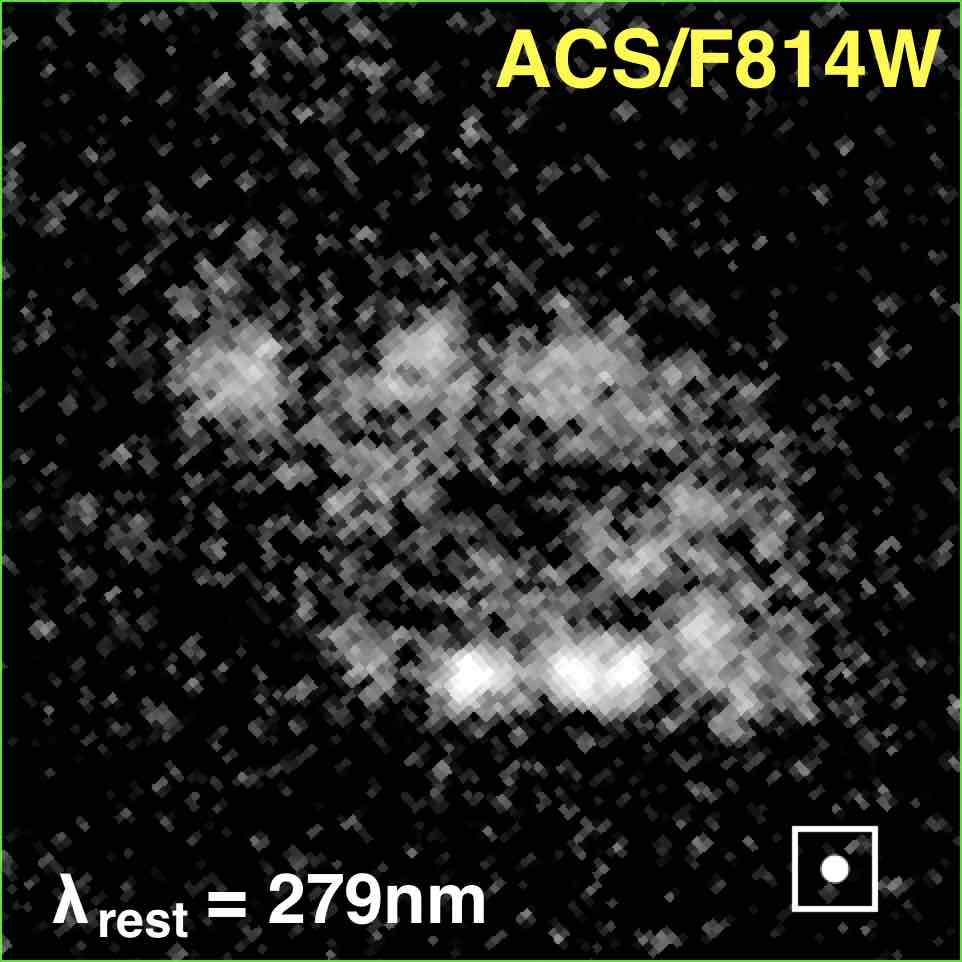}
    \includegraphics[width=3.35cm]{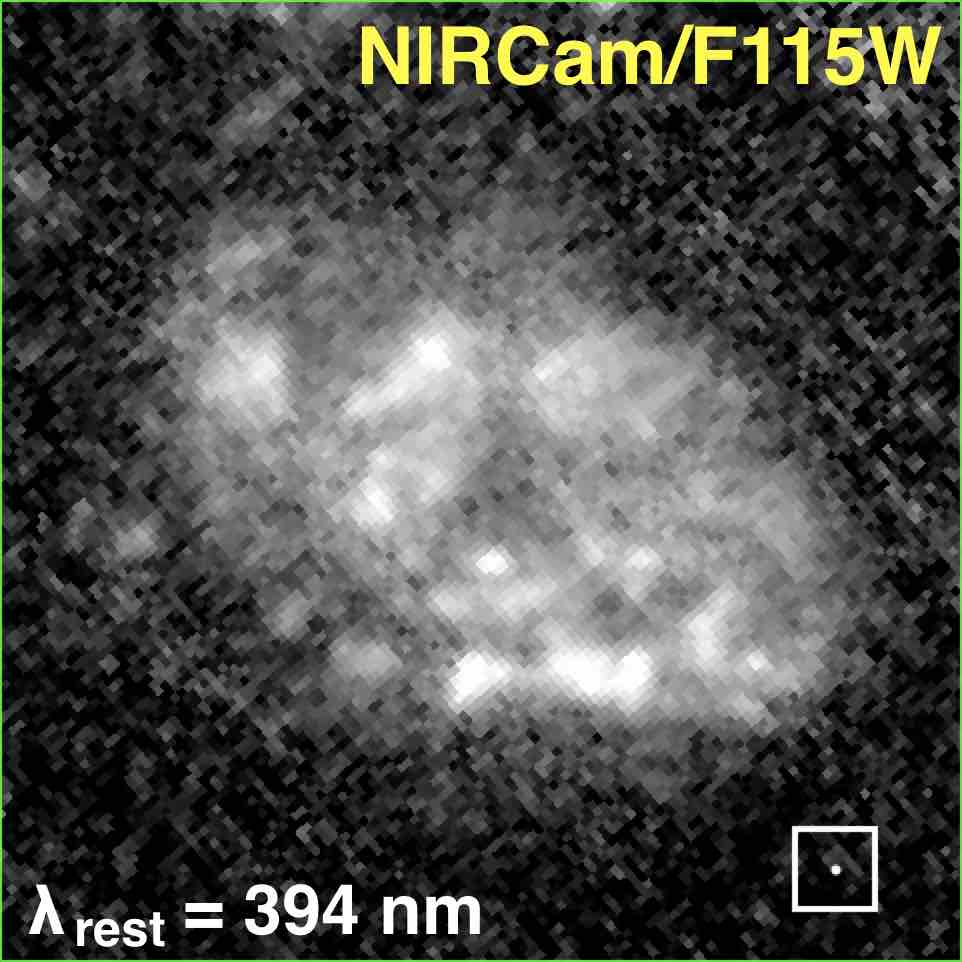}
    \includegraphics[width=3.35cm]{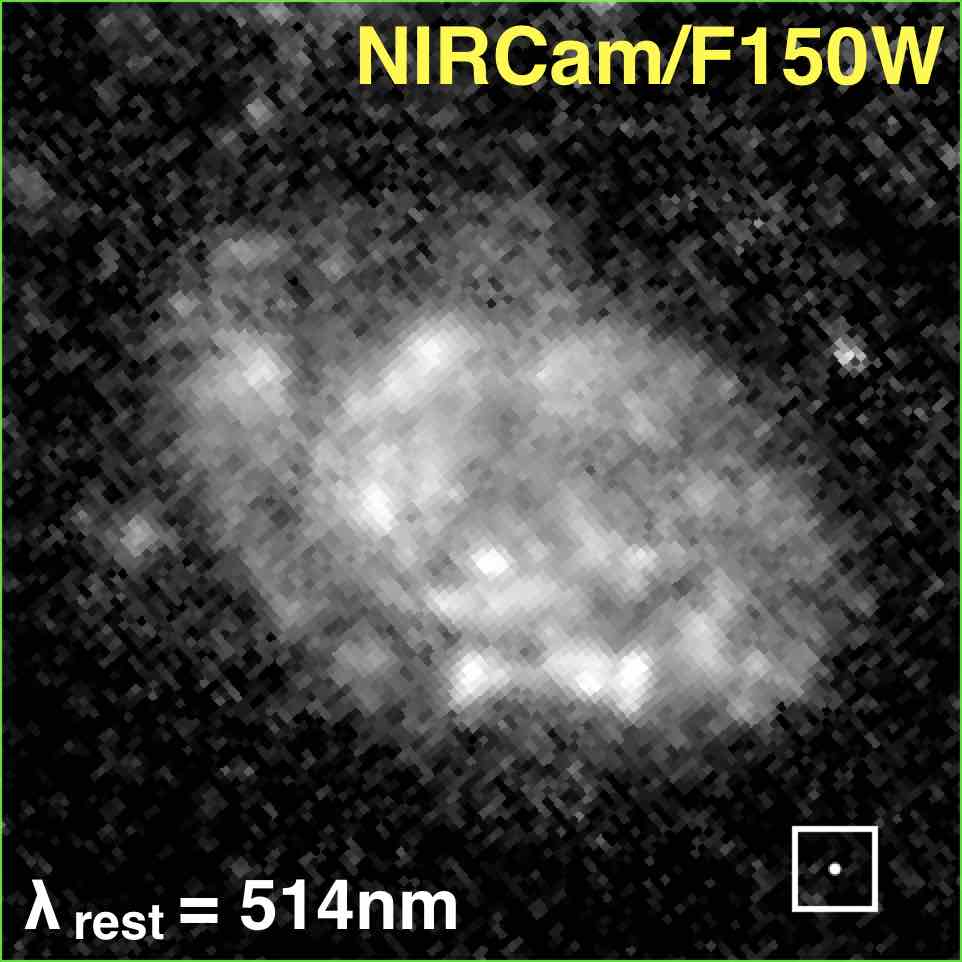}
    \includegraphics[width=3.35cm]{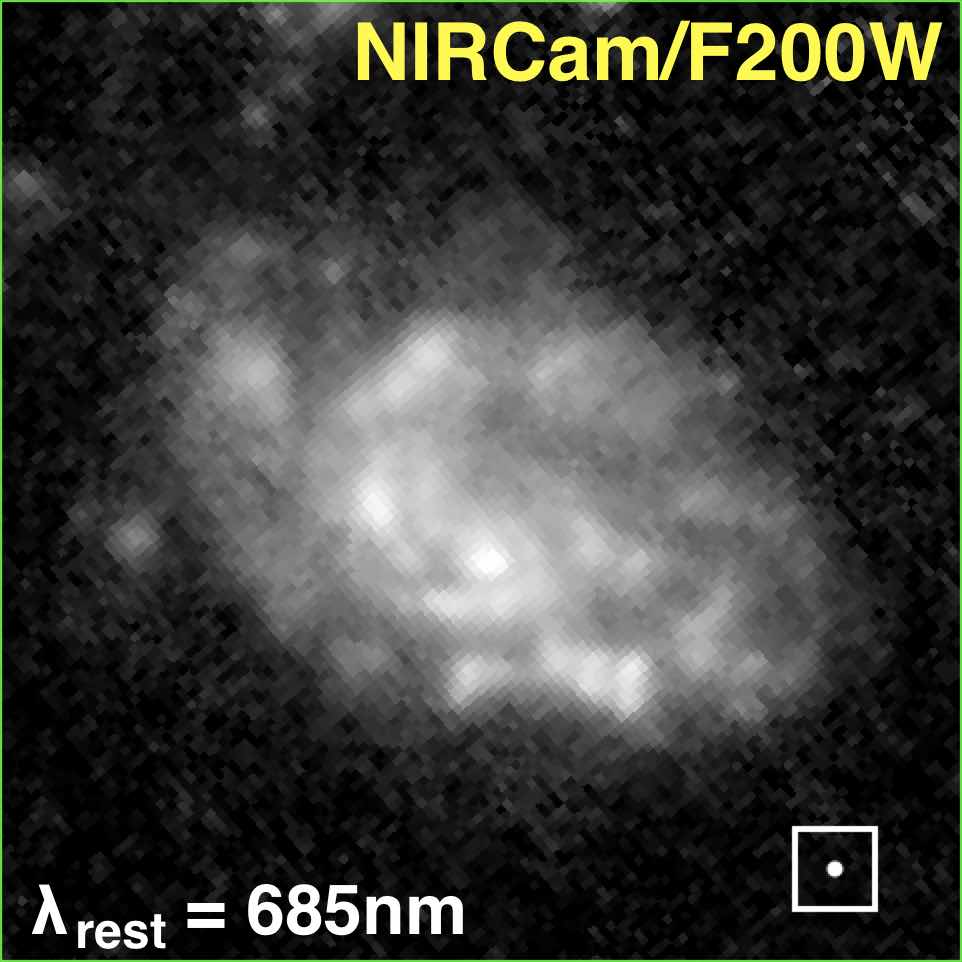}
    \includegraphics[width=3.35cm]{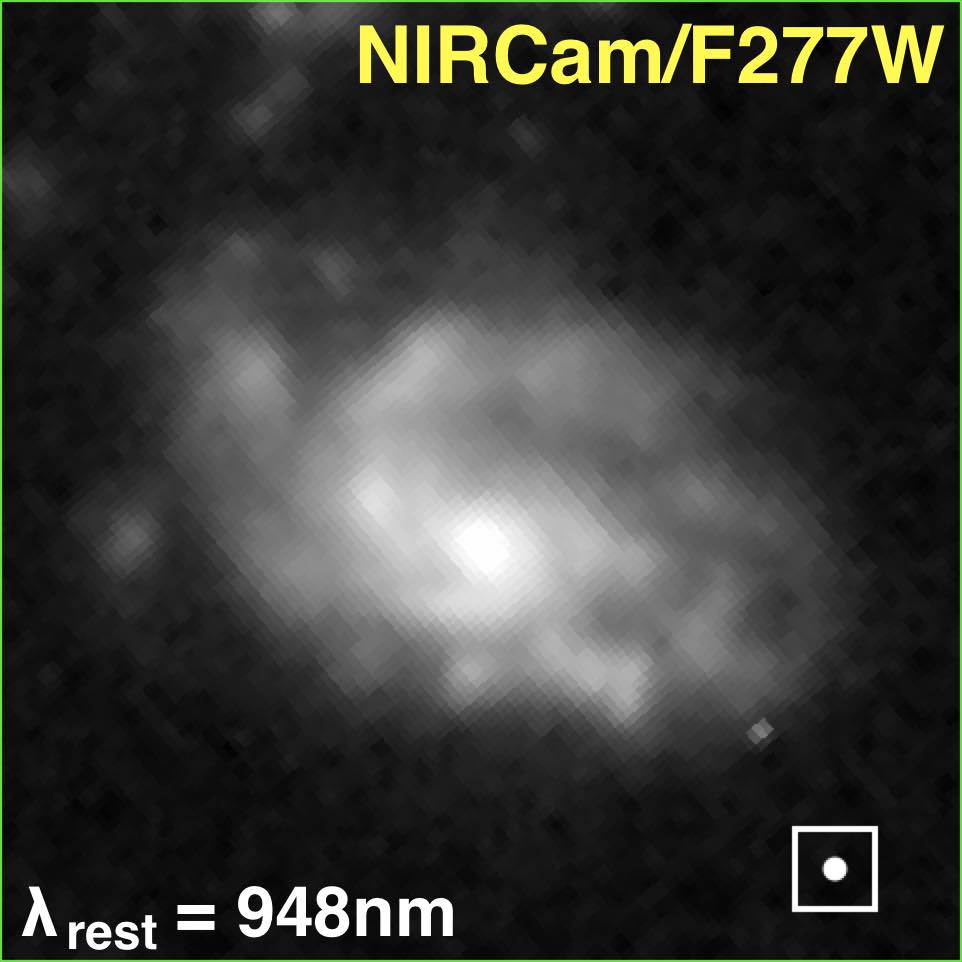}
    \includegraphics[width=3.35cm]{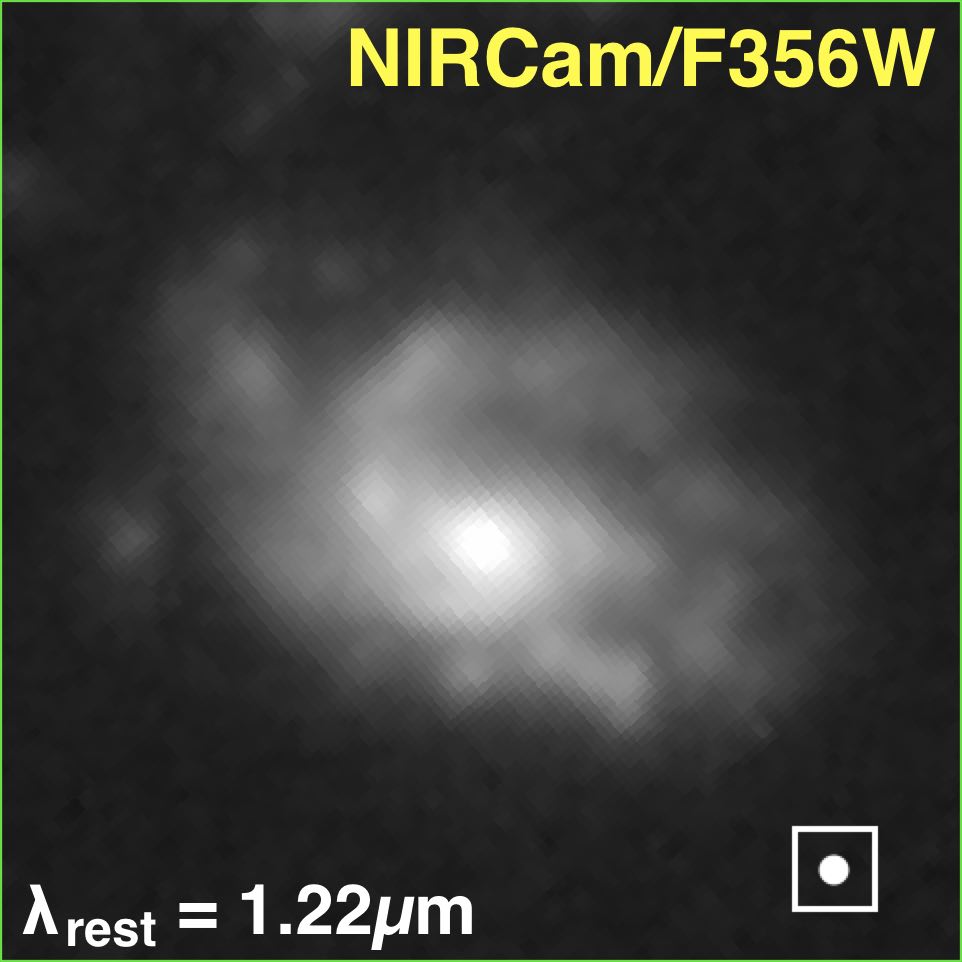}
    \includegraphics[width=3.35cm]{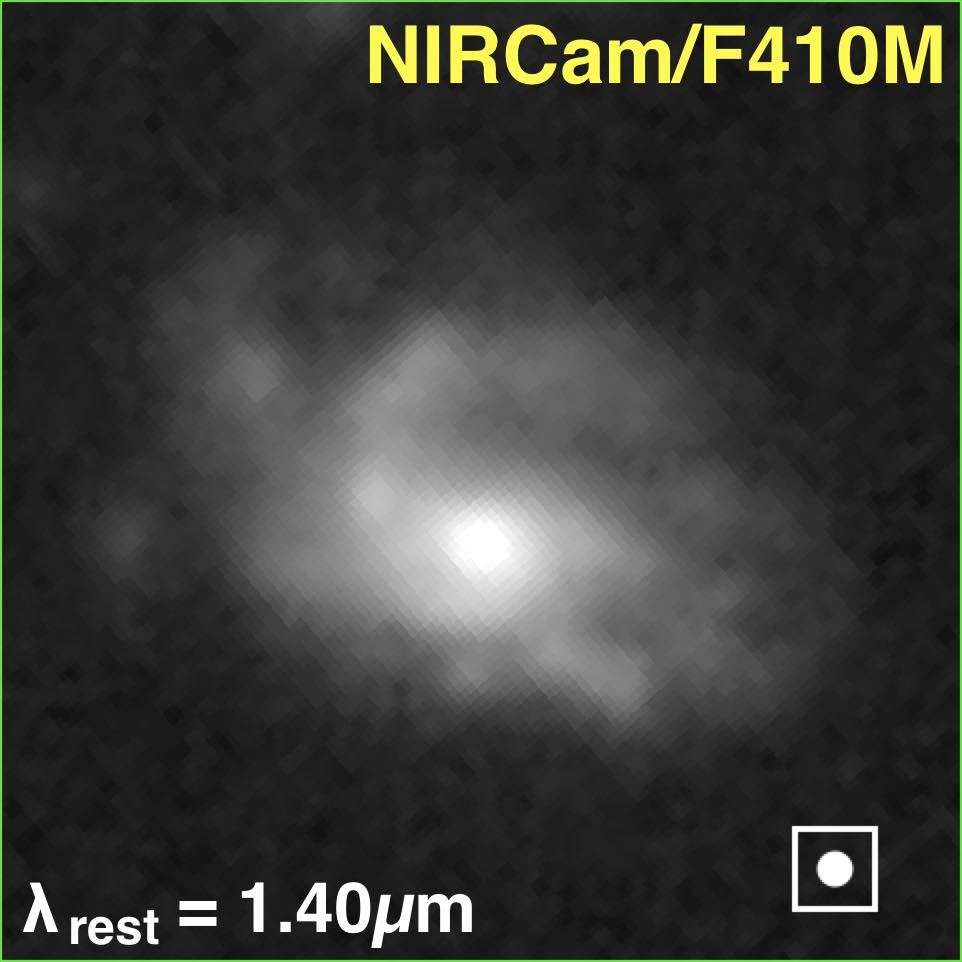}
    \includegraphics[width=3.35cm]{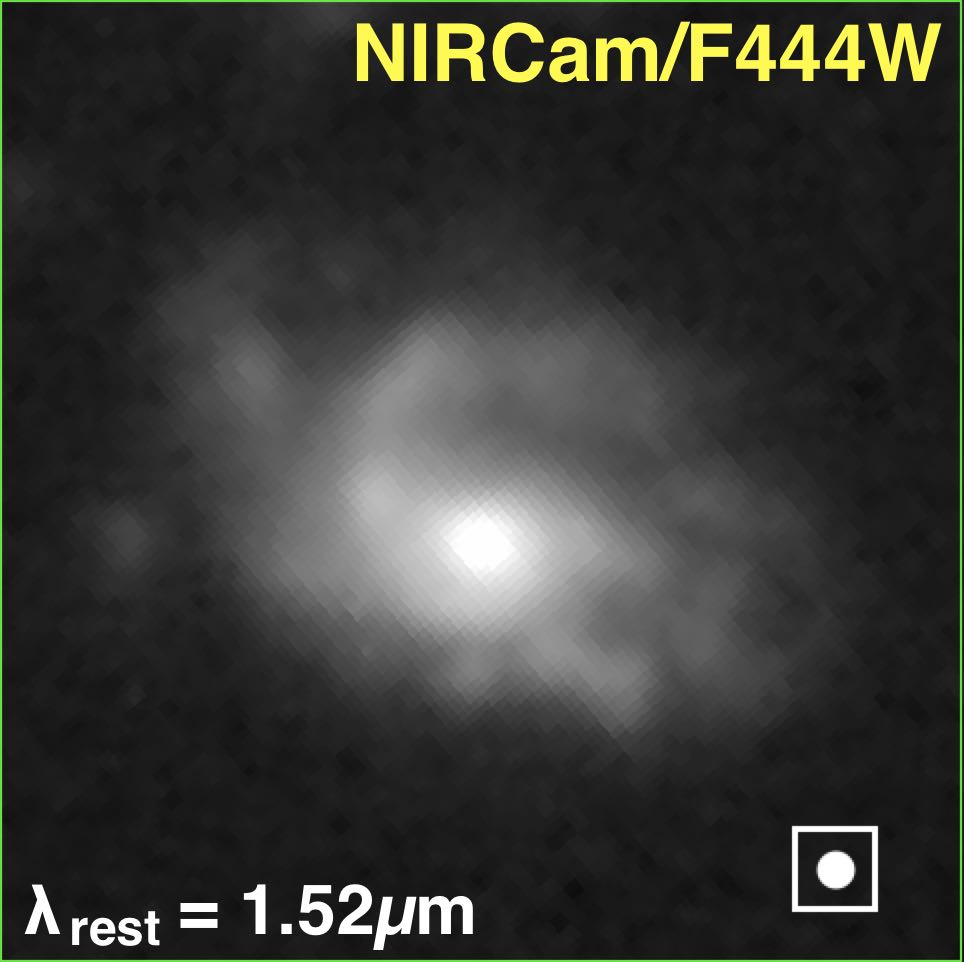}
    \caption{Cutouts of HST/ACS and JWST/NIRCam images of the DSFG ID15371 at $z_{spec} = 1.921$. Cutout size: $3.6'' \times$ $3.6''$. We also indicate the rest-frame wavelength corresponding to each filter (white label). The filled circle in the white box illustrate the PSF size for each filter.}
    \label{fig:cut_id15371}
\end{figure*}

In this Section, we detail the methods used to analyze each galaxy, taking one of them (ID15371) as an example, to better clarify the procedure that we applied to all galaxies.
For each galaxy, we started by creating cutouts in each band (HST/ACS F606W, F814W and JWST/NIRCam F115W, F150W, F200W, F277W, F356W, F410M, F444W). 
We show the cutouts of a DSFG in Fig. \ref{fig:cut_id15371} where one can already see by eye a difference between the disk visible in all bands and the center of the galaxy invisible in the HST/ACS images but getting brighter at longer wavelengths, justifying the need to study each component individually rather than the galaxy as a whole.

The JWST/NIRCam images have a spatial resolution ranging from 0.040$''$ at $1.15\mu$m  up to 0.145$''$ at $4.4\mu$m. The larger $4.4\mu$m PSF allows for a resolution in physical size down to 1.23 (1.12) kpc for a galaxy at redshift 1.5 (3). This means that we were able to spatially resolve galaxy substructures down to a radius $\sim 0.6$kpc. This made the resolution of F444W perfect for this study as we know the sizes of compact star-forming regions and giant clumps to be $\sim 1$kpc (\citealt{gomez-guijarro_goods-alma_2022,rujopakarn_alma_2019,forster-schreiber_constraints_2011}). 

%In Sect. \ref{subsec:sed}, we detail the SED fitting procedure and in Sect. \ref{subsec:prop}, we present the morphological properties we chose to probe; the clumpiness and the lopsidedness. 

\subsection{Measuring galaxy sizes}\label{subsec:sizes}

Several studies have shown that the regions of star-formation, either traced by the dust emission at 1.1mm observed with ALMA or by the radio continuum emission detected by the \textit{Very Large Array} (VLA), are more compact than the optical size of the galaxy (\citealt{puglisi_main_2019,gomez-guijarro_goods-alma_2022,fujimoto_demonstrating_2017,jimenez-andrade_radio_2019,jimenez-andrade_vla_2021}).

With its sensitivity of the near and mid-IR, JWST can detect both the obscured star-forming central part of each galaxy invisible with HST and the less obscured larger system, invisible with ALMA or VLA and bridge the gap. 

To investigate this, we measured the total near-IR half-light radius ($R_{e,NIR}$) of each galaxy in the closest band to $1.6\mu$m rest-frame (F410M or F444W filter depending on the redshift).
This rest-frame wavelength was chosen as it is a known tracer of the stellar mass of galaxies and is not affected by dust attenuation (\citealt{hainline_slar_2011,casey_dusty_2014}). Moreover a recent study using NIRCam/CEERS data showed the excellent agreement between the near-IR size and the stellar mass size of galaxies around cosmic noon (\citealt{van_der_wel_stellar_2023}).
We measured $R_{e,NIR}$ from a curve of growth method, given that in most cases the PSF has a negligible effect (much smaller than any $R_{e,NIR}$). 
The $R_{e,NIR}$ was defined as the radius of a circular aperture, centered at the center of mass (barycenter) of the galaxy, which encompassed half of the total flux density of the galaxy at the considered wavelength and still corrected to take the PSF into account which is important for our few most compact galaxies that approach the size of the NIRCam long wavelength PSF FWHM.
To estimate the uncertainty, we used the fact that we typically have a 5\% uncertainty on the measurement of the total flux of the galaxy (see Sect. \ref{subsec:flux} for more details on the photometry measurements). We also measured the bias introduced when using a circular aperture for edge-on galaxies (like ID23510 in Fig. \ref{fig:cutouts_I}) by comparing the fluxes encompassed in an elliptical aperture and a circular aperture. The difference is about 5\%. Hence, by changing the total flux of the galaxy within 10\% we can estimate the standard deviation on $R_{e,NIR}$ for which 50\% of the total flux is encompassed. 

We also measured the total optical half-light radius ($R_{e,O}$) of each galaxy in the closest band to 550nm rest-frame following the same procedure to compare it to $R_{e,NIR}$.

\subsection{Identification of cores and bulges}\label{subsec:cores}

\begin{figure*}[htb]
    \centering
    \includegraphics[width=0.24\linewidth]{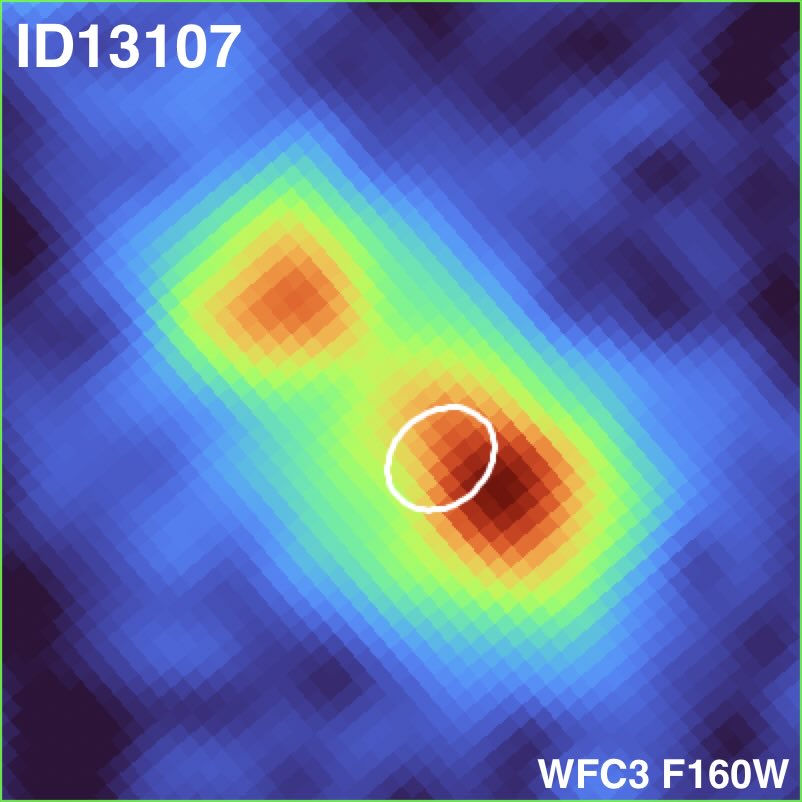}
    \includegraphics[width=0.24\linewidth]{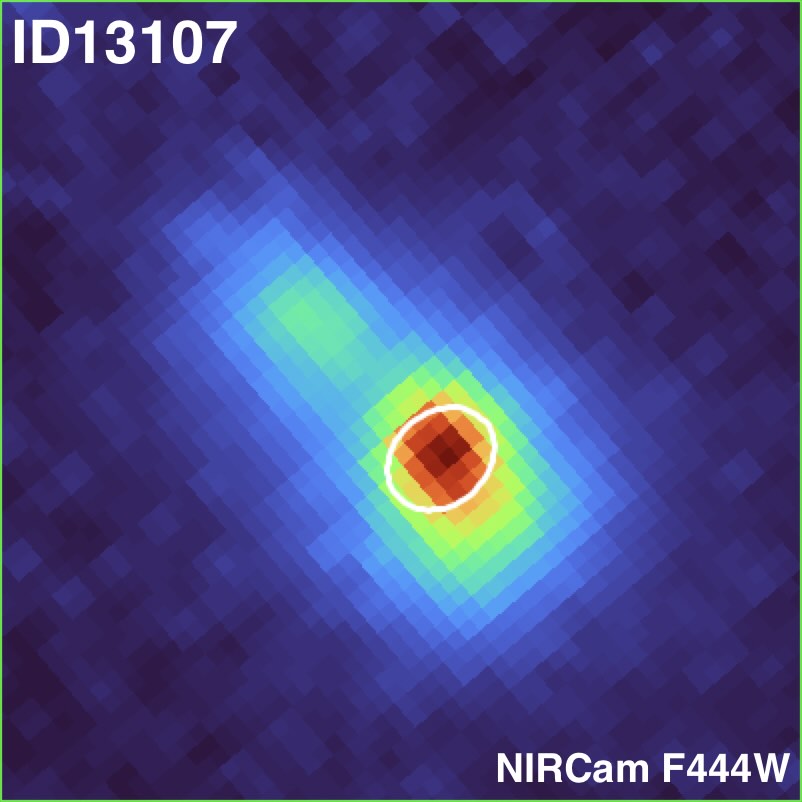}
    \hfill
    \includegraphics[width=0.24\linewidth]{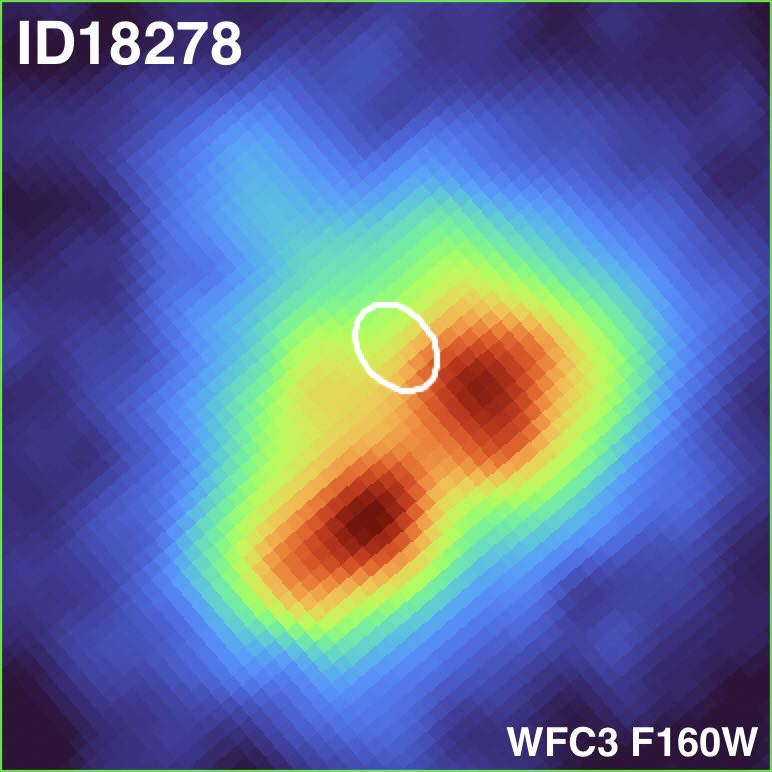}
    \includegraphics[width=0.24\linewidth]{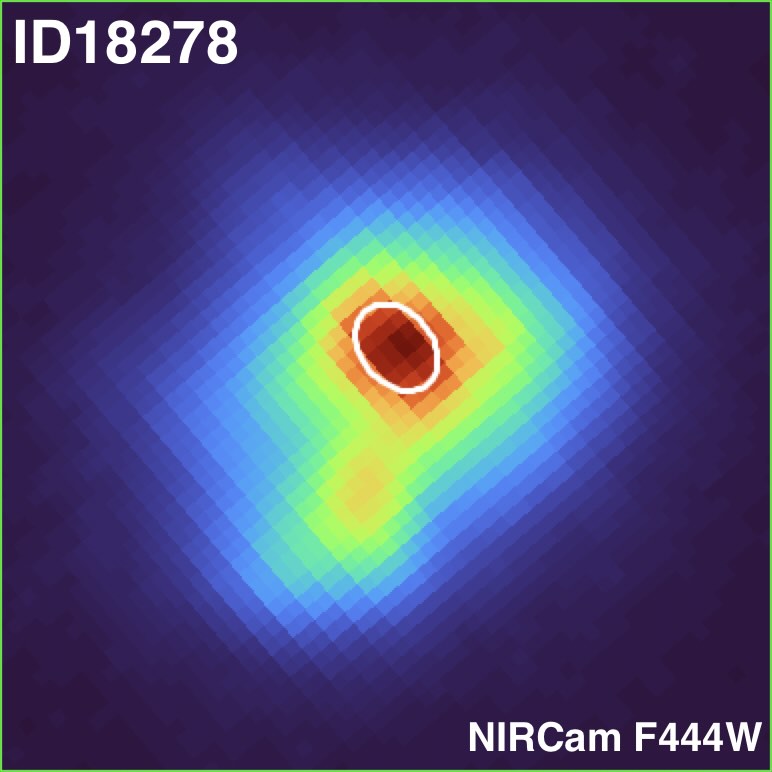}
    \includegraphics[width=0.24\linewidth]{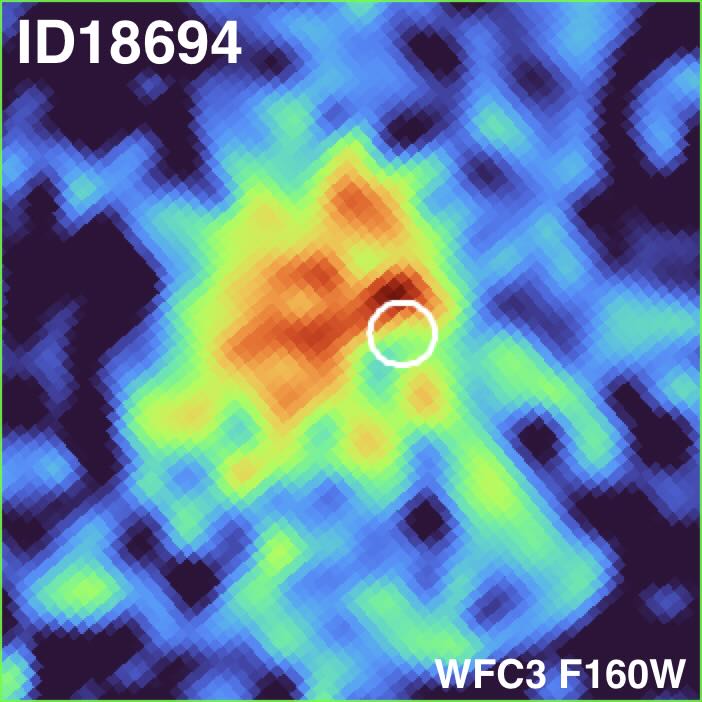}
    \includegraphics[width=0.24\linewidth]{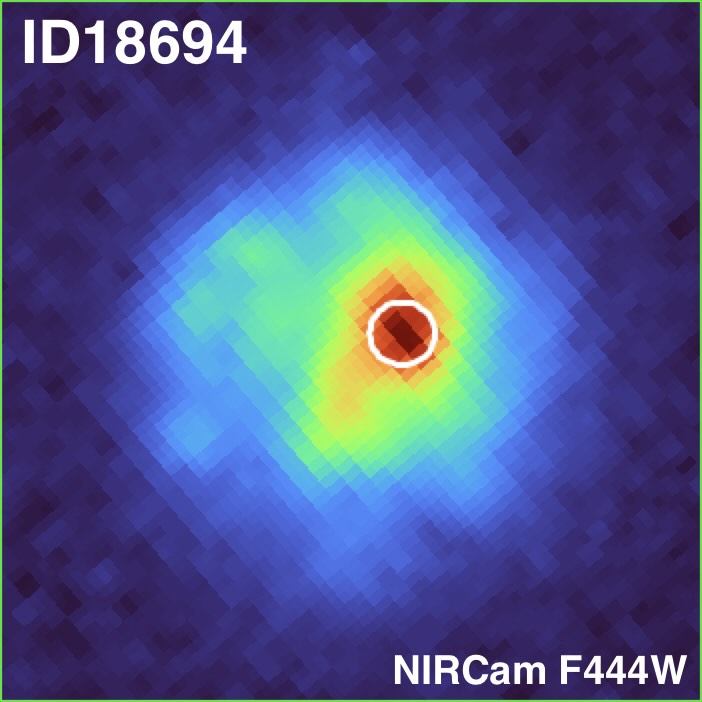}
    \hfill
    \includegraphics[width=0.24\linewidth]{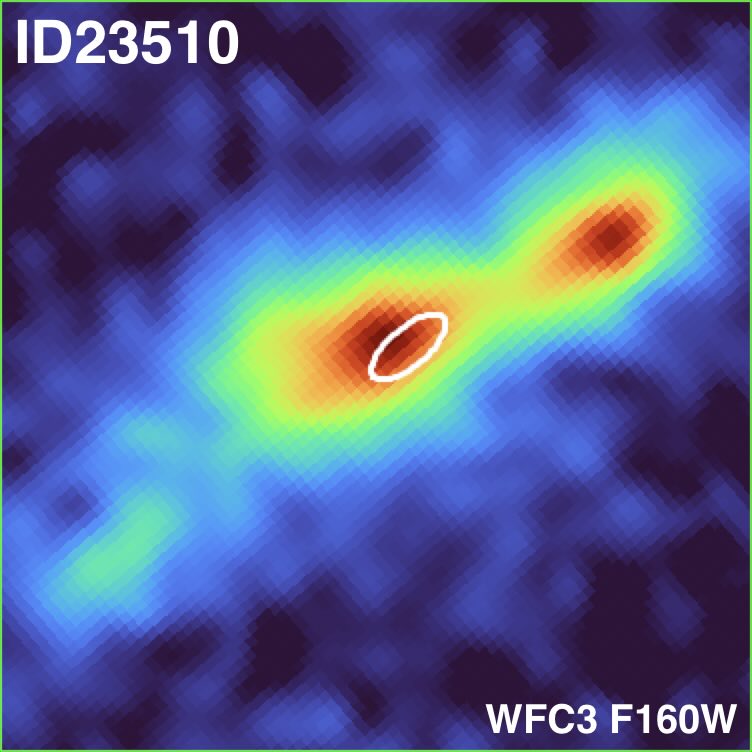}
    \includegraphics[width=0.24\linewidth]{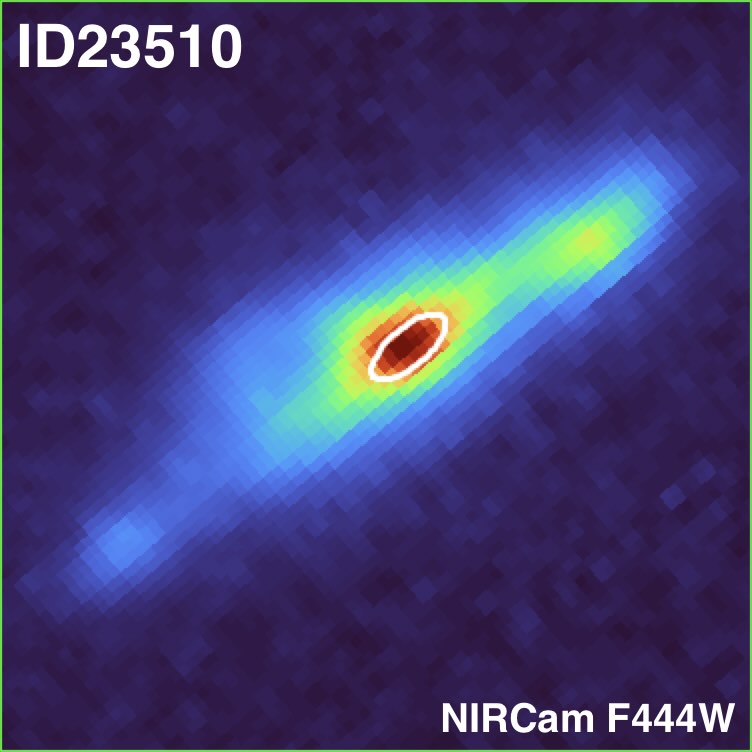}
    \includegraphics[width=0.24\linewidth]{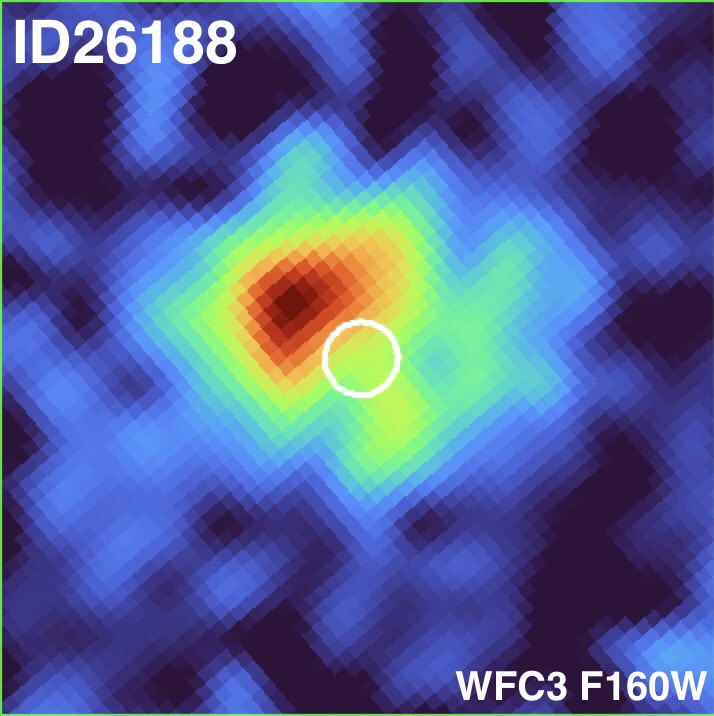}
    \includegraphics[width=0.24\linewidth]{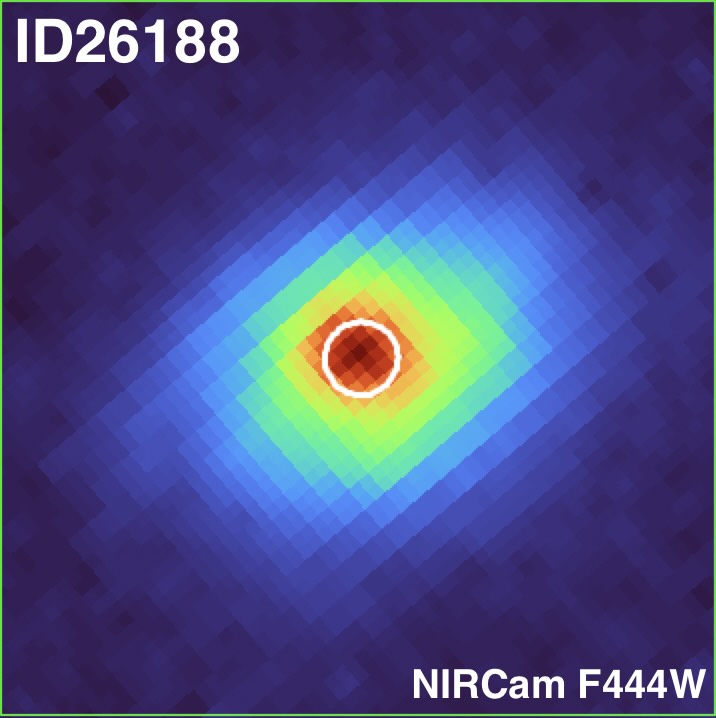}
    \hfill
    \includegraphics[width=0.24\linewidth]{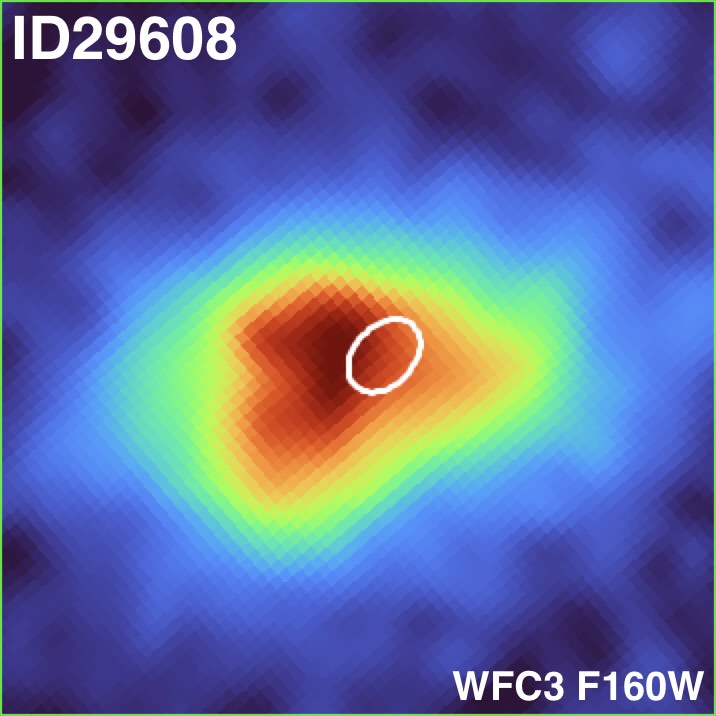}
    \includegraphics[width=0.24\linewidth]{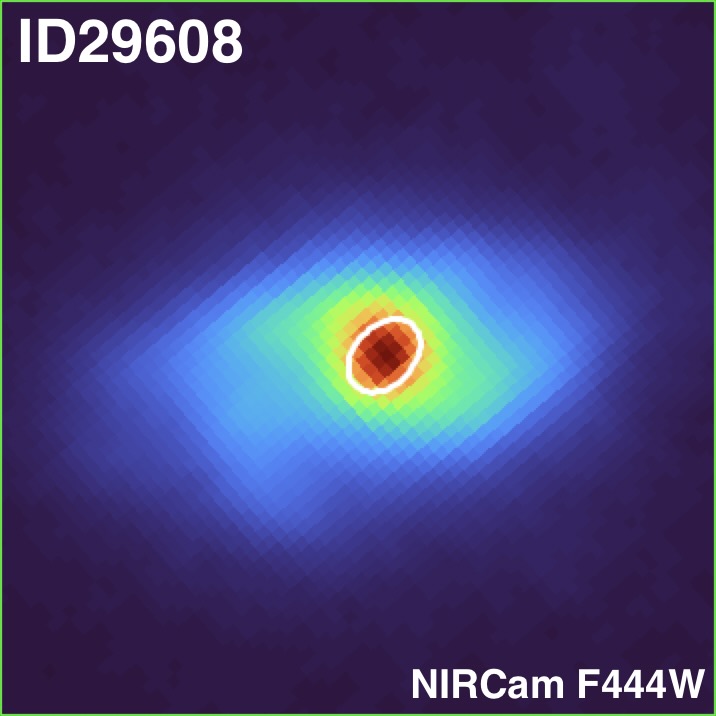}
    \caption{Six galaxies where the F444W peak is hardly distinguishable in the RGB images (Fig. \ref{fig:cutouts_I}, \ref{fig:cutouts_II}, and \ref{fig:cutouts_III}) but is clearly identifiable in the isolated JWST/NIRCam F444W filter (right-hand side of each column). We show the HST/WFC3 F160W cutouts (left-hand side of each column) of the same galaxies for comparison. The white ellipse encompasses the peak identified with the NIRCam.}
    \label{fig:comp_hst_jwst}
\end{figure*}

Depending on the redshift, the F444W filter of the NIRCam probes the rest-frame near-IR between 1.1$\mu$m and 1.8$\mu$m which is a good tracer of stellar mass (\citealt{van_der_wel_stellar_2023}). Hence, inspection of galaxy morphologies in this filter allowed us to search for the center of mass of each galaxy in our sample (or lack there-of) as a well defined peak. By looking at Fig. \ref{fig:cutouts_I}, \ref{fig:cutouts_II} and  \ref{fig:cutouts_III}, the F444W peak is easily identifiable for most of our galaxies. For the 6 galaxies where the peak does not  appear clearly in the RGB images, we show their F444W cutouts in  Fig.\ref{fig:comp_hst_jwst} where the peak becomes clearly identifiable. Hence, we were able to  clearly identify a peak in the flux distribution of this filter for every galaxy in our sample. We then defined a region in each galaxy encompassing the peak (showed as the red dotted ellipse in Fig.\ref{fig:cutouts_I}, \ref{fig:cutouts_II} and \ref{fig:cutouts_III} and the white ellipse in Fig.\ref{fig:comp_hst_jwst}), as the center of mass of the galaxy. The regions are defined by eye as the peak is easily identifiable in every galaxy, the limit of the center of mass being where the flux coming from the red F444W filter stops dominating the RGB (F115W, F200W, F444W) colors.

Generally, a bulge is defined in the literature as a quiescent central component with a high Sersic index (e.g., $n\sim 4$), and is a common component in local massive galaxies. In our study we did not attempt obtaining  Sersic fits of separate components, and, more importantly, we anticipated that in many cases the central concentrations would not be quiescent, actually, most of them were highly star forming and dust attenuated. We thus decided to call the central concentrations cores when they were star forming and bulges when they were quiescent.
They are represented by the regions delimited by the red dotted lines in all galaxies in Fig. \ref{fig:cutouts_I}, \ref{fig:cutouts_II} and \ref{fig:cutouts_III}. 

We emphasize that for most of our sample, it would not have been possible to securely identify the center of mass only based on HST images, and even not at all for at least a third of the sample as shown in Fig. \ref{fig:comp_hst_jwst} where we compare HST/WFC3 F160W and JWST/NIRCam F444W cutouts. Due to the high dust content, the center of mass of some galaxies is hidden in the HST/ACS or WFC3 bands (see, e.g., ID15371 in Fig.\ref{fig:cut_id15371} or Fig.\ref{fig:comp_hst_jwst} as obvious examples). In Fig. \ref{fig:cut_id15371}, the dusty core becomes bright and dominates the galaxy's total emission starting only in the F277W filter. It is however still distinguishable in the shorter wavelength filters (notably F115W and F150W), but the size of the core emission being more compact than the HST/WFC3 IR PSF FWHM, it is not resolved, hence indistinguishable in the HST images, even in the IR bands (see Fig.\ref{fig:comp_hst_jwst} for more examples).
This demonstrates once again the power and usefulness of JWST when it comes to studying high-\textit{z} DSFGs.

\subsection{Lopsidedness}\label{subsec:lop}

Having defined the core or bulge of each galaxy, we considered the rest to be the disk. Hence, we could obtain an evaluation of the lopsidedness for each galaxy. We considered it to be an important property to investigate because a lot of galaxies in our sample are obviously highly lopsided already by visual inspection (see for example ID11887, ID13776, ID18278, ID18694 in Fig. \ref{fig:cutouts_I} and \ref{fig:cutouts_II}). To quantitatively study this phenomenon, we defined two parameters: the eccentricity, defined as: 
\begin{equation}\label{eq:ecc}
    E = \sqrt{\frac{(X_{core} - X_{disk})^{2} + (Y_{core} - Y_{disk})^{2}}{R_{disk}^{2}}},
\end{equation}
where ($X_{core},Y_{core}$) and ($X_{disk},Y_{disk}$) are the coordinates of the core of the galaxy and of its disk respectively, while $R_{disk}$ is the radius of the disk. The center of the core was simply defined as the pixel with the maximum flux density in the F444W filter. The center of the disk was defined as the barycenter of the disk measured in the rest-frame optical band (F150W or F200W depending on redshift). We measured it in the optical and not in the near-IR because the disk is less attenuated than the core, hence brighter than the core at these wavelengths. To not be biased by the core, we applied a circular mask centered on ($X_{core},Y_{core}$) with a radius defined by the closest pixel to the center that has a F444W flux density less than half the core center flux density. Finally, $R_{disk}$ was calculated using a circular aperture centered on ($X_{disk},Y_{disk}$) encompassing half of the disk flux density.
This quantifies the eccentricity of the disk with respect to the core or bulge compared to its size and is a-dimensional.

The other quantity that we defined to probe the lopsidedness of the galaxies is the asymmetry. The asymmetry was calculated for the F444W NIRCam filter as we were trying to probe the mass distribution asymmetries and, as previously mentioned, F444W is the best tracer of the stellar mass distribution. We calculated the asymmetry by rotating each image by $180^{\circ}$ and subtracting it from the original image, the center of rotation was ($X_{core},Y_{core}$) from Eq. \ref{eq:ecc}. The asymmetry is defined as: 
\begin{equation}
    A = \frac{\sum_{i=0}^{N} \vert F_{i} - F_{i}^{180^{\circ}}\vert }{F_{tot}},
\end{equation}
where $F_{i}$ and $F_{i}^{180^{\circ}}$ are the flux of the $i$-$th$ pixel and its $180^{\circ}$ symmetric counterpart with respect to the center of the central core or bulge as defined in Equation \ref{eq:ecc}. $F_{tot}$ is the total flux of the galaxy. Since we worked on background subtracted images, we considered the background asymmetry to be negligible. This quantity describes how smoothly and how symmetrically the stellar mass is distributed around the central core or bulge of the galaxy and is also a-dimensional. Usually, the lopsidedness is probed using a Fourier decomposition (e.g., \citealt{dolfi_lopsidedness_2023,kalita_bulge_2022,jog_lopsided_2009,bournaud_lopsided_2005}). We decided to use a different, simpler method; the asymmetry, that has already been used in gas velocity space and was found to correlate well with the Fourier analysis of stars (\citealt{bournaud_lopsided_2005,matthews_high-resolution_1998}).

\subsection{Clumpiness}\label{subsec:clumpiness}

After identifying the core or bulge of each galaxy, we investigated the surrounding disk-like structures. Some of the galaxies have a smooth disk, others have a much more perturbed and complex disk morphology showing a large number of clumps (see Fig. \ref{fig:cutouts_I}, \ref{fig:cutouts_II} and \ref{fig:cutouts_III}).

We did not embark in a physical study of the clumps in this work. Our goal for this paper is to assess the presence of clumps in the disks and have an idea of how fragmented the disks are. Hence, we did not try to derive any physical properties of the individual clumps. %We designed a simplistic method that was enough for the purpose of this study.
We decided to measure a clumpiness index, defined as the number of clumps in the disk of each galaxy. We counted the number of clumps visually identifiable in the RGB (F115W, F200W, F444W) image, making sure that the bulge or central concentration was not counted as a clump. This number varies from 0 up to 7 for the clumpiest galaxy. To be counted as a clump, the feature had to be compact compared to the size of the galaxy, and either had to have a different RGB color from the surroundings or appear as a local brighter spot. The clumps appear most clearly at the shortest wavelength (F115W or F200W filters), as expected (\citealt{wuyts_smoother_2012}). For ID15371, we identified 4 clumps, they are highlighted by the white ellipses in the left panel of Fig. \ref{fig:id15371}.

\subsection{Spatially resolved photometry} \label{subsec:flux}

To measure physical quantities of our galaxies, we need photometry measurements. The galaxies showed in Fig. \ref{fig:comp_hst_jwst} all have some parts that are bright in the HST/WFC3 image and are invisible in the NIRCam long wavelength filter and some parts that are the other way around. These different parts obviously will have different SEDs. This is also visible in Fig. \ref{fig:cutouts_I}, \ref{fig:cutouts_II} and \ref{fig:cutouts_III} where most galaxies show a wide range of colors. For this reason, we decided to divide our galaxies in several components and study them individually.

\begin{figure}[htb]
    \centering
    \resizebox{\hsize}{!}{
    \includegraphics[width=0.5\linewidth]{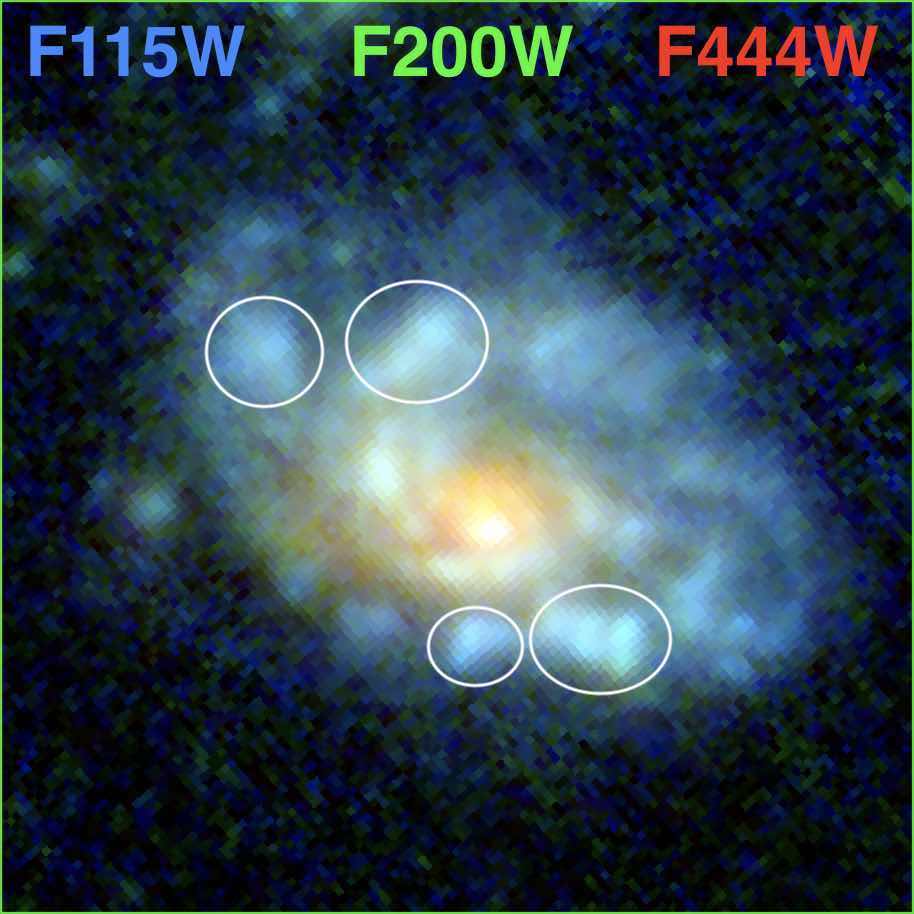}
    \includegraphics[width=0.5\linewidth]{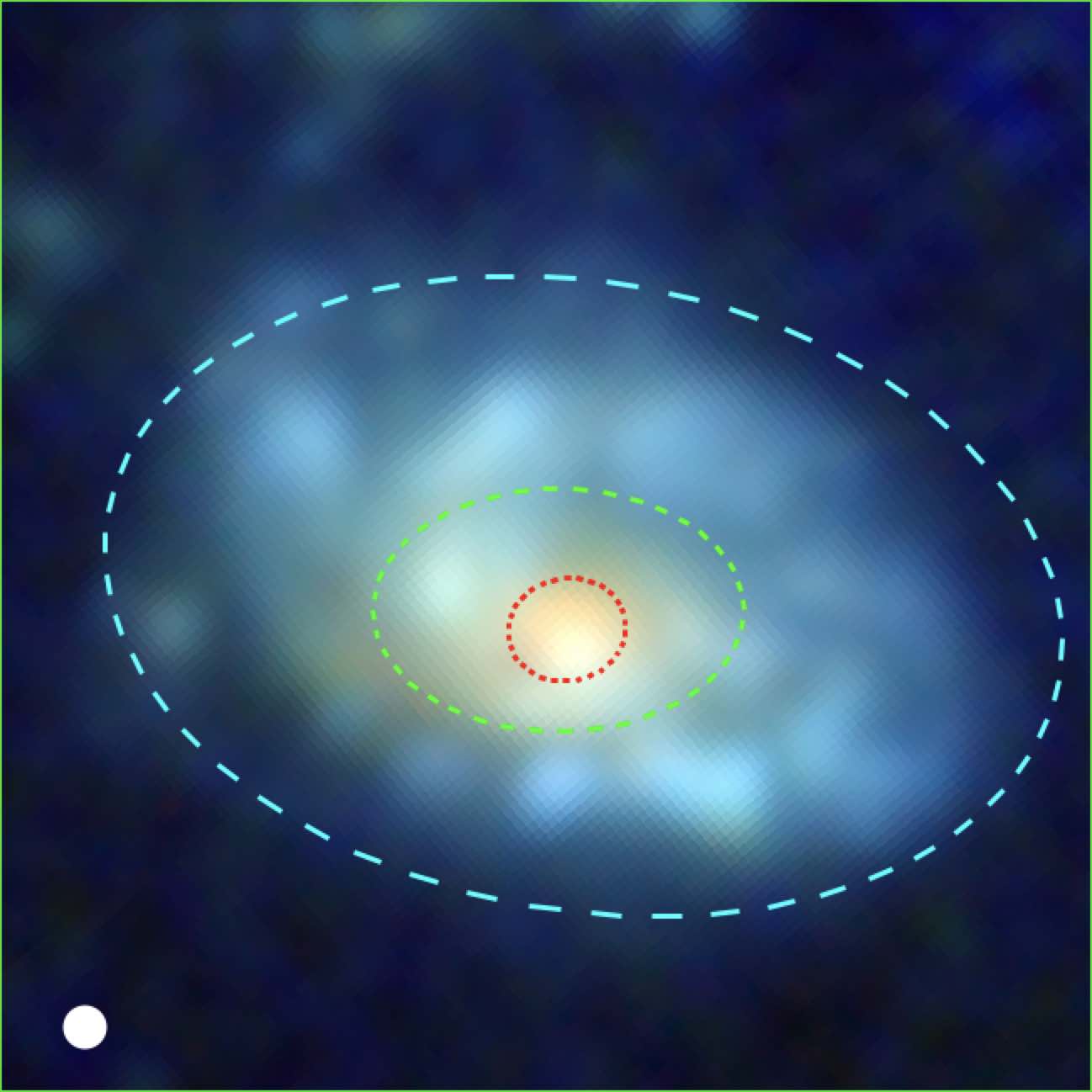}}
    \caption{RGB (F115W, F200W, F444W) image of the galaxy ID15371 ($3.6"\times$ $3.6"$) at $z_{spec} = 1.921$ before (left panel) and after (right panel) PSF matching. In the left panel, the white ellipses show the features we identified and counted as clumps. In the right panel, the colored dotted lines correspond to the division of the galaxy in homogeneously colored regions that we studied separately and the white filled circle shows the PSF size.}
    \label{fig:id15371}
\end{figure}

By dividing each galaxy in sub-galactic components, there was a risk that small regions would get close to the PSF FWHM of some filters. Hence, leading to an underestimation of the flux at the longest wavelengths, and to an artificial deformation of the SED. To avoid this, we decided to work on PSF-matched images using the broader PSF of the F444W filter. The procedure was adopted from \citealt{gomez-guijarro_jwst_2023}, to which we refer for more details. Briefly, PSF-matching was performed on JWST/NIRCam and HST/ACS images by convolving them with a convolution kernel created using the software PyPHER (\citealt{boucaud_convolution_2016}), from PSF images of the different bands built by stacking point sources selected using the software PSFEx (\citealt{bertin_automated_2011}).
In Fig. \ref{fig:id15371}, we show RGB images of the DSFG ID15371 using (F115W, F200W, F444W) before and after PSF-matching (left and right panel respectively). 

We then proceeded to divided our galaxies in several components based on the PSF-matched images. For the simplest cases we only had 2 components: the core or bulge and a homogeneous disk. However, when the disk had several clumps or patches with different colors in the RGB image, we broke it down to several elliptical regions. Each region was designed so that it had, qualitatively, a homogeneous (F115W, F200W, F444W) color. The division of the disk is done by visual inspection. We emphasize that we seek to study each region that has a different color, hence, if several clumps were close and with a similar RGB color, we considered them to be part of the same component. To avoid any bias in our flux measurements, hence in the extracted physical properties, we tried to respect a balance between the size of the component that has to be larger than the PSF FWHM (0.145'') and the homogeneity of the RGB color inside it. %We estimated the human eye to be good enough for this job. 

We emphasize that the components are not necessarily concentric as most of the galaxies are not radially symmetric (see Fig.\ref{fig:comp_hst_jwst} for the least symmetrical galaxies) and are not limited in number. If we observed, for example, two blue disconnected patches in a galaxy, we defined them as two different components and studied them individually. In the case of ID15371, we divided the galaxy in three regions, the red central core or bulge, the bluer disk and an intermediate region that is still part of the disk but close to the red core and with intermediate colors (see Fig. \ref{fig:id15371}).

In terms of rest-frame colors, since our sample of galaxies is distributed across $z \sim 1.5$ to $z \sim 3$, the three filters that we used to make the RGB images probe the rest-frame near-UV/blue ($300-460$nm), the rest-frame green/red ($500-800$nm) and the rest-frame near-IR ($1110-1780$nm) for F115W, F200W and F444W respectively. The scatter in rest-frame wavelength is shorter or equal to the band-width of each filter. This means that we globally probed consistent colors between galaxies when looking at the RGB images. 

In Fig. \ref{fig:cutouts_I}, \ref{fig:cutouts_II} and \ref{fig:cutouts_III}, for each galaxy we overlay the delimitation of the different components we decided to study separately (these RGB images are showed before PSF-matching).

After having defined the regions to study, we measured the flux in each band for each individual region. To do so, we summed the value of each pixel in each region of the science image. The pixels were counted only once, meaning that the flux in the smaller regions (like the red ellipse for ID15371) was not included when calculating the flux of larger regions (like the green ellipse for ID15371, see Fig. \ref{fig:id15371}).

For the properties that we wanted to extract from the SEDs to be reliable, it was crucial that we had reliable uncertainties on the flux measurements.
To estimate these uncertainties, we re-normalized the errors propagated via the Root Mean Square (RMS) images. We defined the uncertainties as: 
\begin{equation}
    df = f_{\lambda,N}\times \sqrt{\sum_{i=0}^{N} \sigma_{i}^{2}},
\end{equation}
where the sum was made on all pixels in the region, $\sigma_{i}$ is the RMS of the pixel $i$, and $N$ the total number of pixel in the region. We decided to define $f_{\lambda,N}$, a normalization factor that takes into account extra noise, for instance, from the correlated signal between pixels that is particularly important for the long wavelength filters that were drizzled from a pixel size of 63mas to 30mas. To calculate this factor, we measured the flux dispersion in $\sim 20$ empty regions of the science image for several apertures in each band. We then compared this value to the RMS calculated from the RMS image in apertures of the same size and the normalization factor is defined as their ratio. To be conservative, we never applied a factor leading to lower uncertainties. These factors are generally small, ($f_{\lambda,N} \sim 1.5$ at most).

\subsection{SED Fitting} \label{subsec:sed}

To characterize our sample of galaxies, we needed to have access to their resolved $M_{*}$ and SFR. To this aim, we fitted each galaxy component SED using the Code Investigating GALaxy Emission (CIGALE, \citealt{boquien_cigale_2019}). 
We used a single declining exponential model also known as ``$\tau$ model" to model the star formation history (SFH) of each galaxy.
We adopted the \cite{bruzual_stellar_2003} model to compute the spectral evolution of single stellar populations with a fixed solar metallicity of Z = 0.02 which is reasonable for $M_{*} \sim 10^{10-12}M_{\odot}$ DSFGs following the Mass-Metallicity relation (\citealt{ma_origin_2016}).
After testing with and without including nebular emissions, we decided not to include them as, for our sample, they lead to higher $\chi ^{2}$ with no noticeable effect on the extracted properties ($A_{V}$, $SFR$, $M_{*}$ and redshift).

However, some galaxies showed possible signature of strong emission lines, visible as green patches or clumps in Figs. \ref{fig:cutouts_I}, \ref{fig:cutouts_II} and \ref{fig:cutouts_III}. When these potential strong emission line were in quiescent regions, including nebular emissions in the fit did not have any effect on the results. When the potential strong emission lines were in star-forming region, including nebular emissions in the fit of the SED only mildly changed the key parameters like the photometric redshift, dust attenuation, SFR or $M_{*}$, keeping them well consistent within error bars. However, adding the nebular emissions allowed us to better constrain the $M_{*}$ and dust attenuation by reducing their uncertainties. Since the effect of nebular emission is mild and in order to keep consistency in our study and be conservative in our uncertainties calculation, we decided not to include them. We underline that studying in detail the emission lines would be going beyond the scope of this paper. We discuss the possible origin of these potential strong emission lines in more detail in Sect. \ref{subsec:em_line}.

We used a modified \cite{charlot_simple_2000} dust attenuation law and the \cite{draine_dust_2007} dust emission models update from 2014 to predict FIR flux densities. The idea behind the modification of \cite{charlot_simple_2000} model is that young stars embedded in their birth cloud suffer from additional attenuation compared to stars that have broken out and escaped into the ISM, and that the attenuation curves associated to the birth cloud and the ISM must be different. In practice, this is modelled
by assuming two different power-law attenuation curves of the form $A(\lambda) \propto \lambda ^{\delta}$: one for the birth cloud with a slope of $\delta _{BC} = -1.3$, and one for the ISM with a slope of $\delta _{ISM} = -0.7$. Because radiation from young stars has to travel through both the birth cloud and the ISM to escape the galaxy, the spectrum of stars younger than 10Myr are attenuated by both the birth cloud and ISM curves. Stars older than 10Myr are only attenuated by the ISM curve (\citealt{boquien_cigale_2019}).

For the redshift, we used the \cite{stefanon_candels_2017} catalog, as well as the latest redshift catalog published by \cite{kodra_optimized_2022}. We encountered three different cases. The first case was if we had a high-quality spectroscopic redshift, then we used it and fixed it. We have 5 galaxies with a spectroscopic redshift. Then, if we had a grism-based redshift from 3D-HST, we downloaded the spectrum and examined its quality, the actual features detected, the redshift probability distribution and defined the redshift and its uncertainty accordingly. We have 10 galaxies for which we find a high-quality grism-based redshift. In the last case, if we only had photometric data, we allowed $(1+z)$ to vary within $\pm 10\%$. We have 7 galaxies with a photometric redshift.

\begin{figure}[htb]
    \centering
    \resizebox{\hsize}{!}{
    \includegraphics{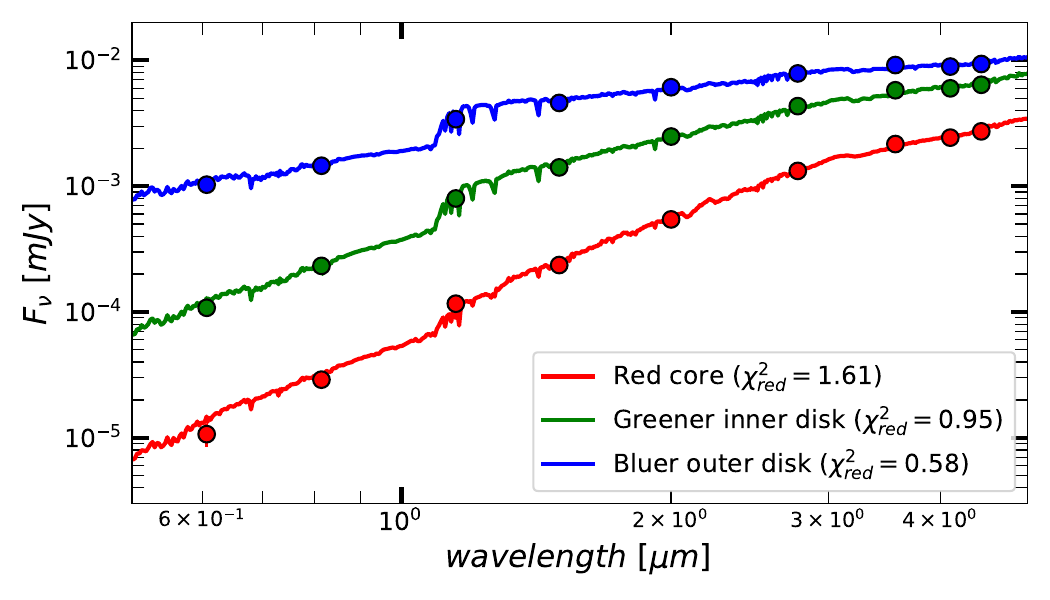}}
    \caption{Best SED models computed by CIGALE (\citealt{boquien_cigale_2019}) for the red core (red line), the blue disk (blue line), and the intermediate region (green line) of the DSFG ID15371 at $z_{spec} = 1.921$ (the same example galaxy shown in Figs. \ref{fig:cut_id15371} and \ref{fig:id15371}). We show in the legend the value of the reduced $\chi ^{2}$ ($\chi _{red}^{2}$) for each SED fit. The dots are the flux measurements from HST/ACS and JWST/NIRCam.}
    \label{fig:sed}
\end{figure}

In Fig. \ref{fig:sed}, we show the best SED models corresponding to each region of our example galaxy defined in Fig. \ref{fig:id15371}.

To be able to extract reliable information from the SED fits, it was crucial to check the fits quality. To be conservative and have reasonable $\chi ^{2}$, we decided to limit the photometric accuracy of each band to $S/N = 20$.
To consider the fit acceptable, we wanted the reduced $\chi ^{2}$ such that $\chi _{red}^{2} \leq 1.67$, which is the reduced critical value corresponding to a significance level of 10\% in the $\chi ^{2}$ test for 8 degrees of freedom. If the CIGALE fit returned higher $\chi ^{2}$ values, it meant that the input flux uncertainties might still be underestimated or that the SED model might not be adapted. In that case, we first tried to adapt the parameter space of the SED to try and get better results, based on information we had from the 3D-HST spectra, the ``super-deblended" catalog and X-ray emissions probing emission lines and AGN activity. We also tried modeling the SFH using a double exponential, allowing addition of a late starburst if necessary. If, despite our tries, the best model still had a large $\chi ^2$, we increased the uncertainties by adding 10\% of the flux to the uncertainty in each band. This can be justified by the fact that our data analysis introduces correlated noise when PSF-matching, that can be particularly important for the smallest regions.

\begin{figure}[htb]
    \centering
    \resizebox{\hsize}{!}{
    \includegraphics[width=0.49\linewidth]{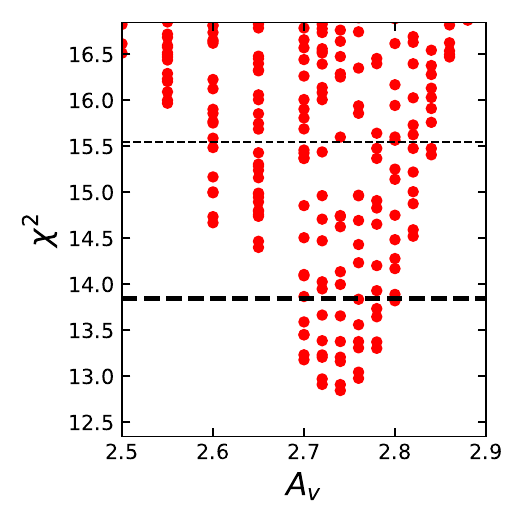}
    \includegraphics[width=0.49\linewidth]{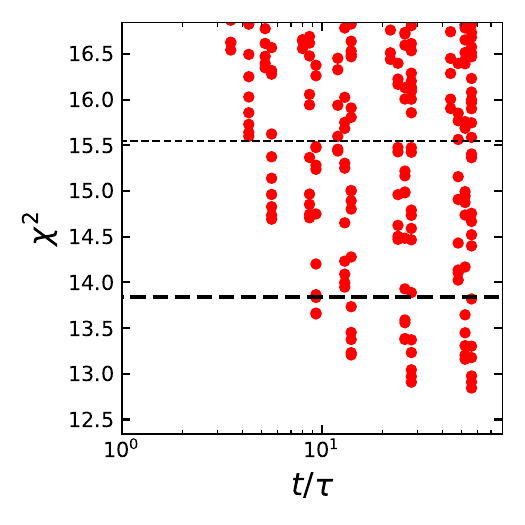}}
    \resizebox{\hsize}{!}{
    \includegraphics[width=0.49\linewidth]{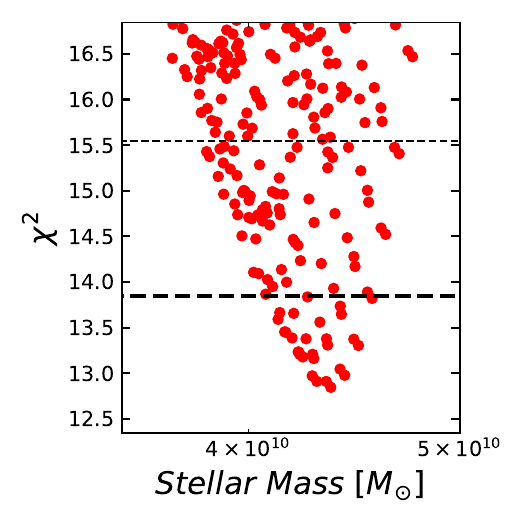}
    \includegraphics[width=0.49\linewidth]{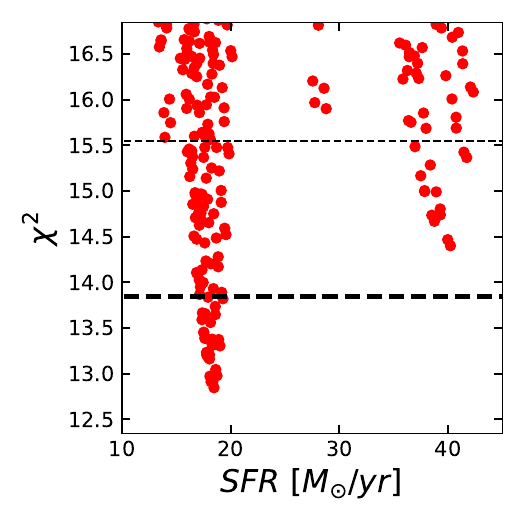}}
    \caption{$\chi ^{2}$ distributions associated with three input parameters of the SED fit, the dust attenuation (upper-left panel) and $t/\tau$ (upper-right panel), and two output properties, the stellar mass (lower-left) panel and SFR averaged over the last 10 Myrs (lower-right panel) produced by CIGALE (\citealt{boquien_cigale_2019}) for the fit of the red core SED of the galaxy ID15371 at $z_{spec} = 1.921$. The thick and thin black dashed lines correspond to the 68\% and 95\% confidence intervals, respectively.}
    \label{fig:sed_params}
\end{figure}

To estimate the robustness of the best SED model and of the extracted physical properties, we studied the $\chi ^{2}$ distributions associated to the 3 main free input parameters: the dust attenuation, the age of the stellar population and the e-folding time. In Fig. \ref{fig:sed_params}, the upper-left panel shows the $\chi ^{2}$ distribution associated with the different input values of the dust V-band attenuation $A_{V}$ of the stellar continuum used to fit the SED of the red core of the DSFG ID15371. The upper-right panel shows the same information for $t/\tau$ with $t$ and $\tau$ being the input values of the age of the oldest stars and the e-folding time of the stellar population used to define the SFH of the galaxy. Likewise, the lower panels show two output physical properties: the $M_{*}$ and the SFR averaged over the last 10Myrs. Taking the width of these distributions at $\chi ^{2}_{min} + 1$ and $\chi ^{2}_{min} + 2.7$ gave us the 68\% and 95\% confidence interval respectively (\citealt{avni_energy_1976}). They are shown by the horizontal thick and thin dashed lines in Fig. \ref{fig:sed_params}. The fact that we observed only a portion of the distribution for $t/\tau$ comes from the fact that the age is getting close to the age of the Universe; allowing a larger $t$ would not make physical sense. However, the clear shape of the distributions allow us to conclude that the red core of the DSFG ID15371 is dusty ($A_{V}$ $\sim$ 2.73), massive ($M_{*} \sim$ 4.2$M_{\odot}$) and weakly star-forming (SFR $\sim$ 18 - 40 $M_{\odot}/yr$ and $t/\tau$ $\gg$ 1).

\begin{figure}[htb]
    \centering
    \resizebox{\hsize}{!}{
    \includegraphics[trim={0.45cm 0.7cm 1.25cm 1.8cm},clip]{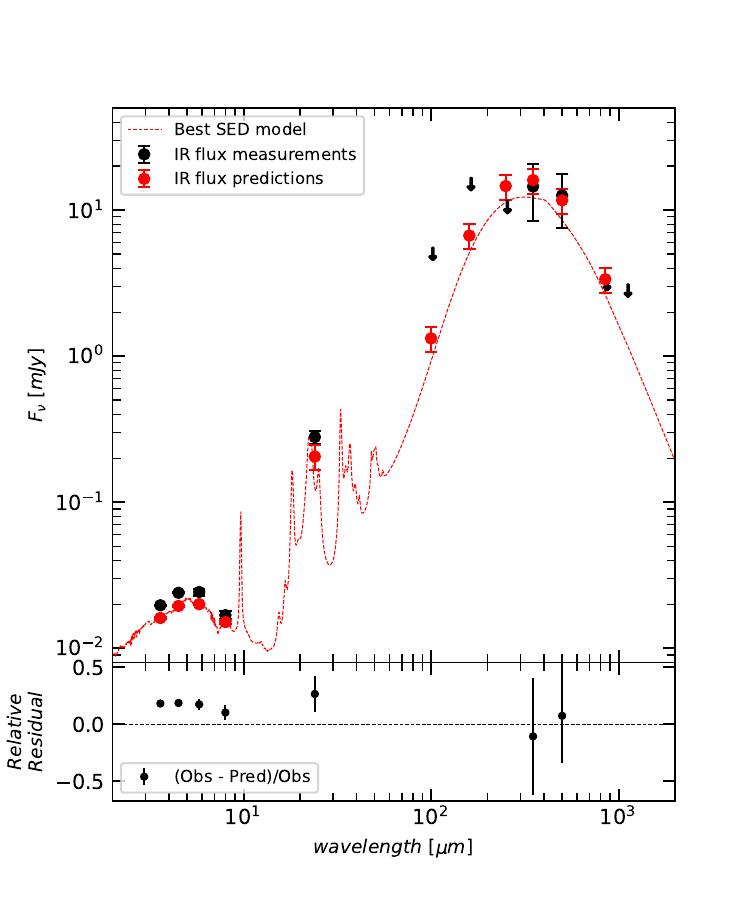}}
    \caption{Total SED of the galaxy ID15371 at $z_{spec} = 1.921$ in red. It was calculated by adding up the CIGALE best SED model of each component. The black points are the near-IR and FIR fluxes with their uncertainties or upper limits (arrows) from the super-deblended catalog (Henry et al. in preparation). From the FIR data points, we have $SFR_{IR} = (150 \pm 15)$M$_{\odot}$ yr$^{-1}$ (Henry et al. in preparation), the CIGALE fits give a consistent total $SFR_{10Myrs} = (197 \pm 125)$M$_{\odot}$ yr$^{-1}$. Given its stellar mass ($M_{*} = 10^{11}$M$_{\odot}$), this galaxy is on the MS.}
    \label{fig:sed_fir}
\end{figure}

As a sanity check, we estimated the SED of the whole galaxies by summing up the SEDs of all the components. We then compared the total SED with the near-IR and FIR flux densities measured in the super-deblended catalog (\citealt{stefanon_candels_2017}, Henry et al. in preparation) to make sure that they were consistent. Having FIR flux densities brighter than predicted by the SED fitting, could be a hint that the galaxy is in a starburst episode or that there is a deeply attenuated component that is not visible even at $4.44\mu$m. It could also be due to the presence of an AGN that boosts the FIR flux, this would be confirmed by a radio excess or an X-ray detection (Henry et al. in preparation, \citealt{stefanon_candels_2017}). We recall that we removed from the sample only one galaxy where we knew that the FIR luminosity was dominated by the AGN luminosity (see Sect. \ref{subsec:reject}) but kept those where the AGN luminosity did not  dominate the FIR luminosity. 
On the contrary, if the SED predicted a FIR flux density brighter than the one measured, it meant that there was a problem in the fitting possibly linked to the grid of the input parameters. In Fig. \ref{fig:sed_fir}, we show the comparison between the total SED of the galaxy ID15371 and the FIR flux densities. For this galaxy, the flux densities are consistent with the predicted FIR SED meaning that there is no hidden component. This is the case for all the galaxies in our sample except one (ID13107, for which we have a FIR detection brighter than the SED model, pointing toward either a deeply attenuated component or an AGN even though there is no AGN signature in X-ray or radio). However for 3 galaxies (ID13098, ID13776 and ID31281), the measured $100\mu$m flux is boosted compared to the SED predicted flux. This is possibly a signature of a hot AGN, 2 of them having an X-ray detected AGN (\citealt{nandra_aegis-x_2015}). 

By observing Fig. \ref{fig:sed_fir}, one can notice that the predicted IRAC fluxes are fainter than the actual measurements. However, this observation is not necessarily true for every galaxy. The fact that when we measured the fluxes in the NIRCam F356W and F444W filters which probe the same band as IRAC CH1 and CH2 we found fainter fluxes in the NIRCam bands for some galaxies is mostly a sign that there is some blending in the IRAC imaging.

A caveat of this SED fitting method is that we used the same SFH and parameters for all regions, some with very different properties. We chose to use the simple tau model because of the meaning of $t/\tau$ regarding the star-forming activity of the galaxy. To address this caveat, we decided to make a two-pass SED fitting, in the first pass, the goal was to separate the star-forming from quiescent components (see below for more details). In the second pass, we fitted the star-forming regions imposing a nearly flat SFH (by imposing $\tau \gg t$) to obtain a good estimate of the recent SFR. For the quiescent components, we imposed $t \gg \tau$ to tackle the degeneracy between the age and the dust attenuation.

\begin{figure}[htb]
    \centering
    %\resizebox{\hsize}{!}{
    \includegraphics[width=0.6\linewidth]{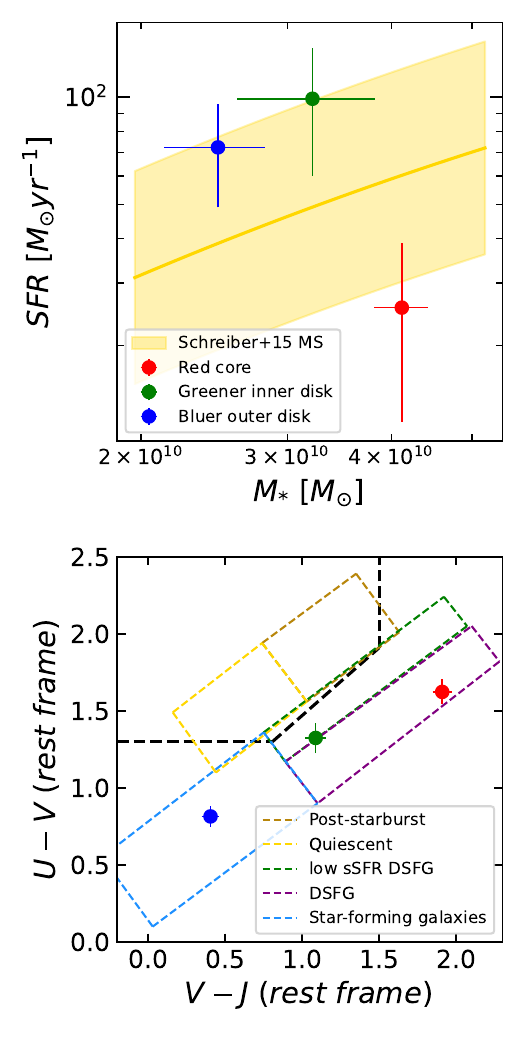}
    \caption{Classification of Galaxy ID15371 components as star-forming or quiescent. Upper panel: Galaxy ID15371 components plotted over the MS (\citealt{schreiber_herschel_2015, huang_exploring_2023}). Lower panel: Galaxy ID15371 components in the UVJ color-color diagram. Quiescent components should lie in the black dotted line delimited zone (\citealt{whitaker_newfirm_2011}). Depending on their properties, galaxy components should lie in one of the colored dotted lines delimited zones (\citealt{zick_mosdef_2018}). See Fig. \ref{fig:id15371} for the definition of the components.}
    \label{fig:res_id15371}
\end{figure}

We estimated that the good quality of the photometry in the rest-frame near-IR and the two-pass SED fitting procedure allowed CIGALE to give robust estimates of the stellar mass ($M_{*}$), the SFR and the dust attenuation ($A_{V}$) of each component. For the galaxy ID15371, in the upper panel of Fig. \ref{fig:res_id15371}, one can see the three components respective $M_{*}$ and SFR plotted on the  MS (\citealt{schreiber_herschel_2015, huang_exploring_2023}). All the components of the DSFG ID15371 have some ongoing star-formation, with the red core having a lower SFR compared to the disk components despite being more massive. Hence, for this galaxy, the disk is more efficient than the core at producing new stars.

Using the best SED models provided by CIGALE, we also estimated the rest-frame U, V and J AB magnitudes. We used \cite{apellaniz_recalibration_2006} U and V filters, and for the J band, we used the 2MASS J relative spectral response curve.
For the DSFG ID15371, we have confirmation in the UVJ color-color diagram that all the components are star-forming (lower panel of Fig. \ref{fig:res_id15371}). Moreover, the three components are aligned on the lower-left to upper-right diagonal of the diagram, which is the signature of a positive gradient of dust attenuation from the outskirts toward the center of the galaxy (\citealt{calzetti_dust_2000}). Indeed, from the SED fitting, we had $A_{V,blue} = 0.75 \pm 0.11  < A_{V,green} = 2.09 \pm 0.23 < A_{V,red} = 2.70 \pm0.11$.

\begin{figure}[htb]
    \centering
    \resizebox{\hsize}{!}{
    \includegraphics{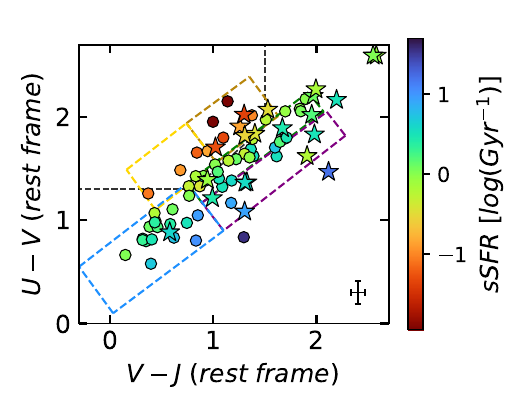}}
    \caption{UVJ color-color diagram. Each marker represents a component of a galaxy. The star-shaped markers are cores or bulges and circular markers are disk components. The color of the marker depends on the sSFR of the considered component. The error bar in the lower-right corner shows the average uncertainty. Quiescent components lie in the black dotted line delimited zone (\citealt{whitaker_newfirm_2011}). Post-starbursts-like components lie in the orange dotted line delimited zone, the quiescent-like in the yellow one, the low sSFR DSFG-like in green one, the DSFG-like in the magenta one, and the star-forming components lie in the blue dotted line delimited zone (\citealt{zick_mosdef_2018}). }
    \label{fig:uvj_ssfr}
\end{figure}

In Fig. \ref{fig:uvj_ssfr}, we display all the components of every galaxies on the $(V-J$, $U-V)$ plane. We recover the sSFR effect: when moving from the lower-right corner to the upper-left corner, the sSFR decreases (\citealt{wang_uvi_2017}). This makes the UVJ color-color diagram ideal to separate star-forming galaxies from quiescent galaxies. We note that the galaxies with sSFR $\lesssim 0.1$Gyr$^{-1}$ are all in the quiescent region defined by \cite{whitaker_newfirm_2011} and delimited by the black dashed line in the Figure. The colored dotted lines delimit the regions defined by \cite{zick_mosdef_2018}. The good agreement between our results and the regions defined in the literature as well as the observed gradients validate our SED fitting procedure.

During our SED fitting procedure, we based the star-forming/quiescent characterization on three indicators. The first one was the UVJ color-color diagram to see if the component was in the quiescent quadrant. The second indicator was the position relative to the MS to see if the component was on or above the MS (SF) or well below it (quiescent) and what was its position compared to the other components of the same galaxy, if the component was at least $\sim 0.6$dex below the other components (like the core of DSFG ID15371, see Fig \ref{fig:res_id15371}), it was also considered quiescent. The last indicator was the value of $t/\tau$, as we used a simple exponential model for the SFH, if $t/\tau \gg 1$, then the peak of star-formation is firmly in the past, and the component is likely on its way to quenching, on the contrary, if $t/\tau \lesssim 1$, the galaxy is still actively star-forming.

The consequence of this separation is that some components that we classified as quiescent were not completely passively evolving and could still be slowly star forming. 

Most of the time, the three indicators were in agreement, however, in some cases the results were ambiguous: the regions where all three indicators were not in agreement represented less than 5\% of all the studied regions, the central component of ID15371 is one of these regions (see Figs. \ref{fig:sed_params} and \ref{fig:res_id15371}). The inconsistency always came from the position of some components in the UVJ color-color diagram (classifying them as SF) compared to their position relative to the MS and their $t/\tau$ value (classifying them as quiescent). The inconsistency of the UVJ color-color diagram can be explained in several ways: the UVJ diagram uses only a part of the information (3 rest-frame bands) contrary to the other probes that uses the full SED. More importantly, real situations exist where the UVJ diagram characterizes correctly the presence of star formation but this star formation is suppressed as exemplified by the sub-MS location (suppressed with respect to the ensemble average given the mass) and by the $t/\tau$ (suppressed with respect to the past SFH of this galaxy). In the rare cases where the $t/\tau$ value did not allow for any conclusion ($t/\tau \sim 1$), we decided based on the MS and the UVJ color-color diagram that were consistently pointing either toward star-formation or quiescence. Hence, in all the cases, we were able to classify the regions as quiescent or star-forming.

To tackle the great diversity of galaxies, we decided to divide them into several classes as defined in the next Section.

\subsection{Classification} \label{subsec:class}

As a result of this analysis, we had 22 vastly different galaxies with various morphologies, colors (see Figs. \ref{fig:cutouts_I}, \ref{fig:cutouts_II} and \ref{fig:cutouts_III}) and resolved star-formation activity. 

The aim of this study is to obtain a better understanding of the formation and evolution of FIR bright DSFGs. One of the most effective way to reach this goal is to look at resolved properties of these galaxies, and especially where the star-formation is taking place. Indeed, by knowing where the quenching and star-formation happen, we have great clues to understand how these galaxies form and evolve.
We found that the variety of features we observe could be meaningfully re-conducted to three galaxy groups. The first group, that we call ``Type I" galaxies are \textit{Star-forming disks with a red star-forming core}, characterized by the fact that all their regions are SF. Some have a complex multi-color clumpy disk morphology in the RGB (F115W, F200W, F444W) image. They all have a dust attenuated red star-forming core. The second group, that we call ``Type II" galaxies are \textit{Quenched disks with a star-forming core}, characterized by a dust attenuated red star-forming core and a quenched disk (in one case, partially quenched). And the last group, that we call ``Type III" galaxies are \textit{Star-forming disks with a quenched bulge}, characterized by a quenched central bulge, while the disk is still star-forming. These are similar to local spirals.

As expected by the selection criteria detailed in Sect. \ref{subsec:reject}, we did not have any fully quiescent galaxy in our sample. For the disks with several components, they usually were all either star forming or quiescent. There was only one galaxy (ID18278) where only a fraction of the disk was quiescent (green region in Fig. \ref{fig:cutouts_II}), we decided to include it to the Type II as the quiescent part encompasses 16\% of the disk stellar mass and could be considered as an early stage of quenching. Four galaxies hosts X-ray AGNs that do not dominate the FIR emissions; 1 is a Type I galaxy (ID30186), 2 are Type II galaxies (ID13098 and ID13776) and the last one is a Type III galaxy (ID23205) (\citealt{nandra_aegis-x_2015}).

\begin{figure}[htb]
    \centering
    \includegraphics[width=0.85\linewidth]{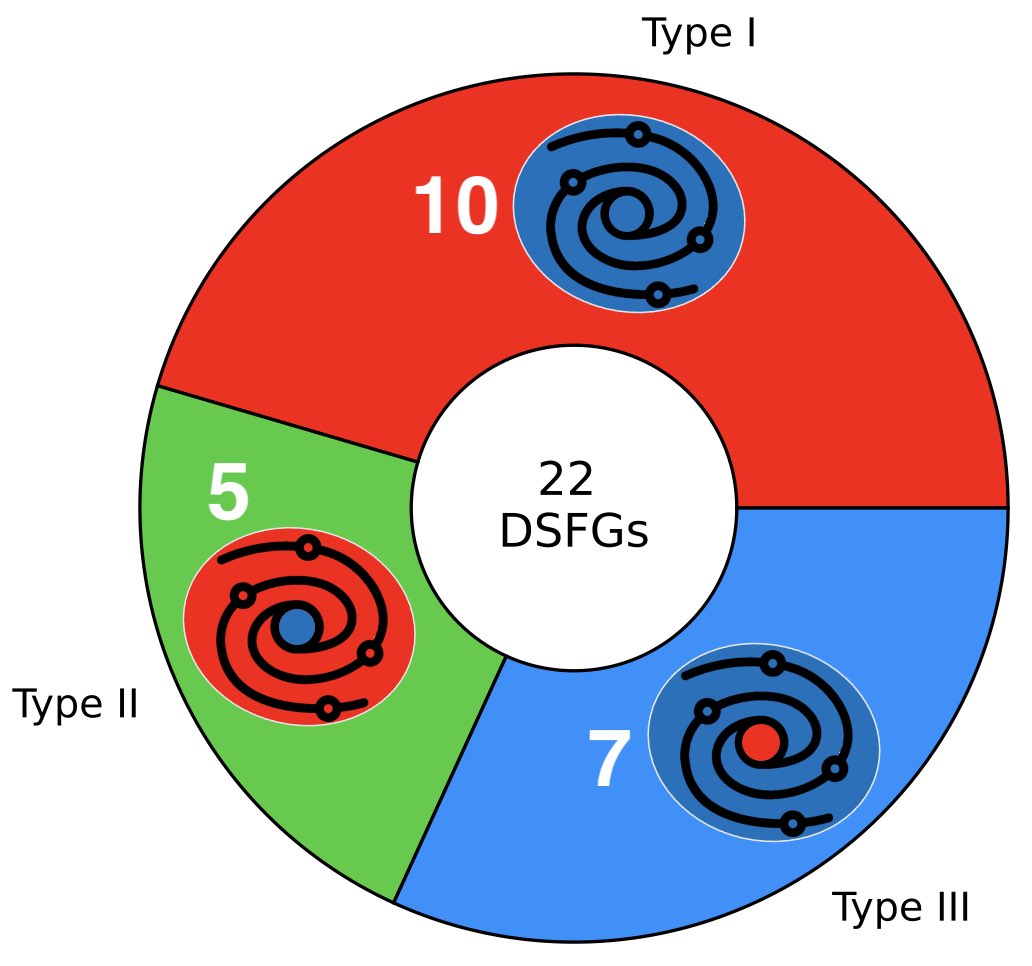}
    \caption{Distribution of the sample of galaxy in the different types based on their resolved star-forming activity. Each wedge size is proportional to the number of galaxies of each Type, written in white. The pictograms illustrate the properties of each Type, the blue and red colors representing star-forming and quiescent regions respectively. We link each Type of galaxy to a color that will be used throughout this paper defined by the wedges color.}
    \label{fig:camembert}
\end{figure}

\begin{figure}[htb]
    \centering
    \includegraphics[width=0.8\linewidth]{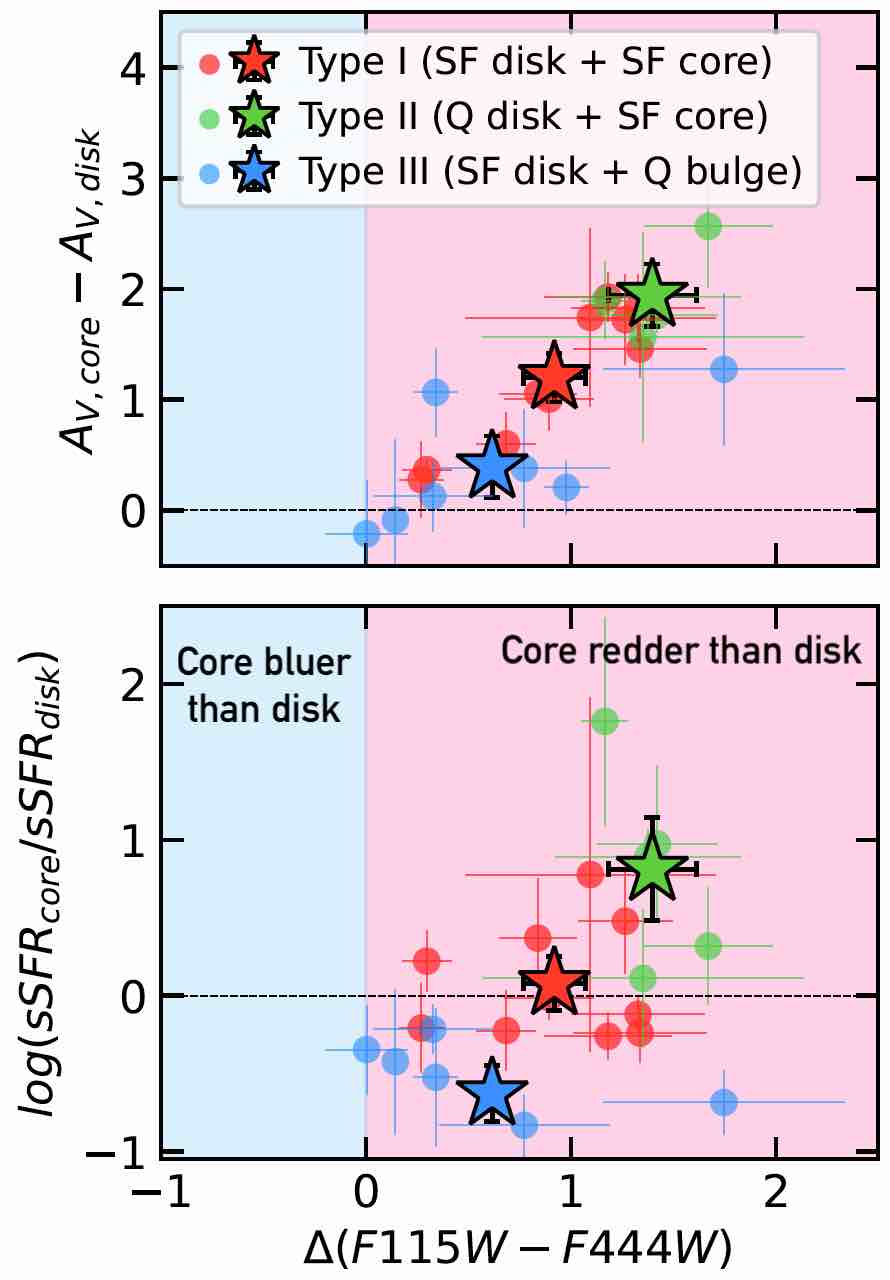}
    \caption{Investigating the main driver of color gradients. $A_{V}$ gradient (upper panel) and sSFR gradient (lower panel) versus color gradient (core - disk, in AB mag), red, green and blue markers are linked to the types defined in the top panel. Circular markers are individual galaxies, star markers are the mean values for each Type of galaxy with their associated error bar indicating the error of the mean.}
    \label{fig:color_grad}
\end{figure}

After having classified our sample of 22 galaxies, we had 10 Type I galaxies, 5 Type II and 7 Type III. The RGB cutouts of our sample are separated following the three types, with Figs. \ref{fig:cutouts_I}, \ref{fig:cutouts_II} and \ref{fig:cutouts_III} showing the Type I, II and III galaxies respectively. This is summarized in Fig. \ref{fig:camembert} where each wedge size is proportional to the number of galaxies of the considered Type. We illustrate each Type with a pictogram, the color red representing quiescent regions and the color blue representing star-forming regions. The color of each wedge is linked to the Type, in all Figures in the rest of this paper, the red markers will represent Type I galaxies, green markers Type II and blue markers Type III. 
Fig. \ref{fig:color_grad} summarizes the properties of each type by looking at the connection of sSFR and $A_{V}$ to color gradients (core - disk, in AB mag). The cores and bulges are systematically redder than disks and there is a strong correlation between the $A_{V}$ gradient and the color gradient (Pearson coefficient = 0.83, p-value = 2e-6), while there is no apparent correlation between the sSFR gradient and the color gradient (Pearson coefficient = 0.27, p-value = 0.23). This means that the color differences that we observe in Figs. \ref{fig:cutouts_I}, \ref{fig:cutouts_II} and \ref{fig:cutouts_III} mostly trace dust density in-homogeneities rather than older or younger stellar population, as it is usually the case with RGB images based on HST observations. This result is consistent with \cite{miller_early_2022} observations.
For example, the Type I galaxies (red markers in the Fig.\ref{fig:color_grad}) do not have a noticeable sSFR gradient ($sSFR_{core} \sim 1.2\times sSFR_{disk}$ on average), but have a strong $A_{V}$ gradient, hence, the fact that the cores of Type I galaxies appear much redder than the disks in Fig. \ref{fig:cutouts_I} is mostly due to their high dust density, while the blue regions are low $A_{V}$ regions. 
For the Type II galaxies, we observe the sSFR gradient we expected, the core is star-forming, while the disk is quenched ($sSFR_{core} \sim 6.5\times sSFR_{disk}$), they have the strongest dust gradient because of their highly dust attenuated core and their quenched disk that has low level of dust attenuation. We note that because of the sSFR gradient, we could expect the core to appear bluer than the disk (because of the younger stellar population in the core); however, we observed the exact opposite. This is because, as already mentioned, the color gradients we observe in Fig. \ref{fig:cutouts_II} are dominated by the dust attenuation gradient.
Eventually, Type III galaxies have low attenuation both in their quenched bulge and star-forming disk, hence have a weak $A_{V}$ gradient. However, their sSFR gradient is strong, and, as expected, of the opposite sign compared to Type II (quenched bulge and star-forming disk, $sSFR_{core} \sim 0.2\times sSFR_{disk}$ on average). In Fig. \ref{fig:cutouts_III} the color gradients mostly trace the age difference between the stellar populations of the (redder) bulge and the (bluer) disk.

\section{Results} \label{sec:results}

In this section, we present the results of the analysis of the 22 galaxies in our sample, distinguishing the three classes we just defined in the previous Section.
We first looked at the properties of the whole galaxies in Sect. \ref{subsec:res_whole} and then at the resolved properties at a sub-galactic level in Sect. \ref{subsec:res_subG}.
In Table \ref{tab:results}, we give the main properties of our sample of 22 galaxies. We note that the total masses and SFRs were computed by summing up the masses and SFRs of every galaxy components, naturally leading to higher uncertainties.

In the following, we compared the behavior of the different types of galaxies. To assess the significance of the trends, we compared the difference between the mean of a property for each Type with the error on the mean. We emphasis that we also checked the median value and we found that taking the mean or the median did not affect the observed trends. In the Figures, each star-shaped marker is the mean and the error bar is the error of the mean (defined as $err_{mean} = rms/\sqrt{N}$ with $rms$ the root mean square of the distribution and $N$ the number of galaxy in each Type).

\subsection{General properties}\label{subsec:res_whole}

\subsubsection{Main-sequence galaxies}\label{subsubsec:main_seq}

\begin{figure}[htb]
    \centering
    \resizebox{\hsize}{!}{
    \includegraphics{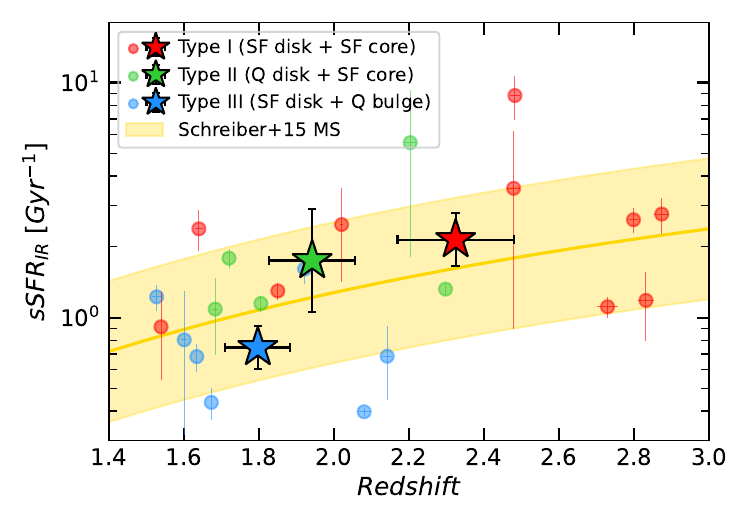}}
    \caption{$sSFR_{IR}$ versus redshift. The yellow shaded region is the MS (\citealt{schreiber_herschel_2015}). We show the error bars for individual galaxies. Circular markers are individual galaxies, star-shaped markers are the mean value for each Type of galaxy with their associated error bar indicating the error of the mean.}
    \label{fig:mass_sfr}
\end{figure}

To characterize the different types of galaxies, we first looked at their typical redshift, $M_{*}$ and $sSFR_{IR}$, as shown in Fig. \ref{fig:mass_sfr} and Table \ref{tab:masses}. The redshifts and $M_{*}$ were extracted from the SED fitting procedure described in Sect. \ref{subsec:sed}, while the $sSFR_{IR}$ was computed by dividing the $SFR_{IR}$ of each galaxy by the sum of the $M_{*}$ of each component with the $SFR_{IR}$ taken from the super-deblended catalog (Henry et al., in preparation).

All of our galaxies have a $M_{*} > 10^{10}$M$_{\odot}$ with an average of $M_{*} = 8.2^{+2.2}_{-1.7} \times 10^{10}$M$_{\odot}$. There is no correlation between the galaxy type and $M_{*}$; all three galaxy types have a similar $M_{*}$, on average (see Table \ref{tab:masses}). %We still could note that Type III galaxies have an average stellar mass slightly higher ($M_{*} = 10.8^{+4.0}_{-2.9} \times 10^{10} M_{\odot}$) than the others, the difference being $1\sigma$, it's still compatible with a weak trend (at most).
By comparing the $sSFR_{IR}$ of our galaxies with the MS of \cite{schreiber_herschel_2015} (see Fig. \ref{fig:mass_sfr}), we confirmed that typically these galaxies are MS galaxies, consistently with Fig. \ref{fig:general_prop} and Sect. \ref{subsec:reject}. The MS sSFR at a fixed redshift was calculated by taking the mean $M_{*}$ of our sample, which is ${<M_{*}>} = 10^{10.92}$M$_{\odot}$.

Moreover, it seems that redshift trend is appearing: the Type I galaxies with their star-forming core and star-forming disk are on average at higher redshift ($z= 2.32 \pm 0.15$) than the Type II galaxies with their star-forming core and quiescent disk ($z = 1.94 \pm 0.11$), that are themselves at a slightly higher redshift than the Type III galaxies ($z = 1.80 \pm 0.09$), analogs to the spiral galaxies we observe in the local universe with a quiescent bulge within a star-forming disk. However, our sample is not complete, meaning that if some galaxies of Type II or III (i.e., with some quiescent components) exist at $z > 2.5$ there is a high probability that, following our selection procedure, they would not make it into our sample. Additionally, the stellar masses between classes are similar and given the high overall SFR, the galaxies cannot be linked as progenitors-descendants as they would grow too quickly, even if the star-formation were to last only 100Myr.

The Type III galaxies, which have a quenched bulge, have the weakest $sSFR_{IR}$ on average ($sSFR_{IR} = 0.75^{+0.18}_{-0.14}$Gyr$^{-1}$ versus $sSFR_{IR} = 2.01^{+0.81}_{-0.56}$Gyr$^{-1}$ for Type I and II galaxies). Despite the incompleteness of our sample, the fact that (1) we do not observe any Type III galaxies with an $sSFR_{IR} > 2$Gyr$^{-1}$ and (2) we observe two times more Type I galaxies at $z>2$ than at lower redshift points toward the idea that Type I galaxies are typically at higher redshift, while Type III galaxies with a quenched bulge are likely to be more evolved than other classes.

\subsubsection{Galaxy near-IR sizes}

The presence of a highly obscured core at the center of most of our galaxies, (see Fig. \ref{fig:cut_id15371} of Fig. \ref{fig:comp_hst_jwst} for examples), could let us believe that we are studying the near-IR counterparts of the ALMA compact star-forming SMGs. Indeed, SMGs are known to be compact, dust obscured and with a high star formation efficiency (e.g., \citealt{puglisi_main_2019, franco_goods-alma_2020, gomez-guijarro_goods-alma_2022, gomez-guijarro_goods-alma_2022-1,puschnig_unveiling_2023}). To asses the possibility of a link between our sample of galaxies and SMGs, we decided to measure their sizes.
%Among others, \cite{hodge_alma_2019} suggested that these galaxies are actually the obscured part of a larger system. 

\begin{figure}[htb]
    \centering
    \resizebox{\hsize}{!}{
    \includegraphics{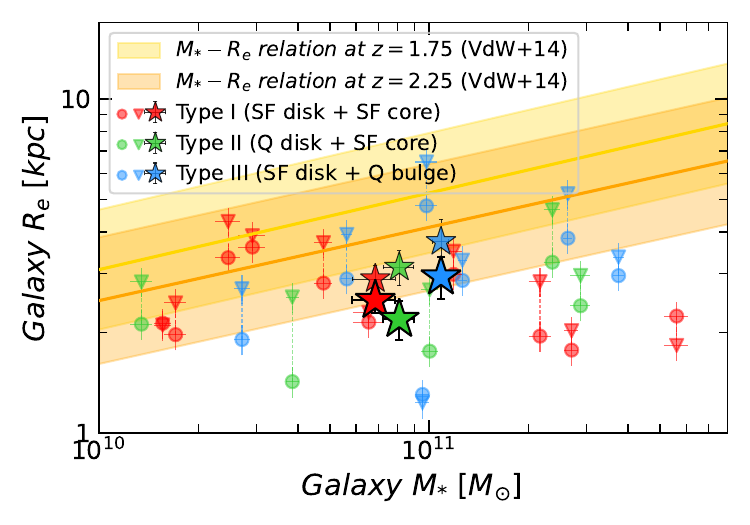}}
    \caption{Optical (triangle and smaller star markers) and near-IR (circle and larger star markers) half-light radius measured in the closest band to 550nm and $1.6\mu$m rest-frame, respectively, versus the total $M_{*}$ of the galaxy. Circular and triangular markers are individual galaxies, star markers are the mean value for each Type of galaxy with their associated error bar indicating the error of the mean. For each individual galaxy, we link their optical (triangles) and NIR sizes (circles) by a dashed line to illustrates the size evolution. The yellow and orange shaded regions illustrate the mass-size relation derived by \cite{van_der_wel_3d-hstcandels_2014} in the optical bands and at the redshift of our sample.}
    \label{fig:red_size}
\end{figure}

In Fig. \ref{fig:red_size}, we show the near-IR sizes ($R_{e,NIR}$) of our galaxies with circular markers. All our galaxies have a $R_{e,NIR}$ between 1kpc and 5kpc with an average of $R_{e,NIR} = 2.5 \pm 0.4$kpc. The $R_{e,NIR}$ seems to be independent from the stellar mass of the galaxy.
We still note that the Type II galaxies and their quiescent disk seem to be on average the most compact galaxies in the near-IR with a typical size of $2.19 \pm 0.30$kpc.

In this Fig., we also compare the $R_{e,NIR}$ of our galaxies to their optical sizes, shown with triangular markers. Most of our 22 galaxies are more compact in the near-IR than they are in the optical, that is, the triangles (optical size) are above the circles (NIR size) in Fig. \ref{fig:red_size}. This demonstrates than in our galaxies, the dust, traced by the near-IR emissions, is more concentrated than the stellar light, traced by the optical emissions.
This is a confirmation of an already well established fact (\citealt{van_der_wel_stellar_2023,gomez-guijarro_goods-alma_2022,jimenez-andrade_vla_2021,puglisi_main_2019,jimenez-andrade_radio_2019,fujimoto_demonstrating_2017}). Indeed, when looking at the $M_{*}-R_{e}$ relation from \cite{van_der_wel_3d-hstcandels_2014} based on rest-frame optical measurements, we found that  $\sim 40\%$ of the $R_{e,NIR}$ are below the $M_{*}-R_{e}$ relation scatter, while $> 80\%$ of the optical sizes of our galaxies are compatible with the $M_{*}-R_{e}$ relation. We note that at higher masses, the difference between the optical $M_{*}-R_{e}$ relation and the $R_{e,NIR}$ of our galaxies seems to get larger.

We note that the Type I galaxies have very comparable optical and near-IR sizes as the mean values for their optical and NIR sizes are compatible within error bars (red stars in Fig. \ref{fig:red_size}). Their star-forming core seem to not be as concentrated as those in the Type II galaxies. However, all these galaxies have a compact star-forming core, we discuss in Sect. \ref{subsubsec:alma} how the Type I and II galaxies might relate to the ALMA SMGs.

When looking at Figs. \ref{fig:cutouts_I}, \ref{fig:cutouts_II} and \ref{fig:cutouts_III}, it appears clearly that our galaxies are not symmetrical. Hence, studying their half-light radius is not enough to get an understanding of the mechanisms at play. We need to investigate their lopsidedness.

\subsubsection{Widespread lopsidedness} \label{subsec:asym}

As one can see in Figs. \ref{fig:cutouts_I}, \ref{fig:cutouts_II} and \ref{fig:cutouts_III}, some galaxies are strongly lopsided (marked with a `\textcircled{\small L}'). They are asymmetric or their red central region is off-centered with respect to the disk. This lopsidedness appears to be quite common among Type I and II galaxies. In Figs. \ref{fig:cutouts_I}, \ref{fig:cutouts_II} and \ref{fig:cutouts_III}, the marked galaxies are the 6 most lopsided galaxies, 3 are Type I (30\% of the group) and 3 are Type II (60\% of the group). The Type III galaxies look much more symmetric, these galaxies have a quenched bulge, they had presumably more time to evolve and stabilize their disk. 

\begin{figure}[htb]
    \centering
    \includegraphics[width=0.5\linewidth]{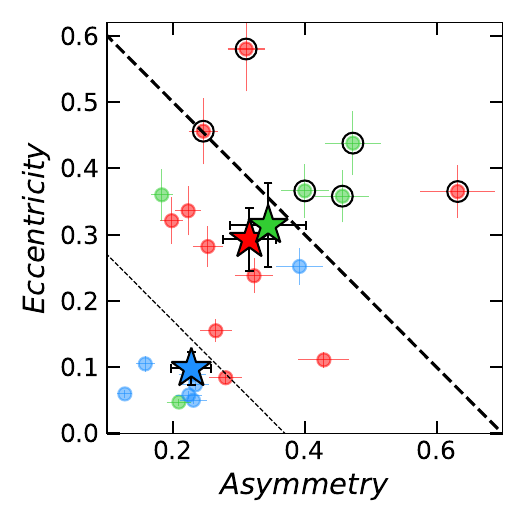}
    \resizebox{\hsize}{!}{
    \includegraphics{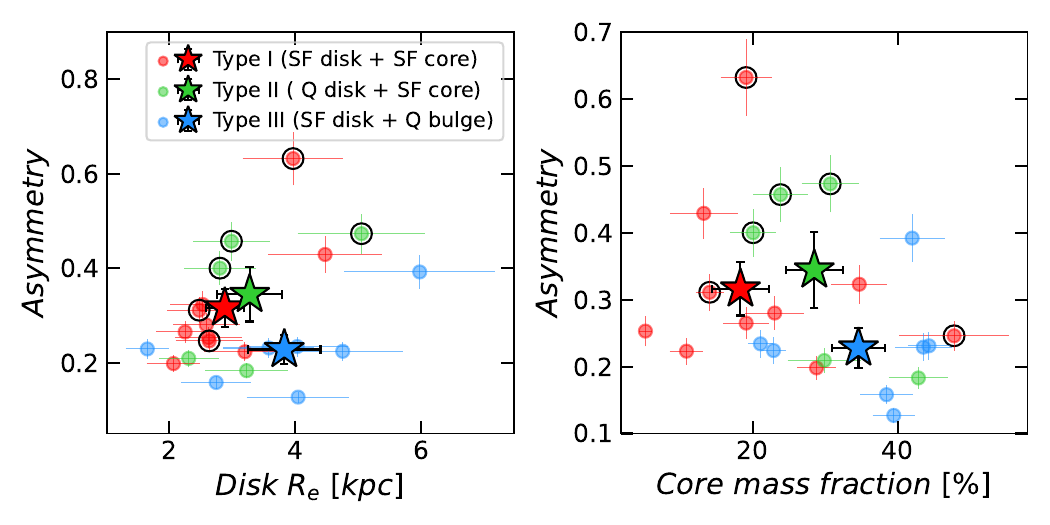}}
    \caption{Eccentricity and asymmetry. Upper panel: Eccentricity versus asymmetry, markers with a black circle are the strongly lopsided galaxies (see Figs. \ref{fig:cutouts_I}and \ref{fig:cutouts_II}), the thin black dotted line delimits weakly lopsided galaxies, the thick black dashed line delimits strongly lopsided galaxies. Lower-left panel: Asymmetry versus disk half-light radius as defined in Sect. \ref{subsec:lop}. Lower-right panel: Asymmetry versus mass fraction in the core or bulge of the galaxy. Circular markers are individual galaxies, star markers are the mean value for each Type of galaxy with their associated error bar indicating the error of the mean.}
    \label{fig:ecc_asym}
\end{figure}

As explained in Sect. \ref{subsec:lop}, for each galaxy we calculated its asymmetry ($A$) and eccentricity ($E$). We show the results in Fig. \ref{fig:ecc_asym}. Type III galaxies appear to be much less lopsided, they have a low eccentricity ($9.8\pm 2.5 \%$) and asymmetry ($22.8\pm 3.0\%$), while Type I and II galaxies, which show comparable lopsidedness, tend to be much more asymmetric ($33.0\pm 3.5\%$) and off-centered ($30.3\pm 4.0\%$) (see upper panel of Fig. \ref{fig:ecc_asym}). The difference has a $4.3\sigma$ and $2.2\sigma$ significance for the eccentricity and asymmetry respectively. In the upper panel of Fig. \ref{fig:ecc_asym}, we show the eccentricity versus the asymmetry. Since the Type III galaxies have, on average, the smallest level of lopsidedness, probably linked to a quiescent bulge growth, we decided to consider them as a proxy for systematic effects. We will consider a galaxy strongly lopsided if it has a lopsidedness significantly larger than an average Type III galaxy. The thin black dotted line shows the threshold to define a galaxy as weakly lopsided ($A + E > 0.37$, this value corresponds to the average $[A+E]+1\sigma$ of Type III galaxies). We have 14 galaxies that are at least weakly lopsided, representing 64\% of the sample. If the galaxies are above the thick black dashed line, meaning that $A+E>0.70$ (this value corresponds to the average $2\times [A+E]+1\sigma$ of Type III galaxies), we consider them as strongly lopsided, we encircled them in Fig. \ref{fig:ecc_asym}, they correspond to the \textcircled{L} marked galaxies in Figs. \ref{fig:cutouts_I} and  \ref{fig:cutouts_II} previously mentioned. We have 6 strongly lopsided galaxies, representing 27\% of our sample. Moreover, the position of the average lopsidedness of Type I and II galaxies in the upper panel of Fig. \ref{fig:ecc_asym}, indicates that being lopsided might be a typical property of these galaxies.

Usually, a strong asymmetry is linked to a strong eccentricity. However, some galaxies with a low level of asymmetry have a highly off-centered disk but we observe a lack of strong asymmetry with low eccentricity.

In the lower panels of Fig. \ref{fig:ecc_asym}, we show the distribution of the asymmetry versus the disk size as defined in Sect. \ref{subsec:lop} and versus the core mass fraction. We observe (1) a lack of galaxies with a compact disk and high asymmetry and vice-versa, (2) a lack of galaxies with a high core mass fraction and high asymmetry and vice-versa and (3) the Type III galaxies, with their low level of asymmetry, are also the ones with the highest core mass fraction on average. This is consistent with the built-up of a quenched bulge.
These observations are consistent with the observation of galaxies in the local universe. Indeed, present-day late-type galaxies with more extended disks and lower central stellar mass density are typically more lopsided than early-type galaxies with smaller disks and higher central stellar mass density (\citealt{dolfi_lopsidedness_2023,varela-lavin_lopsided_2023}).

By studying the general properties of our galaxies, we found that they mostly sit on the MS, are compact in the near-IR and a large fraction of those with a star-forming core is strongly lopsided. To go further in our analysis, we used the unique spatial resolution power of JWST, which allows access for the first time to the properties of sub-galactic regions for cosmic noon galaxies. This resolved properties are crucial to better understand and constrain the morphology and evolution of our DSFGs at cosmic noon.

\subsection{Resolved properties}\label{subsec:res_subG}

\begin{figure}[htb]
    \centering
    \resizebox{\hsize}{!}{
    \includegraphics{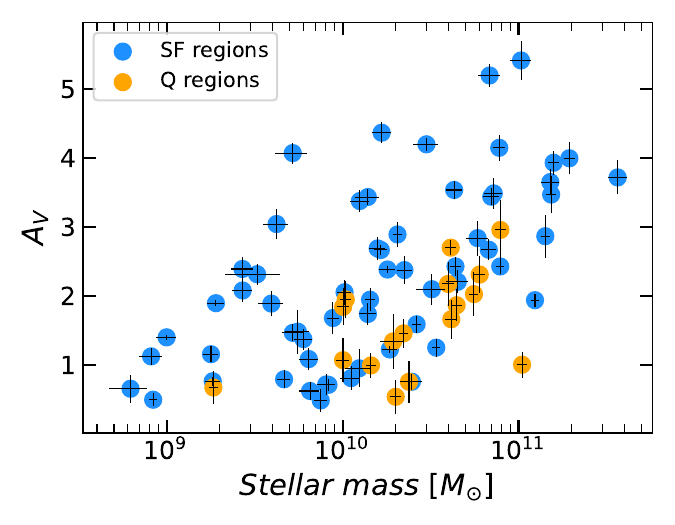}}
    \caption{$A_{V}$ versus $M_{*}$ for each region identified in Figs. \ref{fig:cutouts_I}, \ref{fig:cutouts_II}, and \ref{fig:cutouts_III}. Quiescent regions are in orange, and star-forming regions are in blue.}
    \label{fig:mass_av}
\end{figure}

For each galaxy, each component has been classified either as star-forming or quiescent (see Sect. \ref{subsec:class}). In Fig. \ref{fig:mass_av}, we show that the quiescent regions are massive ($M_{*} \gtrsim 10^{10}M_{\odot}$) and have a relatively low dust attenuation with an average of $A_{V} \sim 1.6$ and a maximum at $A_{V} \sim 3$, while star-forming regions have an average of $A_{V} \sim 2.3$ and a maximum at $A_{V} \sim 5.4$. The fact that quiescent components are less dust attenuated than star-forming components is consistent with the fact that they host a more evolved stellar population where the dust might have been consumed or destroyed.

The star-forming regions follow a correlation (with a Pearson coefficient of 0.62, p-value = 9e-8), the more massive components are more attenuated. This is consistent with the idea that the stellar mass is the main driver of dust attenuation in star-forming galaxies (\citealt{lorenz_updated_2023}).
We note that, even if the dust attenuation of quiescent regions is lower than for star-forming regions, it is still relatively high. Combined with the location of the quiescent regions in the UVJ color-color diagram (see Fig. \ref{fig:uvj_ssfr}), we conclude that some of these quiescent regions might have intrinsic post-starburst UVJ colors that are reddened. This would be consistent with a recent study based on HST data that found that most massive quiescent galaxies at $z \sim 2.4-3.3$ are Post-Starburst galaxies (\citealt{deugenio_typical_2020})

In the following Sections, we look at several resolved properties regarding the cores and bulges (in Sect. \ref{subsubsec:red}) and the disks (in Sect. \ref{subsubsec:uvj}) of our galaxies.

\begin{table}[ht]
    \caption{Average $M_{*}$ of the total and core or bulges of each type of galaxy in our sample.}
    \label{tab:masses}
    \centering
        \begin{tabular}{ccc}
            \hline \hline
            Type & Total $M_{*}$ & Core or Bulge $M_{*}$\\[2pt]
             & ($10^{10} M_{\odot}$) & ($10^{10} M_{\odot}$)\\
            \hline\\
            Type I & $6.86_{-3.19}^{+2.18}$ & $1.26_{-0.53}^{+0.92}$\\[8pt]
            Type II & $8.11_{-5.48}^{+3.27}$ & $2.60_{-1.19}^{+2.19}$\\[8pt]
            Type III & $10.85_{-3.96}^{+2.90}$ & $3.75_{-0.81}^{+1.04}$\\[5pt]
            \hline
        \end{tabular}
\end{table}

\subsubsection{Properties of cores and bulges} \label{subsubsec:red}

We first looked at the red central region of each galaxy, as defined in Figs. \ref{fig:cutouts_I}, \ref{fig:cutouts_II} and \ref{fig:cutouts_III}.

\begin{figure}[htb]
    \centering
    %\resizebox{\hsize}{!}{
    \includegraphics[width=1\linewidth]{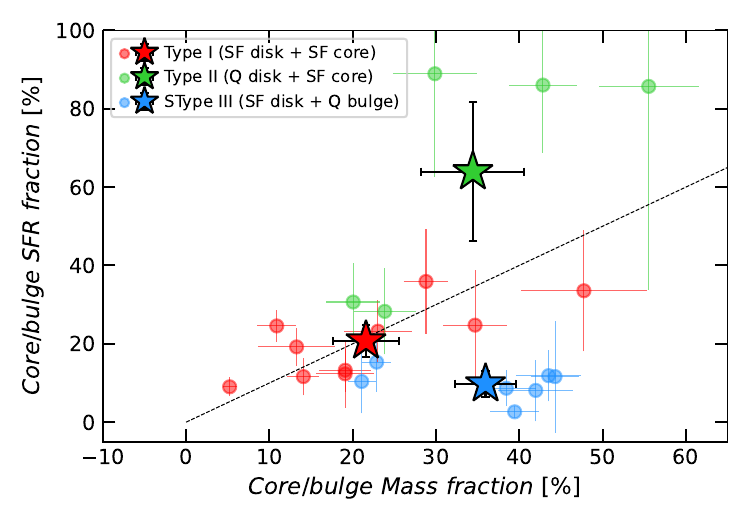}%}
    \caption{Star-forming rate fraction versus $M_{*}$ fraction in the red region. The black dotted line is the 1:1 correlation. Circular markers are individual galaxies, star markers are the mean value for each Type of galaxy with their associated error bar indicating the error of the mean.}
    \label{fig:red}
\end{figure}

We compared the $M_{*}$ and SFR fractions of the red cores and bulges with respect to their host galaxy (see Fig. \ref{fig:red}). As expected from the definition of our types of galaxies, the Type II galaxies have a well defined core with a SFR fraction ($64 \pm 18\%$) significantly greater than their $M_{*}$ fraction ($34.4 \pm 6.2\%$) since the disk is mostly quenched, while the Type III galaxies have a well defined bulge with a $M_{*}$ fraction ($35.9 \pm 3.6\%$) significantly more important than the SFR fraction ($9.8 \pm 3.4\%$) as the bulge is quenched. For Type I galaxies, the red core $M_{*}$ represents only $21.6 \pm 4.0\%$ of the total $M_{*}$ of the galaxy. This fraction is smaller than for the other galaxies of the sample where the central core or bulge represents a similar fraction of $\sim 35\%$ of their total $M_{*}$. This can be linked to the redshift trend, the Type I galaxies being at higher redshift on average, their core could still be at an early stage of growth compared to the other groups. This is also supported by the fact that the cores of Type I galaxies are less massive than the bulges of the more evolved Type III galaxies (See Table \ref{tab:masses}). It would also explain their low $R_{e,IR}/R_{e,O} = 0.89\pm 0.14$, as their $M_{*}$ is much less concentrated in the central region compared to the other two types.

\begin{figure}[htb]
    \centering
    \resizebox{\hsize}{!}{
    \includegraphics{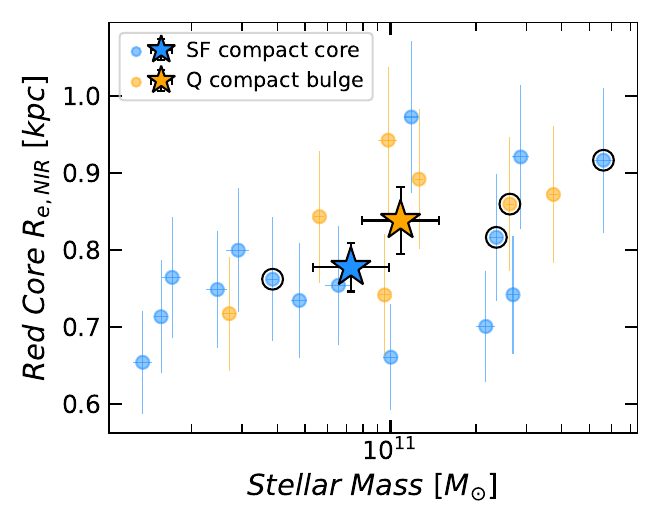}}
    \caption{Half-light radius of compact red cores and bulges versus the stellar mass of the galaxy. Circular markers are individual galaxies, star markers are the mean value for each group of galaxy with their associated error bar indicating the error of the mean. Markers with a black circle are cores hosting an X-ray AGN (\citealt{nandra_aegis-x_2015}).}
    \label{fig:red_clump}
\end{figure}

All of our galaxies have a central core or bulge appearing in the rest-frame near-IR (filter F410M or F444W). For some galaxies, the core or bulge has a clear clump-like morphology, while for others its less clear. We decided to investigate further these cores and bulges by dividing them in two categories, the star-forming cores (from Type I and II galaxies) and the quiescent bulges (from Type III galaxies). We measured the half-light radius of the cores and bulges defined as the radius of a circular aperture encompassing half of the flux of the core, we applied a similar technique as described in Sect. \ref{subsec:sizes} and show the results in Fig. \ref{fig:red_clump}. 

Even if the definition of the core is somehow arbitrary, and that there could be some level of contamination from the disk, the star-forming cores seem to be slightly more compact than the quiescent bulges ($0.78 \pm 0.03$kpc versus $0.84\pm 0.04$kpc, with a $\gtrsim 1\sigma$ significance). We note that the sizes of the star-forming compact cores are compatible with those measured in the sub-millimeter (See Sect. \ref{subsubsec:alma} for more details). In Fig. \ref{fig:red_clump}, the markers with a black circle are the compact cores with an X-ray detection (from \cite{nandra_aegis-x_2015}), possibly tracing an AGN. Three of them are found in star-forming cores including two in the most massive galaxies with the largest star-forming cores. These two AGNs in larger cores ($R_{e,NIR} > 0.8kpc$) are compatible with simulations from this \cite{cochrane_impact_2023}. They found that without AGN feedback, the star-forming core would undergo a compaction event, while the presence of AGN winds would prevent such compaction by evacuating the gas and precipitating the quenching of the core. However, the third star-forming core hosting an AGN is more compact ($R_{e,NIR} < 0.8kpc$), this is more compatible with \cite{ikarashi_very_2017} who found that the most compact cores of SMGs are those where there is both star formation and an AGN. So it seems that, depending on the galaxy, the presence of an AGN could possibly either act as a facilitator of the compaction or prevent it.

% \begin{figure}[htb]
%     \centering
%     \resizebox{\hsize}{!}{
%     \includegraphics[width=0.5\linewidth]{Figures/Figure_13a.jpg}
%     \includegraphics[width=0.5\linewidth]{Figures/Figure_13b.jpg}}
%     \includegraphics[width=0.5\linewidth]{Figures/Figure_13c.jpg}
%     \caption{Cutouts of a galaxy of each Type with the F444W filter. While a central clump-like core is clearly apparent for the ID13776 and ID23205, it is less clear in ID18278. We indicate the rest-frame central wavelength of the filter in parenthesis.}
%     \label{fig:red_im}
% \end{figure}

After analyzing the cores and bulges of our galaxies, we decided to investigate their differences with respect to the disk, especially the reasons of the redness of the core compared to the surroundings.

\subsubsection{Colors and clumps in the disks} \label{subsubsec:uvj}

\begin{figure}[htb]
    \centering
    \includegraphics[trim={0 7.8cm 0 0},clip, width=0.95\linewidth]{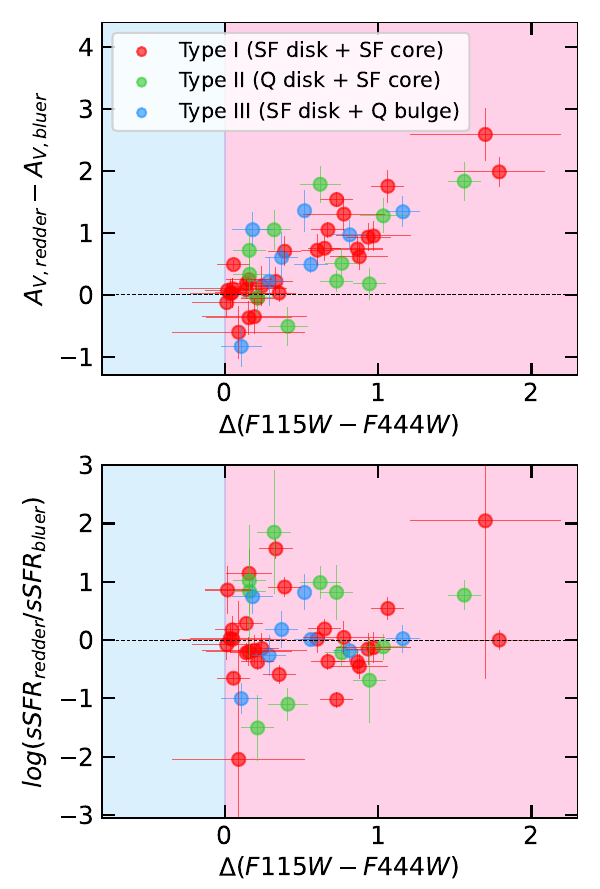}
    \caption{Investigating the main driver of color gradients in disks. Upper panel: Comparison between the $A_{V}$ and the (F115W - F44W) AB mag color between redder and bluer patches within the disks (if the disk has been divided in at least two components in Sect. \ref{subsec:flux}). Lower panel: Comparison between the sSFR and the (F115W - F44W) AB mag color between redder and bluer patches within the disks.}
    \label{fig:uvj_grad}
\end{figure}

In Sect. \ref{subsec:class}, we showed that the main driver of the color gradient between the cores and disks was the dust attenuation. When looking at Figs. \ref{fig:cutouts_I}, \ref{fig:cutouts_II} and \ref{fig:cutouts_III} we noticed that some disks are also highly in-homogeneous in terms of color. To investigate the physical processes responsible for these color variations, we compared the color of different patches in each disk and looked how it relates to the local dust attenuation and sSFR. This is similar to what we did in Sect. \ref{subsec:class} when we investigated the gradients between the cores and the disks. We emphasis that only variations are probed, not gradients, we do not look for radial effects as our galaxies have highly asymmetrical disks. When measuring the variations, we always measured the differences between a redder part of the disk and a bluer part (in other words $\Delta (F115W - F444W) > 0$ in AB mag). We compared all the components of the disks, meaning that if a disk was divided in 3 patches, there are 3 markers in Fig. \ref{fig:uvj_grad} comparing the first and second, second and third and first and third component. In the upper panel of Fig. \ref{fig:uvj_grad}, we identify a correlation between the color variations and the dust extinction variations (Pearson coefficient = 0.78, p-value = 6e-11) consistent with the expectation that the redder regions are those with the greatest $A_{V}$ (\citealt{calzetti_dust_2000}).
However, we did not identify any correlation between color variations and sSFR variations (Pearson coefficient = 0.16, p-value = 0.29, see lower panel of Fig. \ref{fig:uvj_grad}). This two observations demonstrate that the color variations we observe within the disks in Figs. \ref{fig:cutouts_I}, \ref{fig:cutouts_II} and \ref{fig:cutouts_III} are driven by difference in local dust attenuation. 
Indeed, if the color variation was driven by the SFR, the redder regions would always be less star-forming than bluer regions. This would translate into having all the points in the middle panel below 0 as  $sSFR_{redder} < sSFR_{bluer}$. This is not the case at all as half the points are above 0.
Hence, the NIRCam colors at $z \sim 2$ trace dust, red spots are highly extinct, while blue spots are weakly dust attenuated. This is consistent with previous studies based on the NIRCam images (e.g., \cite{miller_early_2022}) .
 
%One can note that we have a few patches for which the $A_{V}$ is greater in the blue patch than in the red patch. They represent $6-8\%$ of the color variations within disks. In the lower panel of Fig. \ref{fig:uvj_grad}, we show that for these regions, the color variations are consistent with the sSFR variations, all the markers in the $A_{V,redder} - A_{V,bluer} < 0$ region also are in the $sSFR_{redder}/sSFR_{bluer} < 1$ region. They correspond to comparison of patches with a similar $A_{V}$, but a strong sSFR variation. Hence, for this small number of patches, the color variations are affected by local sSFR variations.

\begin{figure}[htb]
    \centering
    \resizebox{\hsize}{!}{
    \includegraphics[width=0.5\linewidth]{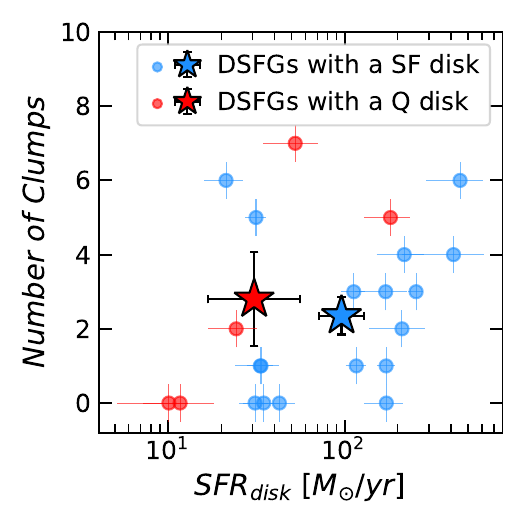}
    \includegraphics[width=0.5\linewidth]{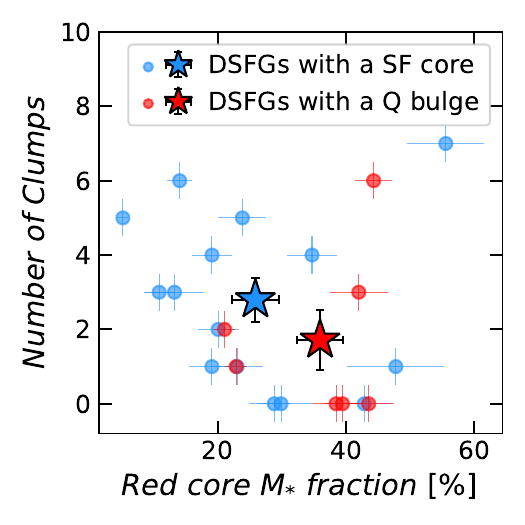}}
    \includegraphics[width=1\linewidth]{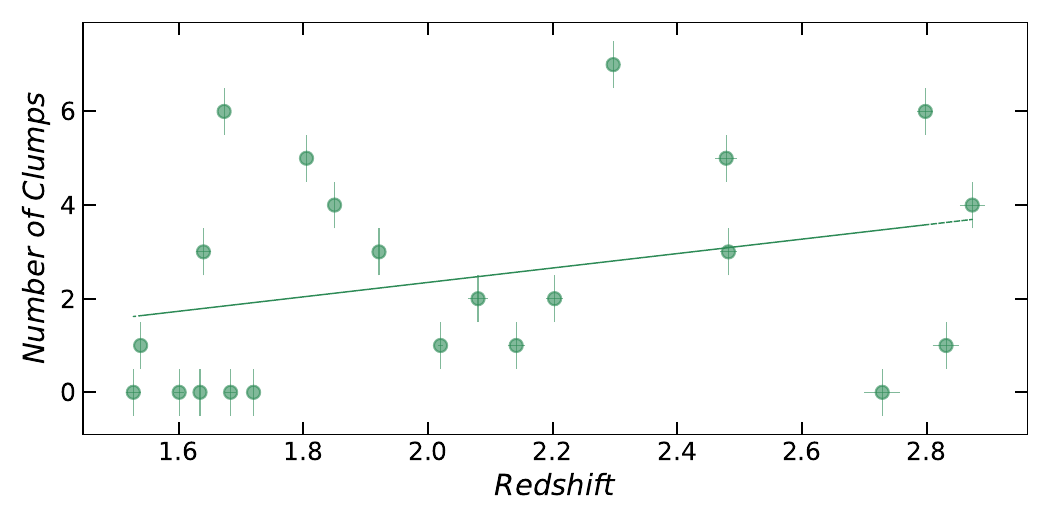}
    \caption{Distribution of the number of clumps in the disk versus its SFR (upper-left panel), the core stellar mass fraction (upper-right panel), and the redshift (lower panel). In the upper panels, galaxies are divided in two groups depending on the star-forming activity of their disk or of their core (left and right panel respectively). Circular markers are individual galaxies, star markers are the mean value for different groups of galaxies with their associated error bar indicating the error of the mean. In the lower panel, the line is a linear regression of the distribution showing an increase of the average number of clumps with redshift}
    \label{fig:clumps}
\end{figure}

As one can see in Figs. \ref{fig:cutouts_I}, \ref{fig:cutouts_II} and \ref{fig:cutouts_III}, some disks are very clumpy.
The clumpiness does not seem to be linked to a particular Type of galaxy. Most of the clumps are observed in the shortest wavelength, consistent with \cite{wuyts_smoother_2012} who state that the number of clumps decreases when moving toward longer wavelengths.
In Fig. \ref{fig:clumps}, we investigate the possible link between the clumpiness of disks and their SFR, the core mass fraction and the redshift of the galaxy. In the upper-left panel, we show the distribution of the number of clumps observed in each disk versus the SFR of the disk (defined as the sum of the SFR of the regions delimited in Figs. \ref{fig:cutouts_I}, \ref{fig:cutouts_II} and \ref{fig:cutouts_III}) separating the star-forming disks from the quiescent disks. We found no apparent correlation between the star-forming activity of the disk and the number of clumps. The fact that we observe clumps in quiescent disks (e.g., ID13107, ID18278 and ID13776 in Fig. \ref{fig:cutouts_II}) is quite surprising as they usually are supposed to be a place of local starburst (\citealt{wuyts_smoother_2012}). The fact that these clumps appear blue in these quiescent disks is mostly due to a low dust attenuation and not a high sSFR, as we showed. Indeed, for ID13776, the disk has $A_{V} = 1.0\pm 0.2$, while the central star-forming core has $A_{V} = 3.5\pm 0.2$. Moreover, we recall that when separating quiescent from star-forming regions, we decided to classify as \textit{quiescent} regions with a particularly low sSFR, even if they were not completely quenched. This could indicate that the clumps observed here could be the result of an unstable and fragmented  disk where the little star-formation remaining is concentrated in these low star-forming clumps, leading to an overall very low sSFR.

In the upper-right panel of Fig. \ref{fig:clumps}, we study the impact of the fraction of stellar mass in the core (in blue) or bulge (in red) on the number of clumps. The galaxies with a quiescent bulge, that we know to be more evolved (see previous Section), have a higher fraction of their mass in their bulge ($35.9\%\pm 3.6\%$) than the galaxies with star-formation in their core ($25.8\%\pm 3.7\%$) with a $\sim 2\sigma$ significance. They also tend to have a smaller number of clump: $1.7\pm 0.8$ clumps on average for a galaxy with a bulge and $2.8\pm 0.6$ clumps on average for a galaxy with a star-forming core ($1.1\sigma$ significance). The figure also shows that when looking at galaxies with a star-forming core (in blue in Fig. \ref{fig:clumps}), the ones with the smallest $M_{*}$ fraction at their core are also the clumpiest. We observe here the stabilizing effect of a growing core or bulge on the disk of a galaxy: a higher central $M_{*}$ fraction leads to a less clumpy disk. Moreover, we do not see any evidence of recent major mergers in our galaxies, suggesting that most of the clumps we observed originate from a fragmentation of a gas-rich, unstable star-forming disk, consistent with \cite{puschnig_unveiling_2023} and \cite{fensch_role_2021} that showed that large scale instabilities in gas-rich galaxies can create such star-forming giant molecular clumps.

In the lower panel of Fig. \ref{fig:clumps}, we show the evolution of the number of clumps. Even if the dispersion is quite large, the linear regression seem to show that, on average, higher redshift galaxies tend to have a larger number of clumps, with an average of $\sim 3.7$ clumps at $z = 3$ and of $\sim 1.6$ clumps at $z = 1.5$. One could argue that the fact that galaxies at higher redshift are clumpier comes from the fact that we probe shorter rest-frame wavelengths, leading to a higher probability of observing clumps (\citealt{wuyts_smoother_2012}). However, the redshift range we are probing here is quite narrow compared to the filters width, and the clumps that we count are the brightests and are visible in several filters. Hence, we estimate that the redshift trend could be real with an decrease of more than $50\%$ of visible clumps in the disks between $z = 3$ and $z = 1.5$. Confirming this trend would require a complete sample as we could be missing galaxies with a low number of clumps at higher redshift. If real, this trend suggests that the clumps either get destroyed within the disk and are not replaced by new clumps, or migrate toward the core and participate to its mass growth possibly triggering enhanced star formation.

%As we discuss in Sect. \ref{subsec:uvj}, the color variations within the disk is due to dust most of the time. But in 1/3rd of the cases, the color variation is due to the sSFR. We know that clumps are usually location of enhanced SF (\citealt{rujopakarn_alma_2019,wuyts_smoother_2012}), they could be partly responsible for these cases, at least in SF disks.

\section{Discussion} \label{sec:discussion}

In this Section, we first discuss the origins of the potential strong emission lines visible in the RGB cutouts in Figs. \ref{fig:cutouts_I}, \ref{fig:cutouts_II} and \ref{fig:cutouts_III} in Sect. \ref{subsec:em_line}. We then investigate how our results compare to previous knowledge from ALMA observations and simulations in Sect. \ref{subsec:prev_knowledge}. In Sect. \ref{subsec:lopsided}, we discuss the origin and evolutionary impact of lopsidedness on high redshift galaxies. Then, in Sect. \ref{subsec:dust}, we discuss how the fragmented disks could be linked to an early stage of bulge formation. Eventually, in Sect. \ref{subsec:quenching}, we discuss possible evolutionary paths that could explain the observation of outside-in quenching galaxies.

\subsection{Origins of potential bright emission lines} \label{subsec:em_line}

When looking at Figs. \ref{fig:cutouts_I}, \ref{fig:cutouts_II} and \ref{fig:cutouts_III}, one can notice that some of the disks have different colors, with a blue and a green part. 
The green clumps or patches are visible in all three types of galaxies. Considering their redshift, they probably are due to bright $H_{\alpha}$ or $[O_{III}]$ emission lines, which are known tracers of star formation. The $H_{\alpha}$ line will fall in the green filter (F200W) for galaxies with a redshift between 1.67 and 2.39, while for galaxies with a redshift between 2.52 and 3.47, it is the $[O_{III}]$ emission line that will fall in this filter. On the 7 galaxies where we identify green patches, 2 are consistent with $H_{\alpha}$ emission from a star-forming region (ID15371 and ID29608) and 3 are consistent with $[O_{III}]$ emission from a star-forming region (ID18694, ID23510 and ID23581). For the 2 remaining galaxies, it is more surprising, as the green patches or clumps are observed in the quiescent disks of Type II galaxies.

For the ID13107 galaxy ($z = 2.21 \pm 0.02$), the green patch is close to the center of the galaxy, it is then possible that the $H_{\alpha}$ line is produced by the accretion disk of an AGN sitting at the center of the galaxy that becomes bright in this region because of a much weaker dust attenuation than in the core. Even though we have no radio or X-ray signature of an AGN in this galaxy, as mentioned before, the predicted SFR from the SED fitting is not enough to explain the FIR flux density observed with \textit{Herschel} for this galaxy. This convinced us that there could be an AGN at the core of this galaxy.

For the ID18278 galaxy ($z = 1.805$), the situation is different, the green patch is in the outer region and composed of clumps. These clumps could have actually been ionized by the hot evolved low-mass stars (\citealt{cid_fernandes_comprehensive_2011,belli_kmos-3d_2017}) with an enhanced $H_{\alpha}$ line due to shocks from the minor merger. Indeed, these clumps are old (age of oldest stars $= 2.5 \pm 0.5Gyr$) and have a very low sSFR, consistent with the ex situ clumps defined in \cite{mandelker_population_2014}.

\subsection{Our results and the current DSFG evolutionary picture} \label{subsec:prev_knowledge}

\subsubsection{Compact ALMA SMGs near-IR counterparts} \label{subsubsec:alma}

In most of our IR-luminous galaxies, a central compact clump-like highly dust attenuated star-forming core is present. While it is nearly invisible in the optical rest-frame, it becomes bright in the near-IR (see Figs. \ref{fig:cut_id15371} and \ref{fig:comp_hst_jwst}). As we showed in the upper panel of Fig. \ref{fig:color_grad}, these cores are surrounded either by a star-forming (Type I) or a quiescent (Type II) disk with much lower dust attenuation. The presence of such dust-attenuated compact cores, could be linked to the compact SMGs observed with ALMA. 

To verify if our galaxies could be SMGs near-IR counterparts, we first measured the sizes of these compact star-forming cores. We found that the average $R_{e,NIR}$ was about 0.78kpc (Fig. \ref{fig:red_clump}). This size is compatible with the sizes measured with ALMA for the compact SMGs : $0.6\pm 0.2$kpc in \cite{zavala_probing_2022}, $\sim 0.73$kpc in \cite{gomez-guijarro_goods-alma_2022-1} or $1-2$kpc across in \cite{rujopakarn_alma_2019}. The NIRCam sizes tend to be slightly larger than the ALMA sizes, this is not due to a spatial resolution issue, but to the heavy dust obscuration of the cores. Then, since we know that compact SMGs at $z \sim 2-3$ are characterized by a SFR $\geq 100$M$_{\odot}$ yr$^{-1}$ (\citealt{gomez-guijarro_goods-alma_2022-1,jimenez-andrade_vla_2021,jimenez-andrade_radio_2019,hodge_alma_2019}), we looked at the SFR of our galaxies. We found that 11 out of the 15 galaxies with a star-forming core have a total SFR compatible with this criteria (see Table \ref{tab:results}) with 5 of them having SFR (or SFR$+\sigma$) $\gtrsim 100$M$_{\odot}$ yr$^{-1}$ in the core alone, 2 having SFR$+\sigma$ $\geq 90$M$_{\odot}$ yr$^{-1}$ in the core and the 4 remaining galaxies having a lower SFR in their core ($\sim 30-60$M$_{\odot}$ yr$^{-1}$). Hence, we identified 7 galaxies as possible SMGs counterpart in the near-IR. As a sanity check, we used the FIR super-deblended catalog in the EGS (Henry et al., in preparation.) to confirm that these galaxies are detected in the SCUBA2/850$\mu m$ band or predicted to be brighter than 1mJy at 1.1mm. We note that the 7 galaxies are equally distributed between Type I and II, we identified these galaxies in Table \ref{tab:results}.

Contrary to what is observed with ALMA, these compact cores are not isolated, they all are surrounded by a larger disk. The fact that there is a huge dust gradient between the core and the disk, as we showed in Sect. \ref{subsec:class} might explain why we do not this the latter in sub-millimeter surveys: the core is bright in the rest-frame near-IR, while the disk is bright in the rest-frame optical. The presence of a disk in all our galaxies confirms the findings of \cite{hodge_alma_2019} and \cite{puglisi_main_2019} who both stated that the compact SMGs could be obscured part of a larger system.

% The fact that some galaxies in our sample have highly extinct cores could link them the so-called HST-dark galaxies. We compared our sample with the HST-dark and HST-faint galaxies in the same field from \cite{perez-gonzalez_ceers_2023}. Our galaxies are in general agreement with the SFG at $z < 4$ in \cite{perez-gonzalez_ceers_2023}, especially with the fact that we observe highly dusty patches out to large radii. Four of the galaxies in our sample are classified as HST-faint (ID16544, ID18694, ID23581 and ID26188).
% All are Type I galaxies, which seems logical because quiescent regions have lower $A_{V}$, hence are brighter in HST. One of them (ID23581) has $A_{V,min} > 3$, hence expected to be HST faint/dark, while the remaining three galaxies have $A_{V,min} \sim 1.5$, which is the average $A_{V,min}$ of the sample. It is more surprising that those three galaxies are HST faint/dark.
% However, these 4 galaxies are actually the galaxies at the highest redshift of the sample ($2.7 < z < 2.9$), with photometric redshift from our SED fitting procedure consistent with the ones from \cite{kodra_optimized_2022} and the ones from the super-deblending (Henry et. al, in preparation). 
% There is a chance that their HST faintness comes more from their higher redshift than their high level of dust (at least for 3 of them).

\subsubsection{Relation to blue nugget simulations}

In the cosmological simulations from \cite{lapiner_wet_2023}, the typical high-redshift and low-mass galaxy is a gas-rich, star-forming, highly perturbed, and possibly rotating system, fed by intense streams from the cosmic web. When the stellar mass is in the ballpark of $\sim 10^{10}$M$_{\odot}$, the galaxy undergoes a major, last, wet compaction into a `Blue Nugget', starting with a compact gaseous star-forming system that rapidly turns into a compact stellar system.
The galaxies that we observe are all above this $\sim 10^{10}$M$_{\odot}$ threshold. However, none of them look like a blue nugget (except possibly ID13098), indeed, if they all have a compact dusty star forming core, they also have a much larger star-forming disk, incompatible with a blue nugget.
A possible explanation of this discrepancy could be that the galaxies are undergoing a rejuvenation event after a blue nugget phase as it is suggested by \cite{lapiner_wet_2023}. However, when comparing the $t_{50}$ of the disks and cores, we found no evidence that the star-forming disks were younger than the cores. If blue nuggets actually exist, the fact that we did not observe any might be due to their low-mass, or low SFR. It also possible that the previous observations were not deep enough to detect the low-luminosity disks. Finally, it may be possible that the most massive galaxies undergo a different quenching mechanism that lower-mass galaxies.

%Thanks to the NIRCam images, we have a view of the entire galaxies compared to ALMA that only see the dusty SF core. We are also able to spatially resolve them in the rest-frame UV and optical up to $z \sim 3$ down to low luminosity, which was impossible using HST. By looking at the morphologies of Type I and II galaxies, we may find some clues on the origin of the dusty SF core. Most of them have a lopsided clumpy disk (see Fig. \ref{fig:cutouts_II} and \ref{fig:cutouts_I}). Since the clumpiness is usually a result of VDI (\citealt{puschnig_unveiling_2023}), this favors a scenario where these clumps migrate, fuelling gas into the core, triggering enhanced star-formation, and as the star-formation rate increases, the core grows in mass and gets more compact due to the deeper gravitational potential well and the absence of AGN feedback (\citealt{cochrane_impact_2023}).

%However, other SMGs counterparts candidates have a smooth clumpless symmetric disk. This favors the scenario of a major merger that could have happened in the past resulting in a compact core at the center of mass. As the features created by a major merger are short lived, we do not necessarily expect to see direct signatures of a recent merger.

\subsection{Origin of lopsidedness in cosmic noon DSFGs} \label{subsec:lopsided}

Galaxy lopsidedness has not so far attracted much attention at high redshift, probably because of a lack of spatial resolution or incomplete data since the most obscured part of the galaxies are not visible with pre-JWST telescopes. However, the spatial resolution of the NIRCam shows that it is a common feature of DSFGs around cosmic noon. Indeed, we showed in Sect. \ref{subsec:asym} that being lopsided seem to be the typical morphology of Type I and II galaxies (see Figs. \ref{fig:cutouts_I}, \ref{fig:cutouts_II} and \ref{fig:ecc_asym}).
\cite{bournaud_lopsided_2005} investigated the origins of lopsidedness in field galaxies and concluded that it is very unlikely the result of internal mechanisms but rather linked to the history and environment of the galaxies. With the NIRCam images, we have access to the spatially resolved morphology of these galaxies, and can try to better understand the origin of the lopsidedness.

Among the lopsided galaxies showed in Figs. \ref{fig:cutouts_I} and \ref{fig:cutouts_II}, some have a clear compact central core and a rather homogeneously colored disks (e.g., ID11887, ID13776), others are mostly clumpy galaxies with a less clearly defined core (e.g., ID18694, ID18278). For the first category, it seems that the galaxies have a stable disk, with no major merger features (confirming this would require kinematics). This means that the lopsidedness of these galaxies, is probably due to accretion and minor mergers. This accretion would be happening via streams of cold gas that would asymmetrically feed more generously one side of the galaxy, making it grow larger than the opposite side. Then, the presence of a large number of clumps in the disks could be explained by VDI triggered by the accretion-induced lopsidedness.

However, the fact that Type I galaxies have a star-forming disk and Type II a quiescent disk means that the properties of gas transport in Type I and Type II galaxies are different. In Type I galaxies, the disk acquires its gas via accretion streams or minor mergers and forms stars, but the gas also reaches the core, which is star forming as well. \cite{bournaud_lopsided_2005} showed via simulations that a strong lopsidedness could be the result of gas accretion if it is asymmetric enough and that the lopsidedness from accretion is relatively long-lived ($\sim 3$Gyr), hence easily observable. This has also been confirmed by a recent study based on the TNG50 simulation (\citealt{dolfi_lopsidedness_2023}) where they conclude that the lopsidedness in local galaxies originates from accretion over several Gyr. In Type II galaxies, on the other hand, while the gas keeps going to the core and keeps it SF, the disk is quenched. This would seem to suggest that the gas does not stay in the disk, but goes straight to the center. A possible explanation would be that Type II galaxies have larger inflows or very powerful outflows that blow away and/or shock the gas in the disk (confirming this would require spectroscopy). It could also be that in Type II galaxies the accreted gas has a more radial accretion, with little angular momentum and goes straight into the central regions. Or, for some reason, the gas rapidly looses its angular momentum and abandon the disk and falls into the center. This would, depending on the direction of accretion, feed the lopsidedness. This effect has already been suggested by \cite{kalita_bulge_2022} where they were able to link the lopsidedness of 3 galaxies at $z \sim 3$ in a dense environment to cold gas accretion using Lyman-$\alpha$ emissions. The strong lopsidedness of these galaxies, would then be a tracer of the point of impact of the accretion streams.

For the clumpier galaxies with a less clearly defined core, the disk is star-forming and not homogeneous in terms of dust attenuation. \cite{kannan_discs_2015} showed with simulations that gas-rich disks are able to survive major mergers and that the following enhanced star-formation is not entirely happening in the core of the galaxy, but a substantial fraction takes place in the disk too. This is compatible with our Type I galaxies, the fact that their star-forming disks are clumpy and heterogeneous in terms of dust and sSFR could be a signature of a recent major merger (\citealt{calabro_merger_2019}). Moreover, \cite{kannan_discs_2015} mention that the presence of a gas-rich disk contributes to reducing the efficiency of bulge formation, which is compatible with the non-compact core observed in some of these galaxies. 
Usually major mergers features are short lived, but the clumps we observe could be preserved due to Toomre instabilities. Indeed \cite{fensch_role_2021} showed, via simulations, that a galaxy with a gas fraction greater than 50\% will have strong disk instabilities leading to the formation of long-lived giant clumps and strong nuclear inflow affecting the structure of the galaxy and possibly introducing lopsidedness. It has already been observed in a local galaxy used as proxy for high redshift galaxies (\citealt{puschnig_unveiling_2023}). A major merger could then result in a clumpy galaxy with a perturbed structure, which is what we have in Fig. \ref{fig:cutouts_I} for some Type I galaxies. The color variations between clumps or regions in the galaxies could be tracers of the original galaxy they were a part of before the merging as they trace the dust attenuation. However, we note that a major merger is not necessarily required, indeed, \cite{rujopakarn_jwst_2023} studied a lopsided galaxy at $z\sim 3$ and concluded that its lopsidedness did not originate from interaction with the environment but from internal, large scale instabilities, that could, in the end, form bars or spiral arms.

% The lopsidedness of these galaxies could also be the signature of the bulge angular momentum build-up. Indeed, either via accretion, minor mergers, major mergers, internal instabilities and tidal effects, the lopsidedness will break the disk balance, consequently creating a torque on the bulge of the galaxy resulting in an angular momentum loss.

The significance of the difference of lopsidedness between Type III galaxies and the rest of the sample means that, by some mechanism, galaxies tend to become more symmetric after cosmic noon. Indeed, we recall that below $z = 2$, we have two times more Type III galaxies than Type I galaxies (See Fig. \ref{fig:mass_sfr}). This could be due to increasing virialization with passing of time, and to the stabilising effect of the larger bulge mass fraction (see lower-right panel of Fig. \ref{fig:ecc_asym}).

\subsection{Fragmented disks and bulge growth}\label{subsec:dust}

In Sect. \ref{subsec:class} and \ref{subsubsec:uvj}, we demonstrated that the color variations observed in our galaxies was linked to $A_{V}$ variations. The fact that the core is much more attenuated than the disk is expected because the SFR surface density is higher in the core than it is in the disk, hence are the dust surface and column densities. However, the patchy distribution of dust within the disks is more surprising. 

% From the lower panel of Fig. \ref{fig:uvj_grad}, we observe a correlation between dust density and sSFR for Type II and Type III galaxies (Pearson coefficient = 0.62 and 0.83 with p-value = 0.04 and 0.01 respectively). Meaning that for these galaxies, the patches could be linked to not yet quenched regions in the disks of Type II galaxies and partly quenching disk for Type III. The patches could then find their origin in internal instabilities, or interactions with the local environment. For Type I galaxies, we do not observe this correlation (Pearson coefficient = 0.35, p-value = 0.07). For these galaxies, the patches could be correlated either to metallicity, higher metallicity leading to higher dust column, or to geometry. 

\begin{figure}[htb]
    \centering
    \resizebox{\hsize}{!}{
    \includegraphics{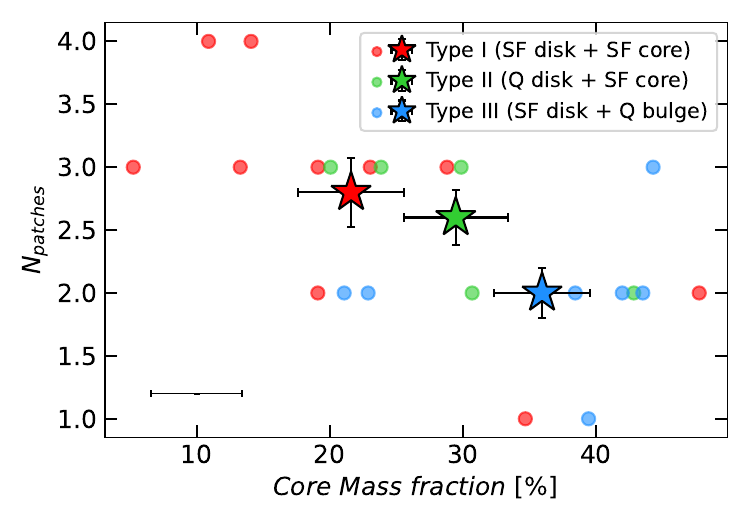}}
    \caption{Number of patches or components of each disk versus the stellar mass fraction in the core or bulge. In the lower-left corner, we show the average error bar for individual galaxies. Circular markers are individual galaxies, star markers are the mean value for each Type of galaxy with their associated error bar indicating the error of the mean.}
    \label{fig:npatch}
\end{figure}

We investigated the possible origin of the patchy distribution by looking for correlations between the greatest difference in $A_{V}$ in each disk and the redshift, the fraction of stellar mass in the core or bulge, the fraction of SFR in the core or bulge, the lopsidedness and the environment. We found no correlation (all p-value $> 0.2$). We then looked for a correlation between the number of components of each disk (as defined in Figs. \ref{fig:cutouts_I}, \ref{fig:cutouts_II} and \ref{fig:cutouts_III}) and the same parameters. The only correlation we found, that is visible in Fig. \ref{fig:npatch}, is with the mass fraction in the core (Pearson coefficient of -0.60, p-value = 0.003), the number of components gets smaller when the mass is more concentrated in the core of the galaxy. This is especially true for the galaxies with a star-forming core, Type I and II, which have a Pearson coefficient of -0.67 and a p-value of 0.006. Type III galaxies have a Pearson coefficient of 0.14 with p-value of 0.76. This correlation is consistent with the one we observed for the clumps (See Sect. \ref{subsubsec:uvj} and Fig. \ref{fig:clumps}). The fact the number of patches and clumps decreases at higher core mass fraction suggests that as the clumps migrate through the disk, they feed the central core and participate to the bulge growth, and, that as the central gravitational potential well gets deeper, the disk is stabilized, the VDI are destroyed, and the galaxy can have a smoother spiral-like disk. This scenario would be consistent with the simulations from \cite{hopkins_what_2023} that showed that a well defined dynamical center is necessary to stabilize the disk and put an end to bursty star-formation. We note that we are also in agreement with the new JWST results from \cite{kalita_bulge_2022}, pointing to an increased galaxy fragmentation with decreasing bulge and core mass fraction. Moreover, this correlation is consistent with the evolution of the lopsidedness discussed in the previous Section. Our observations tend to show that at higher redshift, DSFGs are typically lopsided, have a fragmented clumpy disk and seem to be in an early stage of bulge formation, while at lower redshift, they tend to get more symmetrical, less fragmented and clumpy and their core and bulge grows in mass, bringing stability to the disk.

However, if this (anti-)correlation justifies why we do not see patches in Type III galaxies, it does not  clear up the mystery of their origin. We would need spectroscopy to understand better what is happening in those disks, and even there the mystery would remain of why the disks are so in-homogeneous in dust attenuation, whether it is due to metallicity or geometry differences (and why these would persist over homogeneous patches within a disk, as opposed to simple radial gradients, for example). 

\subsection{Evolutionary paths for outside-in quenching galaxies} \label{subsec:quenching}

The Type II galaxies (see Sect. \ref{subsec:class} and Fig. \ref{fig:cutouts_II}) have an unusual behavior. They have a compact star-forming core embedded in a quiescent disk, and represent $\sim 23\%$ of the galaxies of our sample, so are relatively common. \cite{kalita_bulge_2022} studied such galaxies in a crowded environment at $z \sim 3$ and linked the quiescence of the disk to its strong lopsidedness, which rapidly fuels the gas to the core of the galaxy. In our sample of Type II galaxies, three have a strong lopsidedness, one is only weakly lopsided and has an off-centered core, while one is not lopsided at all. This means that even if lopsidedness can be a driver of outside-in quenching, it is not the only one.

Based on our observations, we have three possible scenarios that could explain the observed suppression of star-formation in the disk.

The first scenario is the one developed by \cite{kalita_bulge_2022} with the lopsidedness either coming from a major merger strong enough to result in this off-centered core or from asymmetric accretion of gas via streams and minor mergers, feeding the disk preferentially on one side. The strong lopsidedness resulting from this would be enough to explain the quenching of the disk would greatly facilitate the transportation of the gas toward the core (\citealt{fensch_role_2021}). The major merger event leading to a quenched disk has been observed in the local universe. Indeed, \citealt{chandar_arp_2023} demonstrated that the local ultra-luminous IR galaxy Arp220 is composed of a central starburst and a larger quiescent disk. They showed that the starburst has been triggered by a major merger. The galaxy is classified as shocked post-starburst galaxy, which is a stage prior to post-starburst. In that case, it appears that shocks induced by the merger forced the outer disk in this galaxy to turn quiescent.

The second scenario is a wet compaction event leading to an apparent outside-in quenching. ID13098 is in the correct range of stellar mass and redshift to be in a `blue nugget' phase (\citealt{lapiner_wet_2023,dekel_formation_2009, tacchella_evolution_2016}) where the galaxy undergo a wet compaction caused by gas-rich mergers or smoother gas streams, leading to an episode of high central star-formation and outside-in quenching. The presence of the low-luminosity quiescent disk might indicate that the compaction is not completely done yet. If it is a blue nugget, the outside-in quenching may not be final as when the gas has been consumed at the center and the bulge has grown, a star-forming ring can form in the disk via accretion of new gas-rich material from the inter-galactic medium leading to an inside-out quenching in the post-blue nugget phase.

The last scenario is an actual outside-in quenching linked to the strong lopsidedness but not resulting from a major merger. In Fig. \ref{fig:mass_sfr}, we showed that the Type I galaxies are the most star-forming and at the higher redshift on average. They also have a stellar mass consistent with the Type II galaxies. This means that there could be an evolutionary path between Type I and Type II galaxies driven by VDI and lopsidedness. The idea is that the star-forming clumps of the Type I galaxies could migrate toward the center of mass of the galaxy (\citealt{mandelker_population_2014}). By doing so, they would fuel strong gas nuclear inflow creating a compact star-forming core (\citealt{fensch_role_2021}). On their way to the center of the galaxy, the clumps would accrete the gas of the disk and could leave a completely gas deprived disk and a compact star-forming core. While growing, the star-forming core or bulge would prevent the formation of new clumps in the disk by stabilizing it (\citealt{hopkins_what_2023}), while the lopsidedness could be conserved due to the large scale instabilities. In this scenario, Type II galaxies would then be observed in a process of outside-in quenching.

While the ID13098 galaxy could be in a blue nugget phase, amongst the four remaining Type II galaxies, two have clumpy heterogeneous disks (ID13107 and ID18278, see Fig. \ref{fig:cutouts_II}), the different properties of the patches favors the idea of asymmetric accretion streams and minor mergers as the source of lopsidedness. The ID13776 galaxy has a clumpy and homogeneous disk, but highly off-centered. The eccentricity of this galaxy could originate from asymmetric accretion or from a major merger strong enough to shift the disk. In a similar way, it is hard to conclude for the last galaxy (ID21190), which is not lopsided and seems to have a smooth homogeneous disk. Its properties could favor the last scenario.

\subsection{The role of environment} \label{subsec:environment}

\begin{figure}[htb]
    \centering
    \resizebox{\hsize}{!}{
    \includegraphics{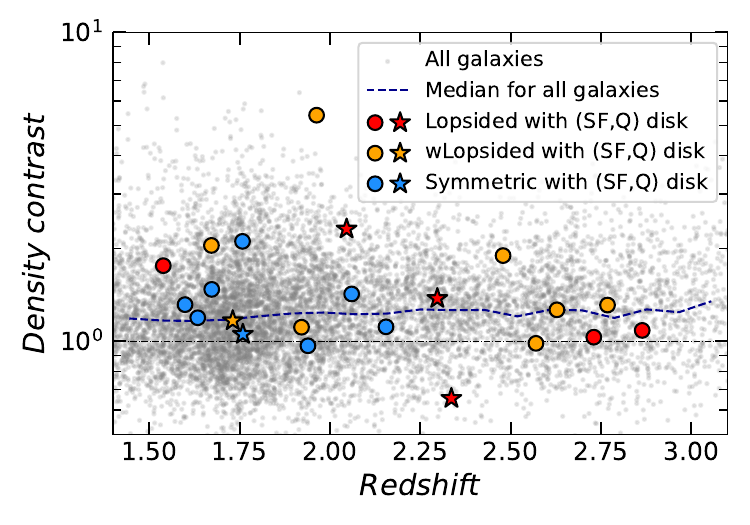}}
    \caption{Density contrast of galaxies versus redshift. The density contrast is defined as the number density enhancement with respect to the average density in the vicinity of the galaxy (\citealt{chartab_large-scale_2020}). The grey scatter is the general population of H-band mag AB $\leq 26$, the black dotted line is 1, the blue dashed line is the median density contrast of the general population in redshift bins. The star-shaped markers are galaxies undergoing outside-in quenching (Type II). Circular markers are the Type I and III galaxies (SF disk). The colors of the markers trace the lopsidedness.}
    \label{fig:env}
\end{figure}

A way to discriminate between the scenarios of outside-in quenching and the origin of the lopsidedness of galaxies could be to look at their local environment. To this aim, we used the environment density measurements from \cite{chartab_large-scale_2020}. They measure the density contrast of galaxies with a magnitude brighter than 26 AB mag in the H-band. The density contrast is defined as the number density enhancement with respect to the average density in the vicinity of the galaxy (local density/background density). In Fig. \ref{fig:env}, we compare the local density contrast of our sample with the general population of galaxies in the EGS field. The star markers are the Type II galaxies, undergoing outside-in quenching. They do not sit in any particular kind of environment, they are relatively close to the median of the general population showed by the blue dotted line. This seems to confirm that outside-in quenching can happen both in dense environment via major mergers and accretion but also in less dense environment via internal effects. The galaxy in the lowest density environment is ID13098 that we discuss in Sect. \ref{subsec:quenching}, probably in a blue nugget phase. The fact that this galaxy is relatively isolated favors the scenario of wet compaction as the origin of its outside-in quenching. For the other galaxies, the local density is insufficient to discriminate between scenarios as they do not sit in strongly over or under crowded environments.

The color of the markers in Fig. \ref{fig:env} traces the lopsidedness of the galaxies. There is no obvious difference between the lopsided galaxies and the general population. We do not see any signature that could link the lopsidedness to the local environment. The circular markers showing a weakly lopsided galaxy in a high density environment is ID30186. This galaxy is the brightest galaxy of a group of $\sim 16$ members at $z_{spec} = 1.85$, and is undergoing a major merger and is surrounded by quiescent intra-halo light (\citealt{coogan_z185_2023}).

Discriminating further between the different scenarios would require spatially resolved spectroscopy to study the kinematics of each of these galaxies, and especially of their disk, to see if it is rotating, which would favor accretion and minor mergers, or if they are dominated by dispersion velocity favoring the scenarios of major mergers and VDI.

\begin{table*}[ht]
    \caption{Main properties of our sample of 22 DSFGs.}
    \label{tab:results}
    \centering
        \begin{tabular}{cccccccc}
            \hline \hline
            ID & Redshift & Mass & Mass$_{core}$ & SFR & SFR$_{core}$ & Type & Morphology\\
            (CANDELS) & & $log_{10}(M_{*}/$M$_{\odot})$ & $log_{10}(M_{*}/$M$_{\odot})$ & M$_{\odot}$ yr$^{-1}$ & M$_{\odot}$ yr$^{-1}$ & & \\
            \hline
            11887 & $1.539$\tablefootmark{b} & $10.5 \pm 0.1$ & $9.59 \pm 0.07$ & $38 \pm 24$ & $5 \pm 3$ & Type I & Lop\\
            13098 & $1.720$\tablefootmark{a} & $10.6 \pm 0.1$ & $10.22 \pm 0.04$ & $73 \pm 63$ & $62 \pm 9$ & Type II & wLop\\
            13107 & $2.21 \pm 0.02$ & $10.1 \pm 0.1$ & $9.43 \pm 0.06$ & $29 \pm 21$ & $10 \pm 2$ & Type II & Lop\\
            13776 & $2.297$\tablefootmark{b} & $11.4 \pm 0.1$ & $10.86 \pm 0.05$ & $340 \pm 214$ & $114 \pm 64$ & Type II & SMGc+Lop\\
            13932 & $2.48 \pm 0.02$ & $10.2 \pm 0.1$ & $8.91 \pm 0.07$ & $35 \pm 20$ & $3 \pm 1$ & Type I & wLop\\
            14559 & $2.020 \pm 0.004$\tablefootmark{a} & $10.2 \pm 0.1$ & $9.59 \pm 0.07$ & $44 \pm 25$ & $10 \pm 3$ & Type I & ...\\
            15036 & $1.60 \pm 0.01$\tablefootmark{a} & $10.5 \pm 0.1$ & $10.02 \pm 0.03$ & $36 \pm 22$ & $3 \pm 2$ & Type III & ...\\
            15371 & $1.921$\tablefootmark{b} & $11.0 \pm 0.1$ & $10.61 \pm 0.03$ & $197 \pm 125$ & $26 \pm 13$ & Type III & wLop\\
            16544 & $2.73 \pm 0.03$\tablefootmark{a} & $11.4 \pm 0.1$ & $10.89 \pm 0.03$ & $269 \pm 186$ & $97 \pm 31$ & Type I & SMGc+wLop\\
            18278 & $1.805$\tablefootmark{a} & $11.5 \pm 0.1$ & $10.84 \pm 0.06$ & $254 \pm 168$ & $72 \pm 23$ & Type II & SMGc+Lop\\
            18694 & $2.80 \pm 0.01$ & $11.1 \pm 0.1$ & $10.22 \pm 0.05$ & $415 \pm 244$ & $52 \pm 9$ & Type I & Lop\\
            21190 & $1.68 \pm 0.01$\tablefootmark{a} & $11.0 \pm 0.1$ & $10.48 \pm 0.07$ & $106 \pm 95$ & $94 \pm 20$ & Type II & SMGc\\
            23205 & $2.14 \pm 0.01$\tablefootmark{a} & $11.40 \pm 0.05$ & $10.78 \pm 0.03$ & $138 \pm 87$ & $21 \pm 10$ & Type III & ...\\
            23510 & $2.51 \pm 0.02$\tablefootmark{a} & $10.4 \pm 0.2$ & $9.51 \pm 0.15$ & $141 \pm 106$ & $27 \pm 6$ & Type I & wLop\\
            23581 & $2.83 \pm 0.02$ & $11.3 \pm 0.1$ & $11.01 \pm 0.06$ & $259 \pm 152$ & $87 \pm 38$ & Type I & SMGc+Lop\\
            25604 & $1.673$\tablefootmark{b} & $11.1 \pm 0.1$ & $10.75 \pm 0.02$ & $24 \pm 18$ & $3 \pm 4$ & Type III & ...\\
            26188 & $2.87 \pm 0.02$ & $10.8 \pm 0.1$ & $10.10 \pm 0.05$ & $257 \pm 178$ & $34 \pm 11$ & Type I & wLop\\
            29608 & $1.64 \pm 0.03$ & $10.7 \pm 0.1$ & $9.72 \pm 0.09$ & $342 \pm 181$ & $84 \pm 11$ & Type I & SMGc+wLop\\
            30012 & $1.53 \pm 0.01$\tablefootmark{a} & $11.0 \pm 0.1$ & $10.62 \pm 0.03$ & $49 \pm 42$ & $6 \pm 3$ & Type III & ...\\
            30186 & $1.85$\tablefootmark{a} & $11.7 \pm 0.1$ & $11.29 \pm 0.03$ & $550 \pm 436$ & $136 \pm 57$ & Type I & SMGc+wLop\\
            31125 & $2.08 \pm 0.02$ & $11.60 \pm 0.05$ & $10.90 \pm 0.03$ & $237 \pm 156$ & $25 \pm 18$ & Type III & ...\\
            31281 & $1.634$\tablefootmark{b} & $10.80 \pm 0.05$ & $10.35 \pm 0.03$ & $36 \pm 35$ & $0.9 \pm 0.5$ & Type III & ...\\
            \hline
        \end{tabular}
        \tablefoot{\\
        \tablefoottext{a}{Grism based redshift.}\\
        \tablefoottext{b}{Spectroscopic redshift.}\\
        Type I: Star-forming disk and star-forming core; Type II: Star-forming disk and quiescent (Q) core; Type III: Quiescent core and star-forming disk\\
        ``SMGc'' and ``(w)Lop'' stand for sub-millimeter galaxy counterpart and (weakly) lopsided, respectively.\\}
\end{table*}

\section{Summary} \label{sec:summary}

In this paper, we used the new set of images in the near-IR from JWST/NIRCam in the EGS field from the CEERS collaboration to investigate the formation and evolution of DSFGs at cosmic noon. 
To start, we selected a sample of DSFGs based on their FIR emissions at around cosmic noon ($1.5 < z < 3.0$). We ended up with 22 galaxies in the CEERS field.
We studied each galaxy on a sub-galactic scale by dividing them into different regions based on their NIRCam (F115W, F200W, F444W) colors, taking advantage of the spatial resolution. Using the available photometry from HST and JWST, we ran an SED fitting and derived physical parameters for each galaxy component and classified the galaxies as star forming or quiescent.
We classified the galaxies into different types based on the star-forming activity in their core and disk. We defined Type I galaxies as those with a star-forming disk with a red star-forming core, Type II galaxies as those having a quiescent disk with a star-forming core, and Type III galaxies as those having a star-forming disk with a quenched bulge. We showed that the NIRCam color gradients, unlike the HST ones, are good tracers of the dust gradients, confirming previous studies.

The main results of this study are as follows:
\begin{itemize}
    \item We find that the DSFGs at a higher redshift tend to have a fragmented disk with patches or clumps of different RGB colors that are not radially symmetric. We showed that these galaxies also have a low core mass fraction, and thus they are at an early stage of bulge formation. Our results show that when moving to a lower redshift, the core mass fraction increases, and the bulge growth is associated with a stabilization of the disk, which translates into less patches and clumps in lower redshift galaxies. The NIRCam data clearly point toward bulge formation in preexisting disks.

    \item Sixty-four percent of our galaxies are at least weakly lopsided, and 27\% are strongly lopsided. The fact that such a large portion of our sample is lopsided demonstrates how important and common this feature is and that it should not be overlooked. The lopsidedness of DSFGs could have a major impact on their evolution by inducing VDI and triggering clump formation. It is very likely that the lopsidedness is caused by asymmetric cold gas accretion and/or minor mergers feeding preferentially one side of the disk, which would, depending on the orientation of the accretion, favor either a star-forming or a quiescent disk. Lopsidedness could also be triggered by a major merger disrupting the disk. By selection, our sample is biased toward the most dusty and star-forming galaxies. Hence, the selection of a complete sample of DSFGs, including lower SFR, would allow for testing of whether this property is common to all (D)SFGs or only to the most luminous ones.

    \item Twenty-three percent of the galaxies in our sample have a quiescent disk but a star-forming core. These galaxies were observed in the process of outside-in quenching. Although they are massive, they generally are incompatible with a ``blue nugget" resulting from wet compaction, as predicted by simulations. Their observed outside-in quenching could then find its origins in their strong lopsidedness that favors VDI and rapid transportation of gas toward the center of the galaxy or from large-scale instabilities and clump migration accreting the gas from the disk to feed it to the core.
    
    \item Most of the DSFGs in our sample have a deeply dust attenuated compact star-forming core that can represent most of the total SFR of the galaxy, while only accounting for a small fraction of its stellar mass. We found that half of our DSFGs with a red star-forming core were good SMG counterpart candidates. Our results then confirm that SMGs are usually part of a wider, less obscured system and demonstrate the necessity to combine sub-millimeter data with near-IR imaging to fully grasp the nature of DSFGs.

    \item Interestingly, among the quiescent disks, we find evidence of clump-like structures. These clumps are not (or are very weakly) star-forming, and they are mostly populated by old stars but seem to be too massive to be compared to the globular clusters we observed in the local universe.

\end{itemize}

This work demonstrates the power of JWST in probing, for the first time, spatially resolved galaxies in the near-IR at cosmic noon, where the only available data was the unresolved images from \textit{Spitzer}/IRAC. This allows for reliable studies of quenching and dust attenuation at sub-galactic scales in DSFGs, facilitating understanding of their morphologies and formation and evolution mechanisms, which appear to be more complex than previously thought.

\begin{acknowledgements}
CGG acknowledges support from CNES.
P.G.P.-G. acknowledges support from Spanish Ministerio de Ciencia e Innovación MCIN/AEI/10.13039/501100011033 through grant PGC2018-093499-B-I00.
The authors acknowledge the precious help from the referee who played a non-negligible role into improving the quality of this paper.
\end{acknowledgements}

\bibliographystyle{aa}
\bibliography{ULIRGS_CEERS}
\end{document}